\documentclass{aa}  

\usepackage{textcomp}
\usepackage{graphicx} % 
\usepackage{amsmath} % 
\usepackage{amssymb} % 
\usepackage{bm} % 
\usepackage{upgreek} %
\usepackage{IEEEtrantools} % 
\usepackage{multirow}
\usepackage[colorlinks=true,allcolors=blue,urlcolor=blue]{hyperref}
\usepackage{txfonts}
\usepackage[normalem]{ulem} 
\usepackage{multirow}
\usepackage[dvipsnames]{xcolor}

\newcommand{\sect}[1]{\text{Sect.~\ref{#1}}}
\newcommand{\fig}[1]{\text{Fig.~\ref{#1}}}
\newcommand{\tab}[1]{\text{Table~\ref{#1}}}

\newcommand{\multitd}{\textsc{multi3d}}
\newcommand{\codename}{\textsc{balder}}
\newcommand{\blue}{\textsc{blue}}
\newcommand{\mtd}{\textlangle3D\textrangle}
\newcommand{\marcs}{\textsc{marcs}}

\newcommand{\stagger}{\textsc{stagger}}
\newcommand{\atmo}{\textsc{atmo}}

\newcommand{\kms}{\mathrm{km\,s^{-1}}}
\newcommand{\teff}{T_{\mathrm{eff}}}
\newcommand{\lgg}{\log{g}}

\newcommand{\vsini}{\varv\,\sin\iota}

\newcommand{\feh}{\mathrm{\left[Fe/H\right]}}
\newcommand{\abrat}[2]{\mathrm{\left[#1/#2\right]}}

\newcommand{\lgt}{\log{\tau_{500}}}
\newcommand{\lgr}{\log{\tau_{\mathrm{R}}}}
\newcommand{\dex}{\mathrm{dex}}
\newcommand{\lyalpha}{\mathrm{Ly\upalpha}}
\newcommand{\halpha}{\mathrm{H\upalpha}}
\newcommand{\hbeta}{\mathrm{H\upbeta}}
\newcommand{\hgamma}{\mathrm{H\upgamma}}
\begin{document} 

\title{Effective temperature determinations
of late-type stars based on 3D non-LTE Balmer line formation}
\titlerunning{3D non-LTE Balmer line formation}
\author{A.~M.~Amarsi\inst{1}
\and
T.~Nordlander\inst{2,3}
\and
P.~S.~Barklem\inst{4}
\and
M.~Asplund\inst{2}
\and
R.~Collet\inst{5}
\and
K.~Lind\inst{1,4}}
\institute{Max Planck Institute f\"ur Astronomy, K\"onigstuhl 17, 
D-69117 Heidelberg, Germany\\
\email{amarsi@mpia.de}
\and
Research School of Astronomy and Astrophysics, 
Australian National University, Canberra, ACT 2611, Australia
\and
ARC Centre of Excellence for 
All Sky Astrophysics in 3 Dimensions (ASTRO 3D), Australia
\and
Theoretical Astrophysics, Department of Physics and Astronomy, 
Uppsala University, Box 516, SE-751 20 Uppsala, Sweden
\and
Stellar Astrophysics Centre, Department of Physics and Astronomy, Aarhus
University, Ny Munkegade 120, DK-8000 Aarhus C, Denmark}

\abstract{Hydrogen Balmer lines are commonly used as spectroscopic
effective temperature diagnostics of late-type stars.
However, reliable inferences require accurate model spectra,
and the absolute accuracy of 
classical methods that are based on
one-dimensional (1D) hydrostatic model atmospheres
and local thermodynamic equilibrium (LTE) is still unclear.
To investigate this, we carry out
3D non-LTE calculations for the Balmer lines, 
performed, for the first time, over
an extensive grid of 3D hydrodynamic \stagger~model atmospheres.
For $\halpha$, $\hbeta$, and $\hgamma$~we find significant
1D non-LTE versus 3D non-LTE differences (3D effects):
the outer wings tend to be stronger in 3D models, 
particularly for $\hgamma$, 
while the inner wings can be weaker in 3D models,
particularly for $\halpha$.
For $\halpha$, we also find significant 
3D LTE versus 3D non-LTE differences (non-LTE effects):
in warmer stars ($\teff\approx6500\,\mathrm{K}$)
the inner wings tend to be weaker in
non-LTE models, while at lower effective temperatures 
($\teff\approx4500\,\mathrm{K}$)
the inner wings can be stronger in non-LTE models; 
the non-LTE effects are more severe at lower metallicities.
We test our 3D non-LTE models 
against observations of well-studied benchmark stars. 
For the Sun, we infer concordant effective temperatures
from $\halpha$, $\hbeta$, and $\hgamma$; however the value
is too low by around $50\,\mathrm{K}$~which 
could signal residual modelling shortcomings.
For other benchmark stars, our 3D non-LTE models generally reproduce
the effective temperatures to within $1\sigma$~uncertainties. 
For $\halpha$, the absolute 3D effects and non-LTE effects
can separately reach around $100\,\mathrm{K}$, in terms of 
inferred effective temperatures. For metal-poor turn-off stars,
1D LTE models of $\halpha$~can underestimate effective temperatures
by around $150\,\mathrm{K}$.
Our 3D non-LTE model spectra are publicly available,
and can be used for more 
reliable spectroscopic effective temperature determinations.}

\keywords{radiative transfer --- line: formation --- line: profiles ---
stars: atmospheres --- stars: late-type}

\maketitle

%-------------------------------------------------------------------------------
\section{Introduction}
\label{introduction}

Late-type stars can be described by a number of
atmospheric parameters including effective temperature $\teff$,
surface gravity $\lgg$, and metallicity $\feh$\footnote{$\abrat{A}{B}=
\left(\log{N_{\mathrm{A}}}/\log{N_{\mathrm{B}}}\right)_{*}-
\left(\log{N_{\mathrm{A}}}/\log{N_{\mathrm{B}}}\right)_{\odot}$.}.
These parameters need to be precisely and accurately constrained
before reliable surface chemical compositions can be
obtained, and are needed to 
infer fundamental stellar parameters such as mass, radius, and age.
Furthermore, these fundamental parameters, 
together with chemical compositions, are essential to constrain,
for example, nucleosynthetic yields, 
Galaxy structure and formation, and Galactic chemical evolution.

For dwarfs and giants with $4500\lesssim\teff/\mathrm{K}\lesssim8000$, 
\ion{H}{I} Balmer lines are useful spectroscopic diagnostics for the
effective temperature
\citep[e.g.][]{1960AnAp...23..278C,1962ApJ...135..790S,
1981A&amp;A...100...97G,1993A&amp;A...271..451F,1994A&amp;A...285..585F,
2011A&amp;A...531A..83C,2013MNRAS.429..126R}.
For such stars, the emergent Balmer lines are characterised
by pressure-broadened wings, with large sensitivity to 
the gas temperature.
Furthermore, the wings have the same pressure dependence as the dominant
source of continuous opacity, H$^{-}$, such that,
after continuum-normalisation, the wings
are only weakly sensitive to the surface gravity 
\citep[e.g.][Chapter 13]{2008oasp.book.....G}.
Lastly, \ion{H}{I} is by far the dominant species in the photospheres
of such stars, meaning that the line wings
are only weakly sensitive to the surface metallicity
and helium abundance.
Consequently, the emergent Balmer lines can be used to measure
effective temperatures
to better than $100\,\mathrm{K}$~\citep[][Table 4]{2002A&amp;A...385..951B}.

The usefulness of Balmer lines as effective temperature diagnostics
is potentially limited by errors in the line formation models.
The line broadening theory, for Stark broadening 
\citep{1994A&amp;AS..104..509S,1999A&amp;AS..140...93S}
and self-resonance broadening 
\citep{2000A&amp;A...355L...5B,2008A&amp;A...480..581A},
is now on firm footing 
\citep[with a corresponding {$\teff$}~error
of the order {$20\,\mathrm{K}$};][Table 4]{2002A&amp;A...385..951B}.
It has thus been suggested that the
two factors limiting effective temperature determinations 
are 
the use of one-dimensional (1D) hydrostatic model atmospheres
\citep[e.g.][]{2005ARA&amp;A..43..481A,2009A&amp;A...502L...1L},
and
the assumption that the gas is in local thermodynamic equilibrium
\citep[LTE; e.g.][]{2004ApJ...609.1181P,2004ApJ...610L..61P,
2007A&amp;A...466..327B}.

The problem with using 1D hydrostatic model atmospheres
lies in their inability to describe stellar surface convection,
which is a 3D and dynamic phenomenon.
Thus, various
mixing-length parameters \citep{1958ZA.....46..108B,1965ApJ...142..841H}
are usually employed in 1D simulations,
to account for convective energy transport,
alongside microturbulence and macroturbulence parameters
\citep[e.g.][Chapter 17]{2008oasp.book.....G},
to account for the line broadening effects of 
the photospheric convective velocity field and
temperature inhomogeneities.
In contrast, ab initio 3D hydrodynamic stellar atmosphere simulations
naturally encapsulate the physics of convection,
and thus have no need to employ any such free parameters
\citep[e.g.][]{2000A&amp;A...359..729A,
2009LRSP....6....2N,2015A&amp;A...573A..89M}.

Of the various parameters present in 1D analyses, the 
(dimensionless) mixing-length ($\alpha_{\text{MLT}}$)~has the
most influence on emergent Balmer lines,
via its influence on the model temperature stratification.
\citet{1993A&amp;A...271..451F,1994A&amp;A...285..585F}
and \citet{2002A&amp;A...385..951B}
independently analysed observations of emergent Balmer lines
and calibrated $\alpha_{\text{MLT}}\approx0.5$;
while a degeneracy between the mixing-length and the $y$~parameters
\citep[the latter describing the temperature profile
within convective elements; e.g.][]{1965ApJ...142..841H}
influences this calibration, the degeneracy is such that
both low $\alpha_{\text{MLT}}$~and low $y$~are favoured.
This reflects that the formation of Balmer line wings is
biased towards the hot convective upflows.
In contrast, calibrations 
that~account for the overall effects 
of convective energy transport
(i.e.~that consider the role of both upflows and downflows),
for example by matching the adiabatic 
entropy or the mean temperature structure of the deep convection 
zone \citep[e.g.][]{1999A&amp;A...346..111L,
2014MNRAS.445.4366T,2015A&amp;A...573A..89M},
or by considering emergent continuum fluxes 
\citep[e.g.][]{1999ASPC..173..217S},
give significantly higher values,
typically in the range
$1.6\lesssim\alpha_{\text{MLT}}\lesssim2.1$~after adopting
standard values for $y$~and the other mixing-length parameters.

Earlier works have already illustrated 
the importance of adopting 3D model atmospheres when
modelling Balmer line formation.
For example,
\citet{2011A&amp;A...531L..19T,2013A&amp;A...559A.104T}
demonstrated how 3D hydrodynamical modelling
could resolve the problem of
the surface gravity distribution of DA white dwarfs,
wherein the surface gravities of cooler DA white dwarfs
determined from Balmer line modelling
based on 1D model atmospheres and mixing-length theory
were up to {$0.2\,\dex$}~higher than expected 
(in white dwarfs, the Balmer lines are pressure-sensitive).
For late-type stars,
\citet[][]{2009A&amp;A...502L...1L} used
a differential 1D LTE versus 3D LTE comparison 
to demonstrate how 1D modelling is unable to predict the
same emergent Balmer line shapes as 3D modelling,
for any choice of mixing-length.
They further showed that, compared to 3D LTE,
effective temperatures determined by 
1D LTE model emergent Balmer lines
have errors reaching up to $300\,\mathrm{K}$,
depending on the atmospheric parameters and the Balmer line
in question.

Departures from LTE are an added complication.
Using theoretical and semi-empirical 1D model atmospheres,
and the most complete model hydrogen atom to date,
\citet{2007A&amp;A...466..327B} 
showed that collisional processes are not necessarily efficient enough
to make LTE valid for the Balmer line wings.
Compared to 1D non-LTE modelling, 
effective temperatures that are determined
using LTE models of the $\halpha$~lines
are potentially susceptible to errors reaching of the order $100\,\mathrm{K}$.
The predicted departures from LTE are generally expected
to be larger in 3D hydrodynamic model atmospheres, 
where they are driven by the steep horizontal temperature
gradients associated with overshooting
convective upflows and downflows.

Motivated by these problems, we present a study of Balmer line
formation on a grid of 3D hydrodynamic model atmospheres and 
3D non-LTE radiative transfer. 
We describe the methodology in \sect{method}.
We present the results of our simulations in
\sect{results}, and present fits to 
well-studied benchmark stars in \sect{benchmark}.
We discuss the effective temperatures inferred 
for the benchmark stars in \sect{discussion}.
We present 3D non-LTE model spectra for the astronomy community to use
for effective temperature determinations in \sect{access}.
We summarise our findings in \sect{conclusion}.

%-------------------------------------------------------------------------------
\section{Method}
\label{method}

\begin{figure*}
\begin{center}
\includegraphics[scale=0.31]{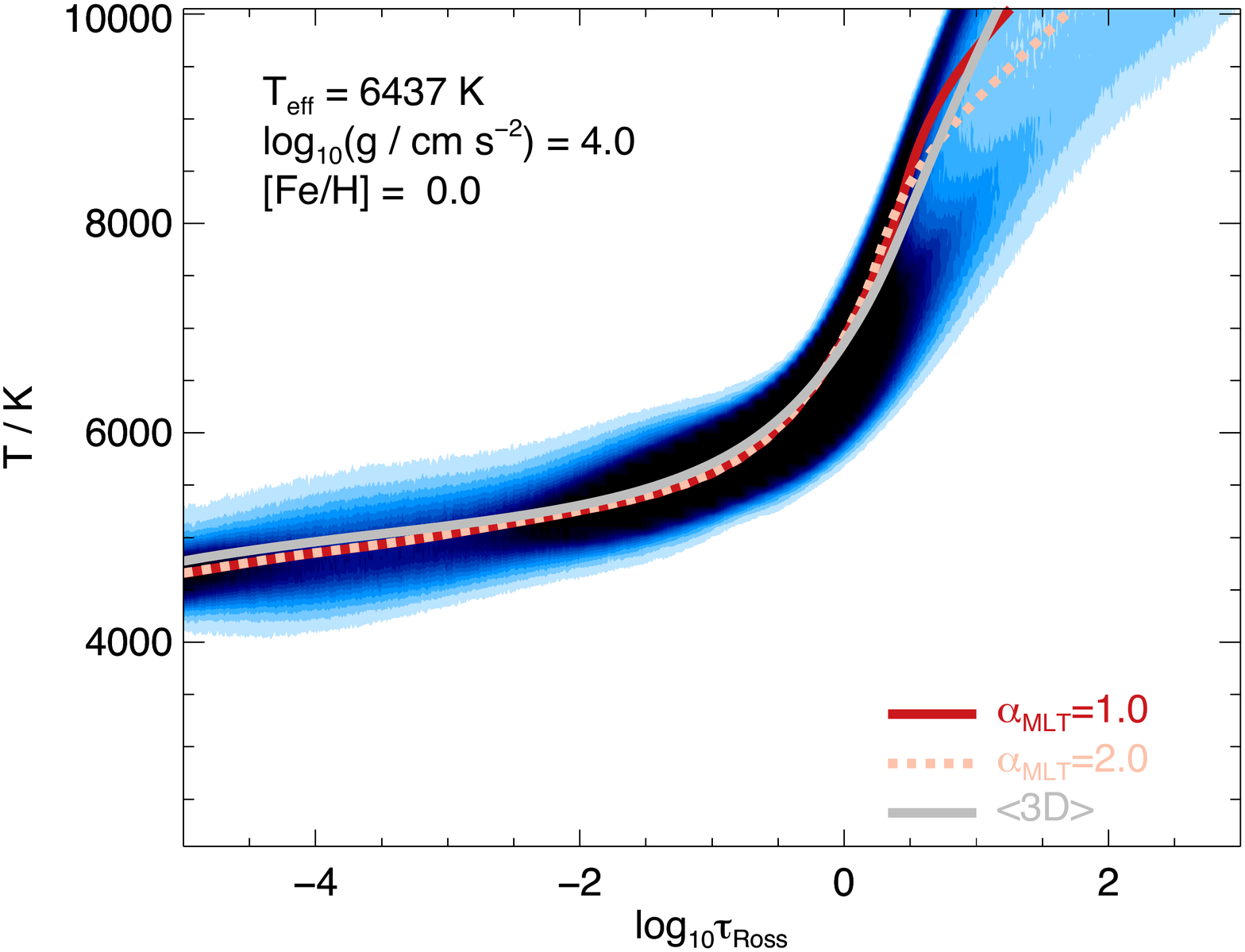}\includegraphics[scale=0.31]{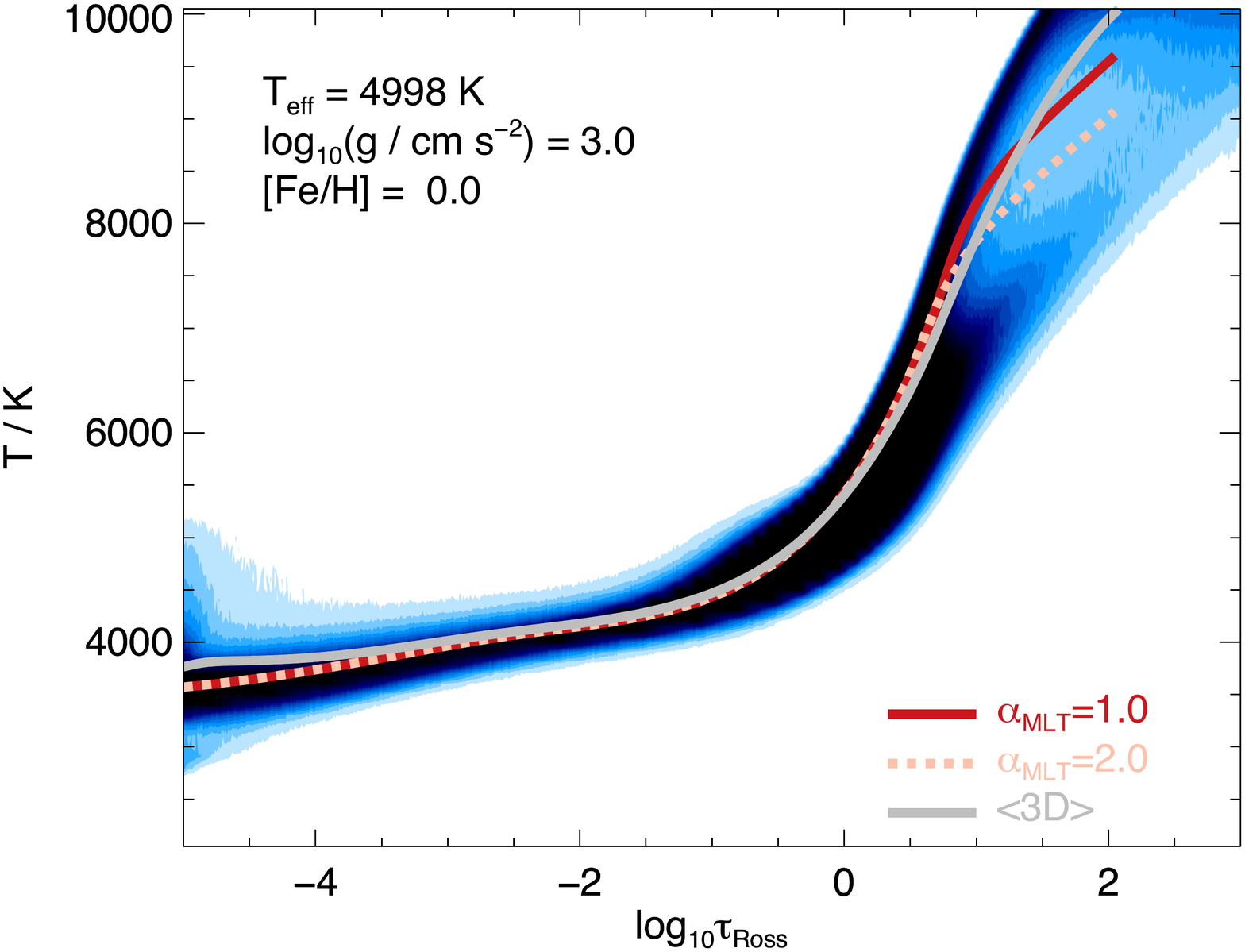}
\includegraphics[scale=0.31]{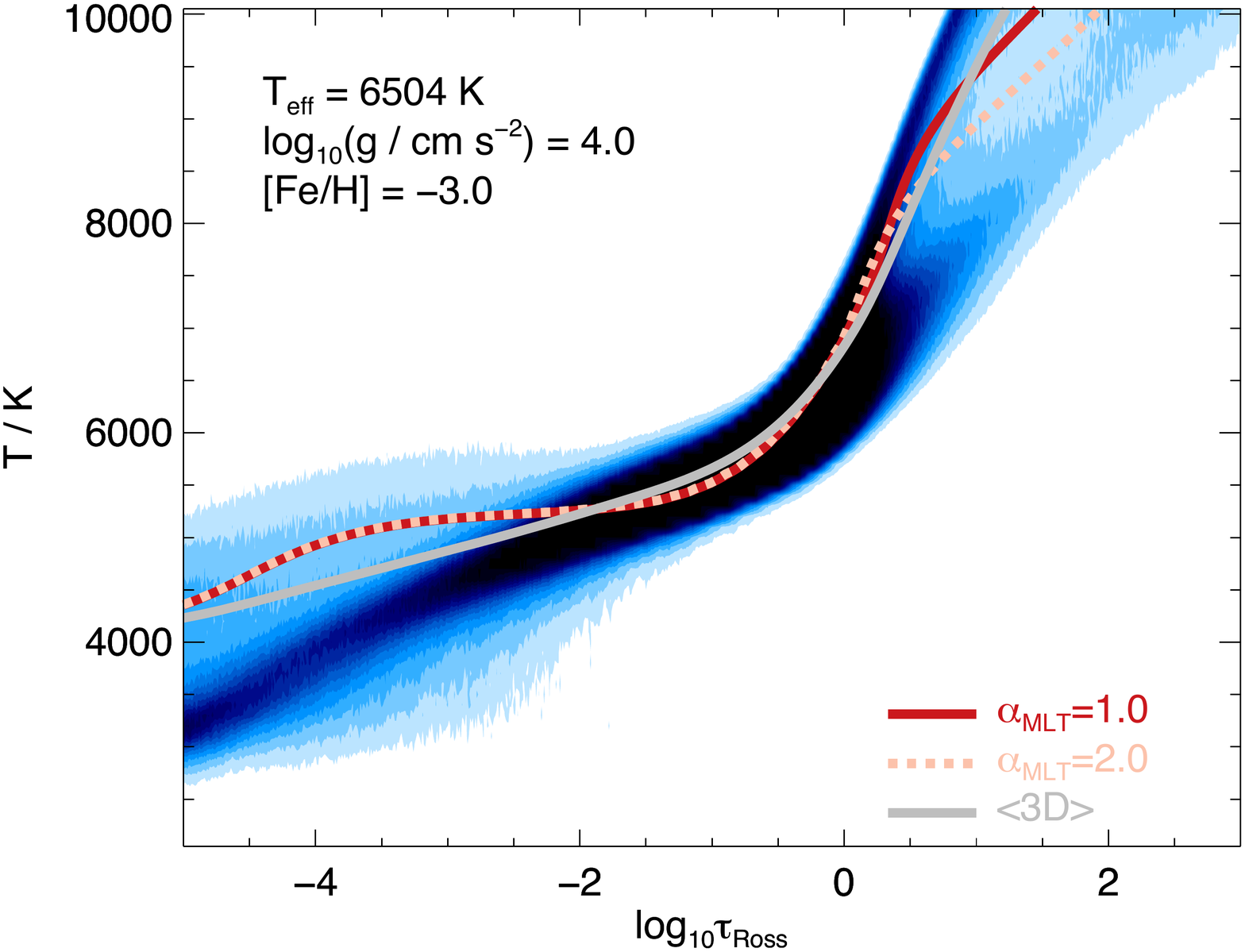}\includegraphics[scale=0.31]{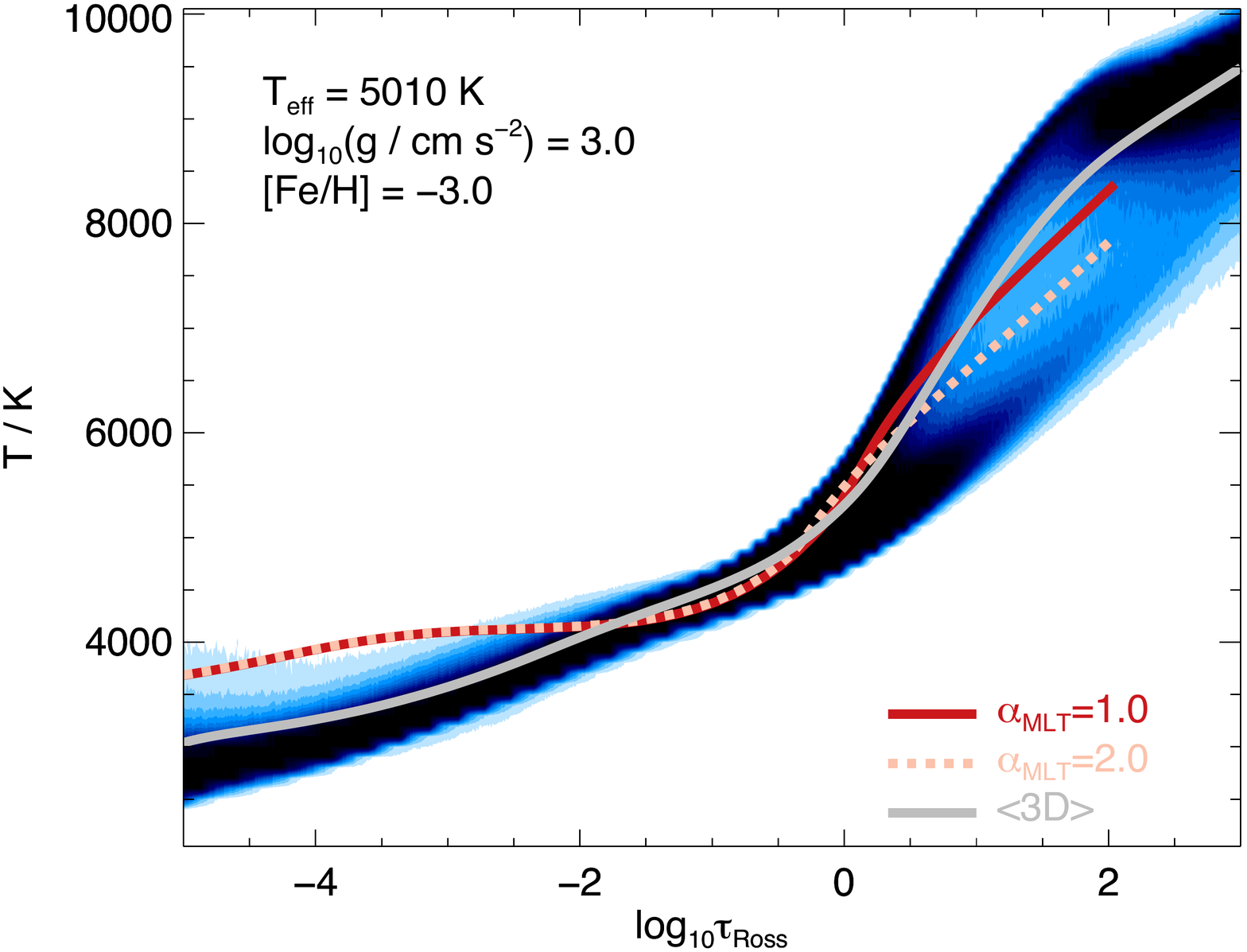}
\caption{Gas temperature-vertical optical depth
distributions in 3D hydrodynamic models
of turn-off stars (left) and 
sub-giants (right)
with solar metallicity (top) and $\feh=-3.0$~(bottom).
Darker shading indicates a larger density of grid-points.
The 1D model atmosphere temperature stratifications,
for two different mixing-lengths, are overplotted,
as are the \mtd~temperature stratifications obtained
from averaging on surfaces of equal $\lgt$.}
\label{figure_temp1}
\end{center}
\end{figure*}

\subsection{Model atmospheres}
\label{method_atmospheres}

\subsubsection{\stagger-grid}
\label{method_atmospheres_stagger}

Calculations were performed on $169$~models of 
the \stagger-grid of 3D hydrodynamic model atmospheres.
We illustrate the gas temperature distributions
of some typical model atmospheres in \fig{figure_temp1}.
We refer to \citet{2013A&amp;A...557A..26M}~for 
a comprehensive description of the simulations.
Here, we only provide an overview of the pertinent details.

The model atmospheres are 
characterised by three atmospheric parameters:
$\left(\teff,\lgg,\feh\right)$.
They assume standard chemical compositions:
solar abundances from \citet{2009ARA&amp;A..47..481A}
scaled by $\feh$, with an implicit enhancement to $\upalpha$-element abundances
of $+0.4\,\dex$~for $\feh\leq-1.0$.
The extent of the \stagger-grid in each dimension
is not regular \citep[see][Fig.~1]{2013A&amp;A...557A..26M}.
The maximum extent in each dimension is
$4000\lesssim\teff/\mathrm{K}\lesssim7000$ in steps 
of roughly $500\,\mathrm{K}$,
$1.5\leq\log\left(g / \mathrm{cm\,s^{-2}}\right)\leq5.0$ in steps
of $0.5\,\dex$, and $-4.0\leq\feh\leq0.0$~in steps of 
$1.0\,\dex$; calculations were also performed
on models with $\feh=+0.5$.

The original hydrodynamic simulations 
used a Cartesian mesh with $240^{3}$~grid-points.
For this work, the
snapshots were down-sampled in the two horizontal directions
by a factor of three. Further, the optically thick layers
($\lgr\gtrsim2$)~were trimmed,
and the thermodynamic quantities
were interpolated onto a new depth scale
roughly equally spaced in the mean temperature so as
to better resolve the steep, continuum-forming regions.
The final number of grid-points was $80\times80\times220$.
Calculations were performed on typically five snapshots of each model
spanning the entire sequence,
so as to obtain temporally-averaged emergent spectra.
This number of snapshots is sufficient to obtain effective
temperatures to better than $10\,\mathrm{K}$~precision (precluding any 
modelling or observational errors, which are typically much larger
than $10\,\mathrm{K}$).

\subsubsection{\atmo-grid}
\label{method_atmospheres_atmo}

To understand the effect of using a consistent and realistic
treatment of convection, 
calculations were also performed
on a grid of 1D hydrostatic \atmo~model 
atmospheres \citep[][Appendix A]{2013A&amp;A...557A..26M}.
These were calculated on the same 
$\left(\teff,\lgg,\feh\right)$~nodes
as the 3D grid we described in \sect{method_atmospheres_stagger}.
The \atmo~simulations used the same equation of state and
radiative transfer solver (angle quadrature, opacity-binning scheme, 
and numerical solver) as used by the \stagger~simulations,
to facilitate a differential 1D versus 3D comparison.

We illustrate the impact of the mixing-length
on the 1D model atmospheres in \fig{figure_temp1}.
Lowering the mixing-length tends to steepen
the temperature stratification in the deeper layers 
where the continuum forms.
Since the Balmer lines are sensitive to the choice of mixing-length
\citep[e.g.][]{1993A&amp;A...271..451F},
line formation calculations were thus
performed for three different mixing-lengths:
$\alpha_{\text{MLT}}=1.0$, $1.5$, and $2.0$.
Other mixing-length parameters were set to standard values 
\citep[{$y=0.076$, $\varv_{\text{conv}}=8.0$}; e.g.][]{2008A&amp;A...486..951G}.
Turbulent pressure was neglected.

Line formation calculations based on 1D model atmospheres
require extra microturbulent and macroturbulent broadening parameters
\citep[e.g.][Chapter 17]{2008oasp.book.....G}
to reproduce the broadening effects of 
the photospheric convective velocity field and
temperature inhomogeneities on lines without 
pronounced wings \citep[e.g.][]{2000A&amp;A...359..729A}.
For simplicity we adopted throughout the study 
a depth-independent microturbulence of $1.0\,\kms$, which is
consistent with the value used to construct the model atmospheres,
and neglected macroturbulence, which has no practical impact
on the Balmer line profiles.

\subsubsection{\marcs-grid}
\label{method_atmospheres_marcs}

To quantify the 3D effects, 
the most differential approach is to
compare results from the \stagger-grid of 3D hydrodynamic model atmospheres
(\sect{method_atmospheres_stagger}) 
with those from the \atmo-grid of 1D hydrostatic model atmospheres
(\sect{method_atmospheres_atmo}).
Nonetheless, we briefly note that calculations were also performed
on an extensive grid of 1100~\marcs~model atmospheres,
with standard chemical compositions and
standard mixing-length 
parameters~\citep[][]{2008A&amp;A...486..951G}.
The calculations on the 1D \marcs~model atmospheres
were performed in exactly the same way as 
those on the 1D \atmo~model atmospheres.

The maximum extent in each dimension
of the \marcs~grid used here is 
$4000\leq\teff/\mathrm{K}\leq6750$ in steps 
of $250\,\mathrm{K}$,
$1.5\leq\log\left(g / \mathrm{cm\,s^{-2}}\right)\leq5.0$ in steps
of $0.5\,\dex$, and $-4.0\leq\feh\leq0.5$~in steps of 
$0.25\,\dex$~for $-1.0\leq\feh\leq0.5$,
$0.5\,\dex$~for $-3.0\leq\feh\leq-1.0$,
and $1.0\,\dex$~otherwise.
The grid is complete in the parameter region of interest,
and more finely-spaced in 
$\teff$~and $\feh$~than the \stagger-grid.
Consequently, the resulting grid of model Balmer lines
were used to estimate interpolation
and extrapolation errors
on the coarser and incomplete \stagger-grid. 
We discuss this in \sect{benchmark_method_interpolation}.

The \marcs-based results of the fits to
benchmark stars that we present in \sect{benchmark}
were found to be very similar to those from the 
1D \atmo~model atmospheres
with the same mixing-length parameters
(i.e.~$\alpha_{\mathrm{MLT}}=1.5$,
$y=0.076$, $\varv_{\text{conv}}=8.0$),
after accounting for 
interpolation and extrapolation errors.
The largest differences in inferred
effective temperatures were for $\hgamma$~and only 
of the order 
$20\,\mathrm{K}$, which is already
much smaller than the uncertainties intrinsic to the method
(mainly in placing the continuum). Consequently, for brevity,
we do not discuss the results from the 
\marcs~grid of model atmospheres in detail.

\subsection{Post-processing line formation calculations}
\label{method_code}

We used our 3D non-LTE radiative transfer code 
\codename~to calculate emergent Balmer line spectra
for all of the models 
(1D and 3D; LTE and non-LTE).
This code is originally based on the
code \multitd~\citep{1999ASSL..240..379B,2009ASPC..415...87L},
but with our own customisations 
\citep[e.g.][]{2016MNRAS.455.3735A,2016MNRAS.463.1518A,
2017MNRAS.464..264A}.
We provide a brief overview of the code here.

The code employs the MALI algorithm \citep{1992A&amp;A...262..209R}
to find the statistical equilibrium
of \ion{H}{I}/\ion{H}{II} simultaneously,
assuming no feedback on the background atmosphere.
The mean radiation field $J$~was determined by 
solving the radiative transfer equation using
a short characteristics integral solver
\citep{2013A&amp;A...549A.126I},
using the eight-point Lobatto quadrature on the 
interval $\left[-1,1\right]$~for the integration
over $\mu=\cos\theta$, where $\theta$~is the angle relative to 
the vertical, and, for non-vertical rays,
an equidistant four-point trapezoidal quadrature on the
interval $\left[0,2\pi\right]$~for the integration
over the azimuthal angle $\phi$.

The equation-of-state (EOS) and background opacities were computed
using our code \blue, which was previously described in
\citet[][Sect.~2.1.2]{2016MNRAS.463.1518A}.
The EOS was recomputed in post-processing
(i.e.~with no feedback
on the background atmosphere)
assuming LTE with corrections $-\Delta\chi$~to the
ionisation potentials $\chi$~to account for Debye shielding.
A more general treatment of perturbations caused by neighbouring
particles exists in the occupation probability formalism
\citep{1987ApJ...319..195D,1988ApJ...331..794H}.
Using the \texttt{HBOP}~code, 
part of the \texttt{HLINOP}~package \citep{2015ascl.soft07008B},
we tested the impact of adopting the occupation probability formalism
on the emergent Balmer lines in 1D LTE. The $\hgamma$~line profiles
become marginally weaker, with normalised
flux residuals of the order $10^{-4}$,
or of the order a few kelvin in effective temperature;
the lower members of the Balmer series are even less affected. 
These errors are small, compared to other uncertainties
inherent in the modelling.
Background line opacities were precomputed on
temperature-density grids for a given chemical composition,
and interpolated onto the model atmosphere at runtime.
On the other hand, background continuous opacities were computed 
at runtime for highest accuracy.

After the last iteration, the final emergent intensities were computed
using an integral solver on
a viewing-angle aligned grid 
\citep[so as to avoid diffusion errors
associated with interpolating the specific intensity;
e.g.][]{2017arXiv170809362P}.
A monotonic cubic interpolation scheme was used to
interpolate the extinction and source-function onto the ray-aligned
grid; we found that the default linear interpolation scheme
originally implemented in \multitd~can lead to errors of
up to $1000\,\mathrm{K}$~in
the worst case ($\hgamma$~in metal-poor red giants).
The astrophysical fluxes were computed 
by disk-integrating the emergent intensities,
using the seven-point Lobatto quadrature on the 
interval $\left[0,1\right]$~for the integration
over $\mu$, and, for non-vertical rays,
an equidistant eight-point trapezoidal quadrature on the
interval $\left[0,2\pi\right]$~for the integration
over the azimuthal angle $\phi$. Since the integrand for the
astrophysical flux is identically null at $\mu=0$,
this amounts to 41 rays in total.
Background line opacities were neglected at this stage.
The spectra were trivially normalised, by
repeating these radiative transfer calculations
without any \ion{H}{I}~line opacities.

To calculate the hydrogen absorption line profiles 
$\phi\left(\lambda\right)$~\citep[which enter 
into the calculation of exctinction coefficients;
e.g.][Chapter 8]{2014tsa..book.....H},
we implemented into our code the Fortran modules \texttt{HLINPROF}~and 
\texttt{HLINOP}~\citep{2015ascl.soft07008B}.
The former module was used for an accurate treatment
of low Balmer absorption line profiles 
with lower state $n=2$~and upper state $n\leq6$,
and is based on \citet{1999A&amp;AS..140...93S}
for Stark broadening, and
\citet{2000A&amp;A...355L...5B} for self-resonance broadening.
The latter module was used for the remaining \ion{H}{I}~lines
(in the solution of the statistical equilibrium); 
it is based on \citet{1960ApJ...132..883G}
and \citet{1973ApJS...25...37V} for Stark broadening,
and \citet{1966PhRv..144..366A} for self-resonance broadening.
More details can be found in \citet[][Sect.~2.1.1]{2007A&amp;A...466..327B}
and \citet[][Sect.~4.1.1]{2016A&amp;ARv..24....9B}.
Complete redistribution was assumed
throughout, which means that the absorption line profiles
are identical to the emission line profiles.

\subsection{Hydrogen model atom and inelastic collisions}
\label{method_atom}

The model atom includes all \ion{H}{I}~states up
to $n=20$, as well as \ion{H}{II}~(so that the 
non-LTE excitation and ionisation balance were solved
together, consistently). 
All lines and continua involving these levels were considered. 
\citet{2004ApJ...609.1181P}~recommended including 
at least the number of levels corresponding to 
the classical Inglis-Teller limit
\citep[e.g.][]{1939ApJ....90..439I,1966JQSRT...6..575V}, for a given star.
Using the semi-empirical solar model atmosphere
of \citet{1974SoPh...39...19H}, 
the limit at $T_{\text{gas}}=5772\,\mathrm{K}$~is
$n\approx11.6$, which indicates that $n=20$~is more than sufficient
for this study of late-type stars.

Being very much a minority species in late-type stellar atmospheres,
H$^{-}$~is not expected to influence the populations
of \ion{H}{I}~or \ion{H}{II}, and was thus treated in LTE.
However, H$^{-}$~is the dominant source
of background opacity in the optical region,
so any departures from LTE in this species would 
have a large influence on the results presented here.
The non-LTE H$^{-}$~problem is highly non-trivial
and is, to our knowledge, yet unsolved
\citep[][]{1984SoPh...93...23L}.
We intend to revisit this problem in a future study.

Since the energies and transition probabilities 
for hydrogen are known to exceptionally high precision,
the main subtlety in constructing the model atom
is in the treatment of inelastic collisions.
We based our construction on that presented in 
\citet[][Sect.~2.1.2]{2007A&amp;A...466..327B}:
\begin{itemize}

\item{Inelastic H+e collisional excitation rate coefficients 
were taken from \citet{2004ApJ...609.1181P}, based on 
the R-matrix method in the close-coupling approximation,
for the transitions between the states with $n\leq7$. 
Rate coefficients for the remaining transitions were
calculated using the semi-empirical formula of
\citet{1980PhRvA..22..940V}.}

\item{Inelastic H+e collisional ionisation rate coefficients
were based on the experimental results of
\citet{1987JPhB...20.3501S}~for the $n=1$~transition,
and \citet{1981JPhB...14..111D}~for the $n=2$~transition,
both via the analytical formula of \citet[][]{2007A&amp;A...466..327B}.
Rate coefficients for the remaining transitions were 
calculated using the semi-empirical formula of
\citet{1980PhRvA..22..940V}.}

\item{Inelastic H+H collisional excitation rate coefficients
were taken from \citet{1955PPSA...68..173B},
based on the Landau-Zener model, for the
lone~transition
between the states with $n_{\text{lo}}=2$~and $n_{\text{up}}=3$.
Rate coefficients for the 
transitions between the states with $4\leq n_{\text{lo}}\leq10$, 
$1\leq n_{\text{up}}-n_{\text{lo}}\leq5$~were taken from
\citet{2004JPhB...37.4493M},
based on a semi-classical
theory for resonant energy exchange in Rydberg atoms.
In contrast to \citet[][]{2007A&amp;A...466..327B},
we used \citet[][Eq.~18]{1991JPhB...24L.127K},
based on the free-electron model in
the scattering length approximation,
for the remaining transitions with $n_{\text{lo}}\geq4$.
These rates were calculated using the IDL packages 
\texttt{MSWAVEF}~\citep{2017ascl.soft01006B},
and \texttt{KAULAKYS}~\citep{2017ascl.soft01005B}.
The remaining transitions,
with $n_{\text{lo}}=2$, and $n_{\text{lo}}=3$,
were neglected (see the end of this section).}

\item{Inelastic H+H collisional ionisation rate coefficients
were taken from \citet{1996PhyS...53..159M}, based on a semi-classical
theory for resonant energy exchange in Rydberg atoms,
for the $4\leq n_{\text{lo}}\leq10$~transitions. A linear fit 
to the (logarithmic) rate coefficients
against transition energy was obtained and extrapolated
to obtain rate coefficients for transitions
with $n_{\text{lo}}\geq11$.  The remaining transitions,
with $n_{\text{lo}}=2$, and $n_{\text{lo}}=3$,
were neglected (see the end of this section).}

\item{Inelastic $\mathrm{H^{+}+H^{-}}$~mutual neutralisation 
rate coefficients were taken from \citet{1986JPhB...19L..31F},
based on a quantum close-coupling treatment,
for the $n_{\text{lo}}=2$~and $n_{\text{lo}}=3$~transitions, 
via the formula of \citet[][]{2007A&amp;A...466..327B}.
More robust calculations were presented
in \citet{2009PhRvA..79a2713S};
however, their results compare well to
those from \citet{1986JPhB...19L..31F},
and \citet{2007A&amp;A...466..327B} demonstrated
that these transitions are not very important 
for the Balmer line wings.}

\end{itemize}
Here H+H refers to collisions with neutral hydrogen in the $n=1$~ground state.
Penning ionisation, involving
collisions with neutral hydrogen in the $n=2$~excited state
\citep[e.g.][]{1967PPS....91..288B}, is
not expected to be important \citep[][]{2007A&amp;A...466..327B}
and was neglected here.

The treatment of
inelastic H+H collisional ionisation from the 
$n_{\text{lo}}=2$~and $n_{\text{lo}}=3$~states,
and also inelastic H+H collisional excitation from 
the $n_{\text{lo}}=2$~and $n_{\text{lo}}=3$~states
into $n_{\text{up}}\ge4$~states, 
are perhaps the main source of uncertainty 
in the non-LTE modelling.
As discussed in \citet[][]{2007A&amp;A...466..327B},
there is presently no satisfactory description for these rates. The 
Drawin recipe \citep{1968ZPhy..211..404D,1969ZPhy..225..483D}
for example predicts very small rates for 
these transitions,
that have only a minor impact on the overall results.
As we mentioned above, we
thus decided to neglect these rates altogether, 
erring on the side of slightly overestimating the departures from LTE.
We refer to \citet[][]{2007A&amp;A...466..327B}
for a detailed discussion of the sensitivity of the non-LTE 
effects on the collisional transitions.

%-------------------------------------------------------------------------------
\section{Results}
\label{results}

\begin{figure*}
\begin{center}
\includegraphics[scale=0.31]{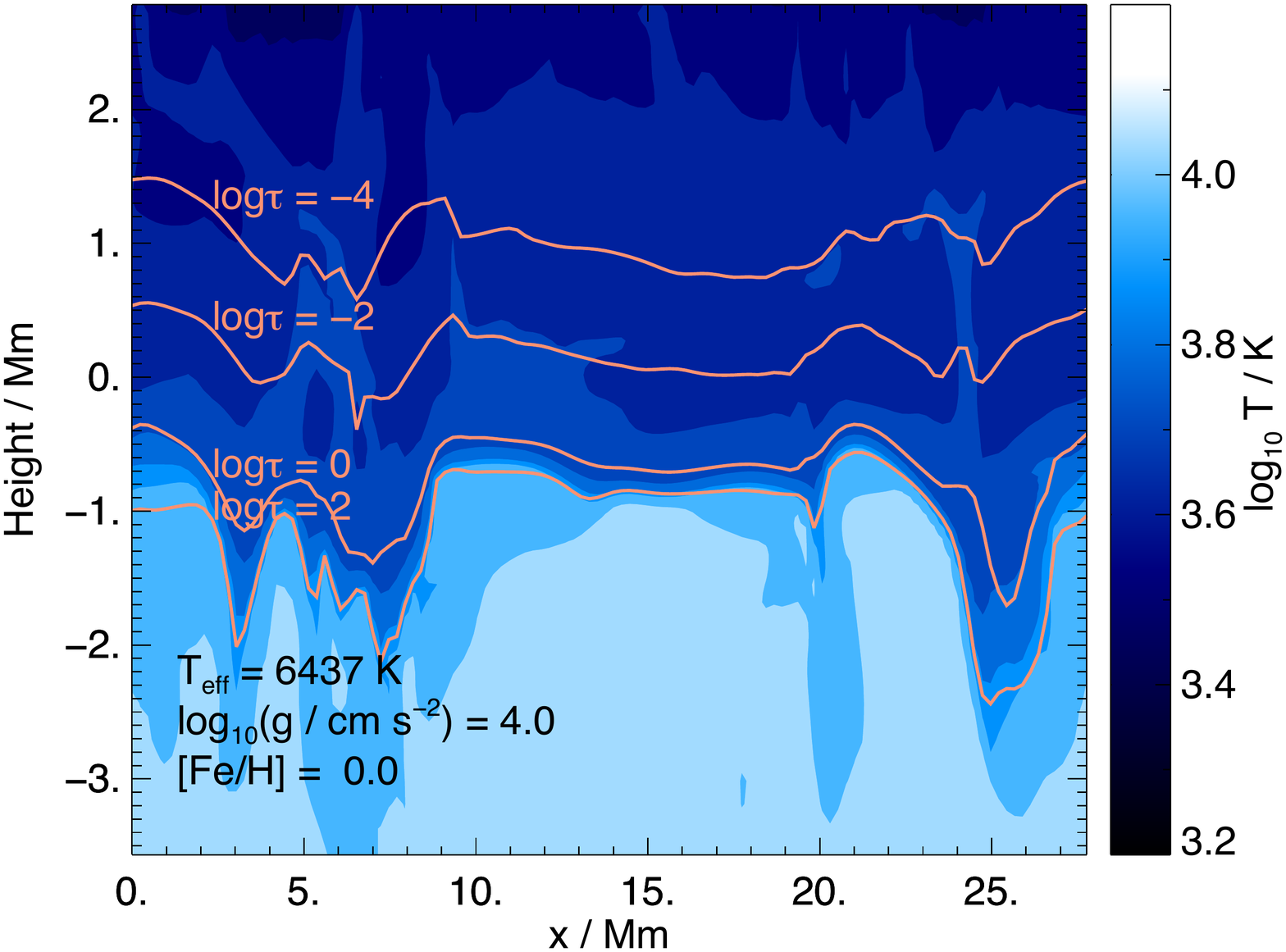}\includegraphics[scale=0.31]{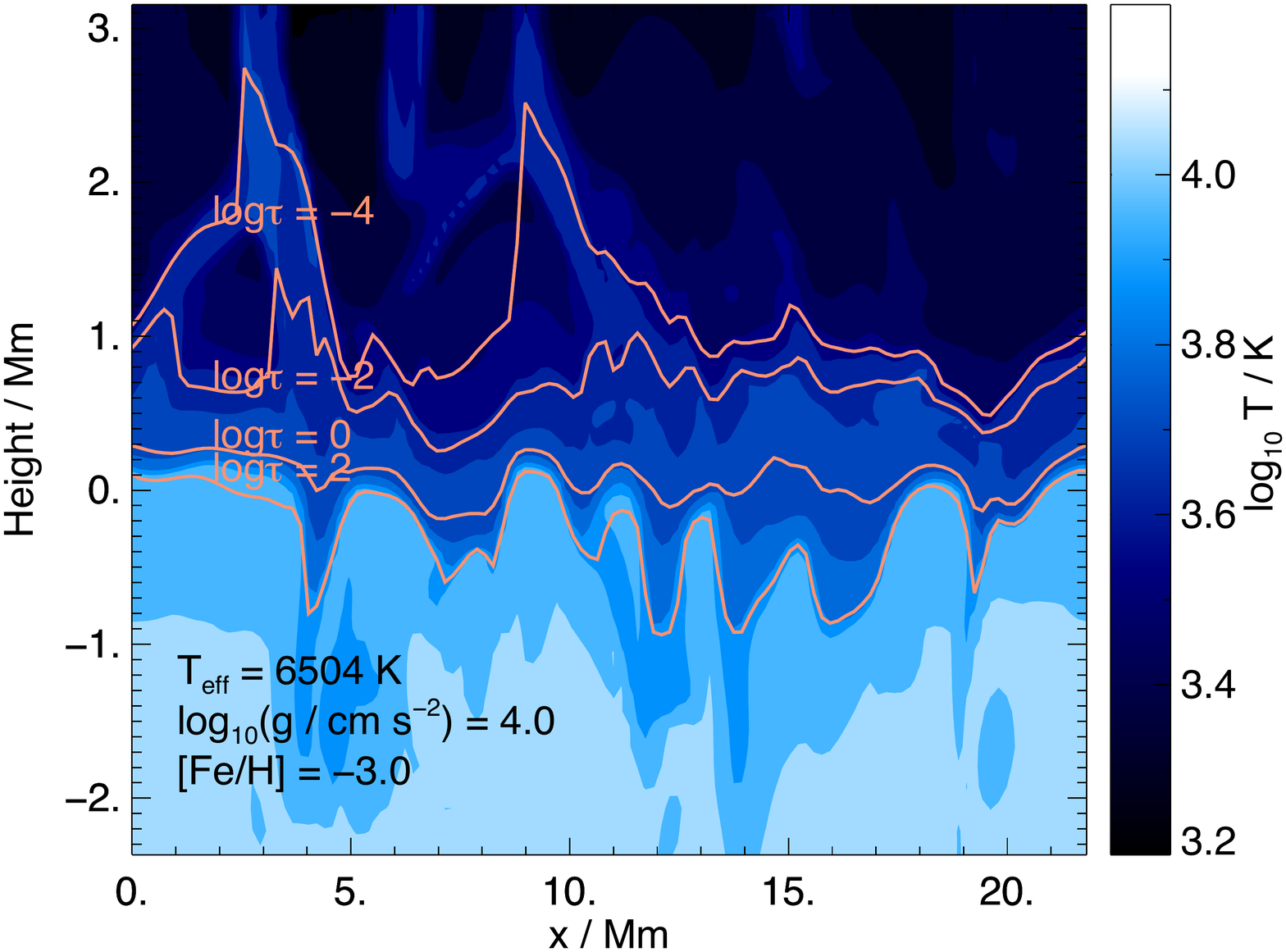}
\includegraphics[scale=0.31]{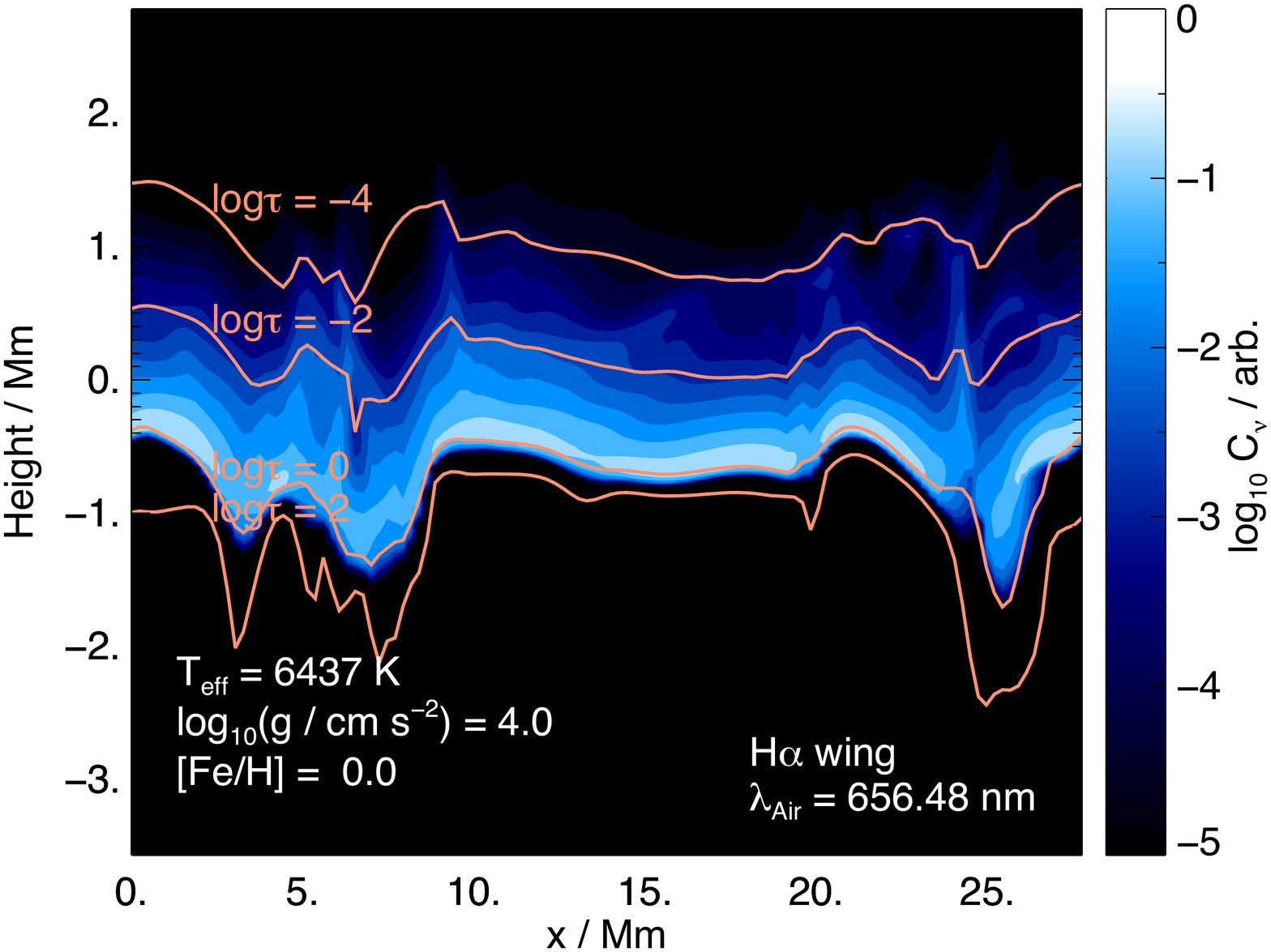}\includegraphics[scale=0.31]{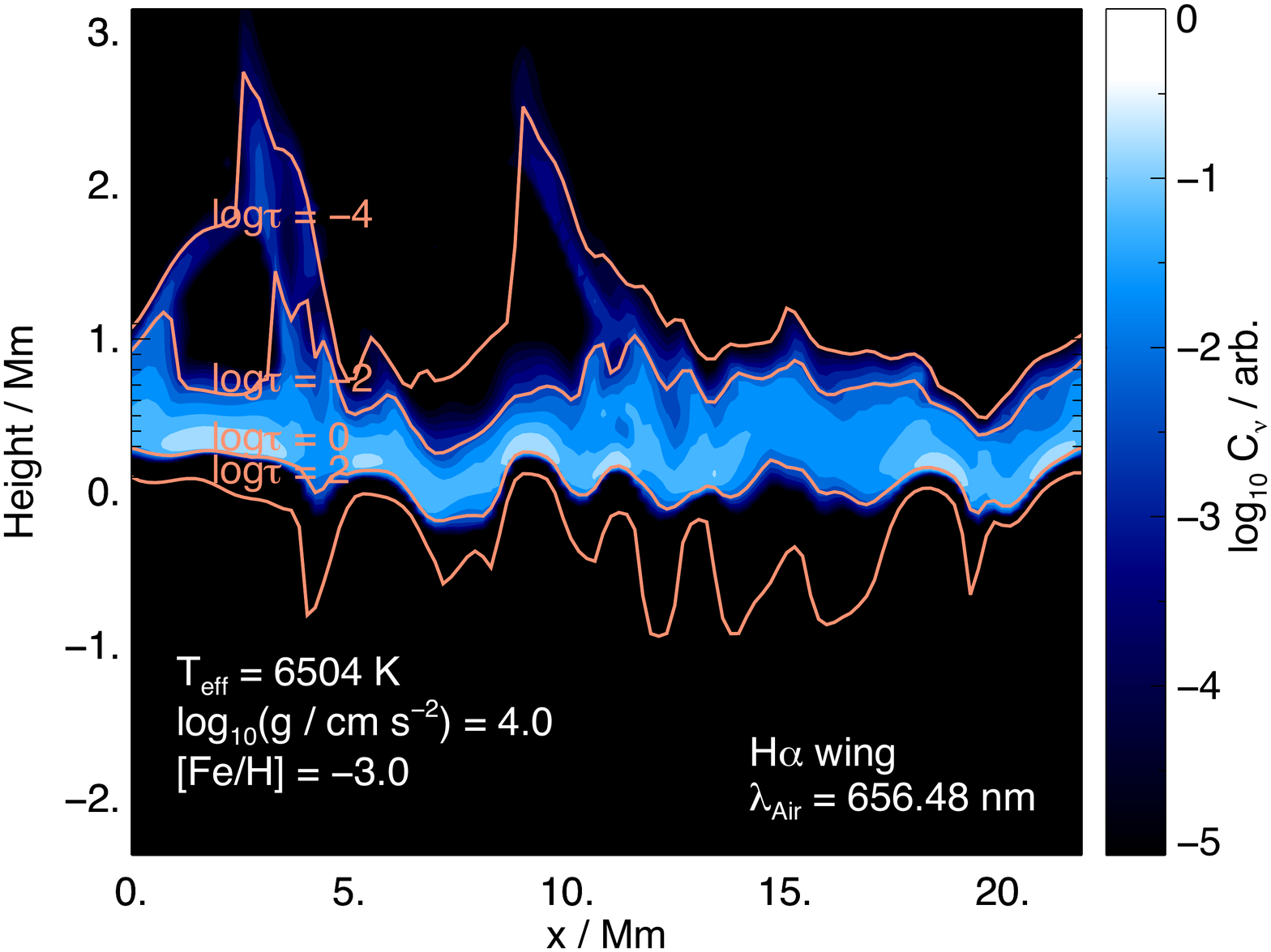}
\includegraphics[scale=0.31]{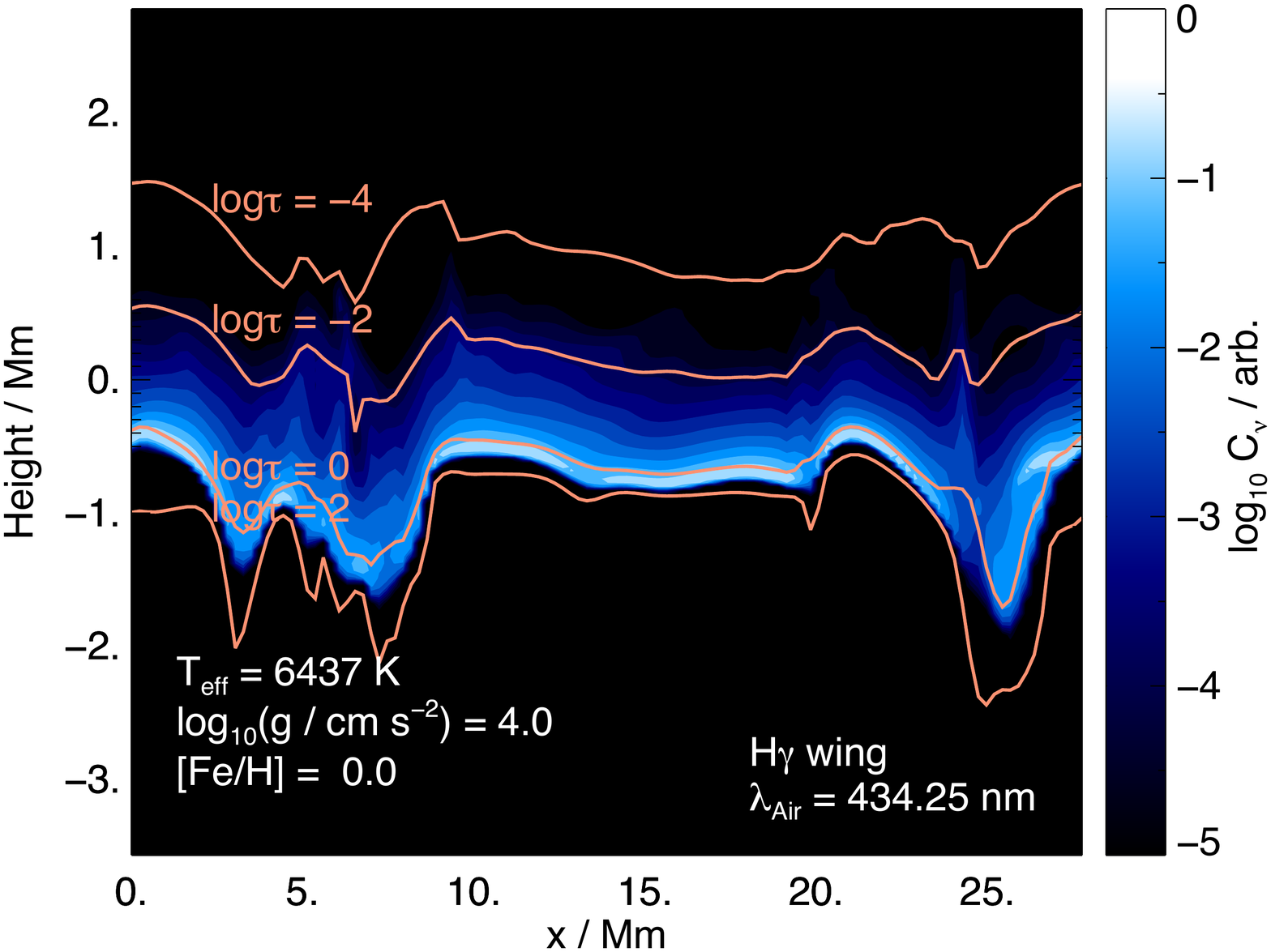}\includegraphics[scale=0.31]{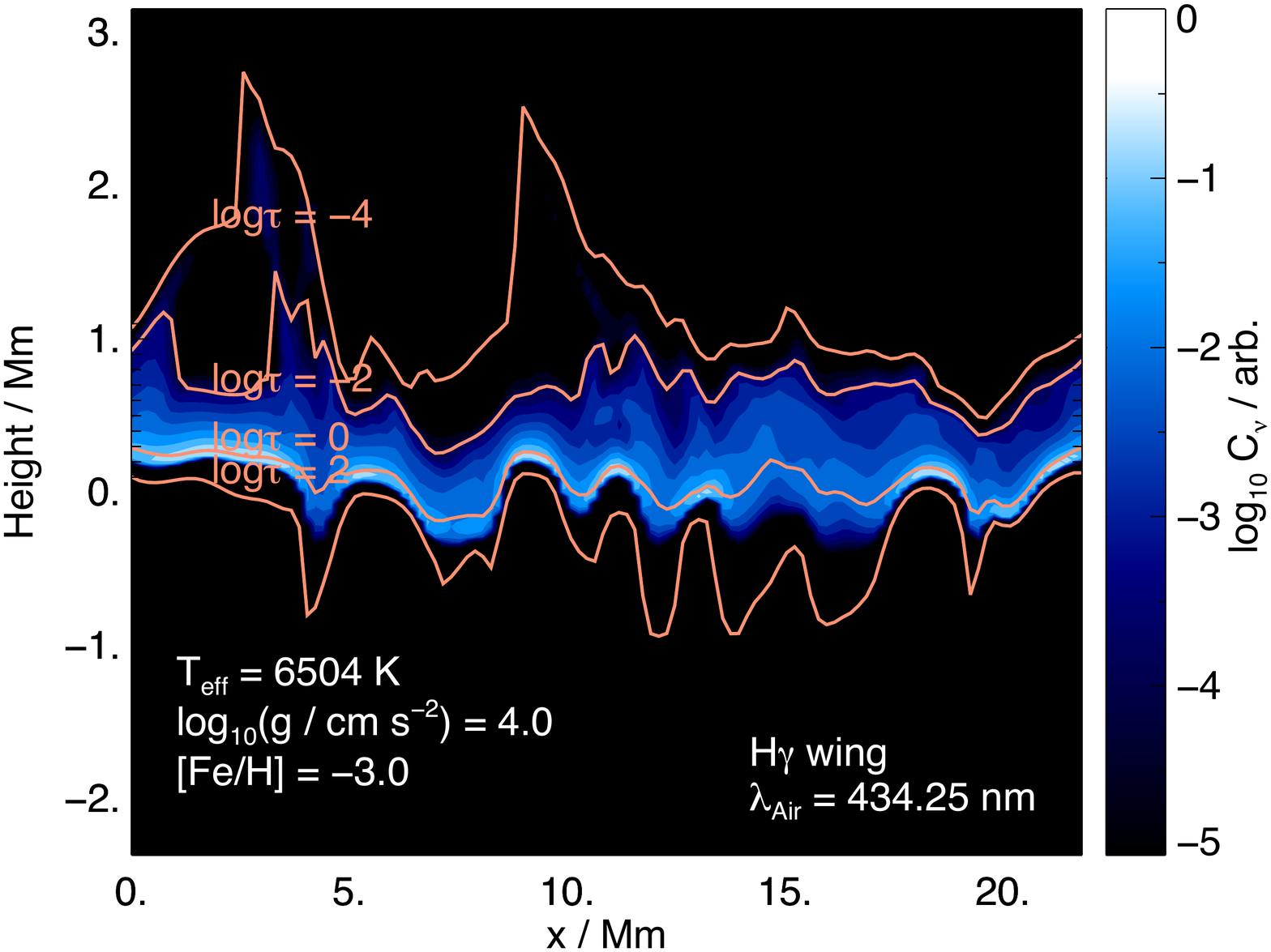}
\caption{Gas temperatures $T$~(first row) and
contribution functions $C_{\nu}$~for the inner wings
($0.2\,\mathrm{nm}$~redward of the line centre)
of the $\halpha$~(second row) and $\hgamma$~(third row) lines, 
in vertical slices of the solar metallicity (left column)
and metal-poor (right column) model
atmospheres shown in \fig{figure_temp1}.
Lighter shading indicates larger temperatures
(first row), and more
emergent flux contribution (second and third rows).
Contours of constant $\lgr$~are overdrawn.
In each plot, the contribution functions 
are normalised such that their maximum values are unity.}
\label{figure_cf1}
\end{center}
\end{figure*}

\begin{figure*}
\begin{center}
\includegraphics[scale=0.31]{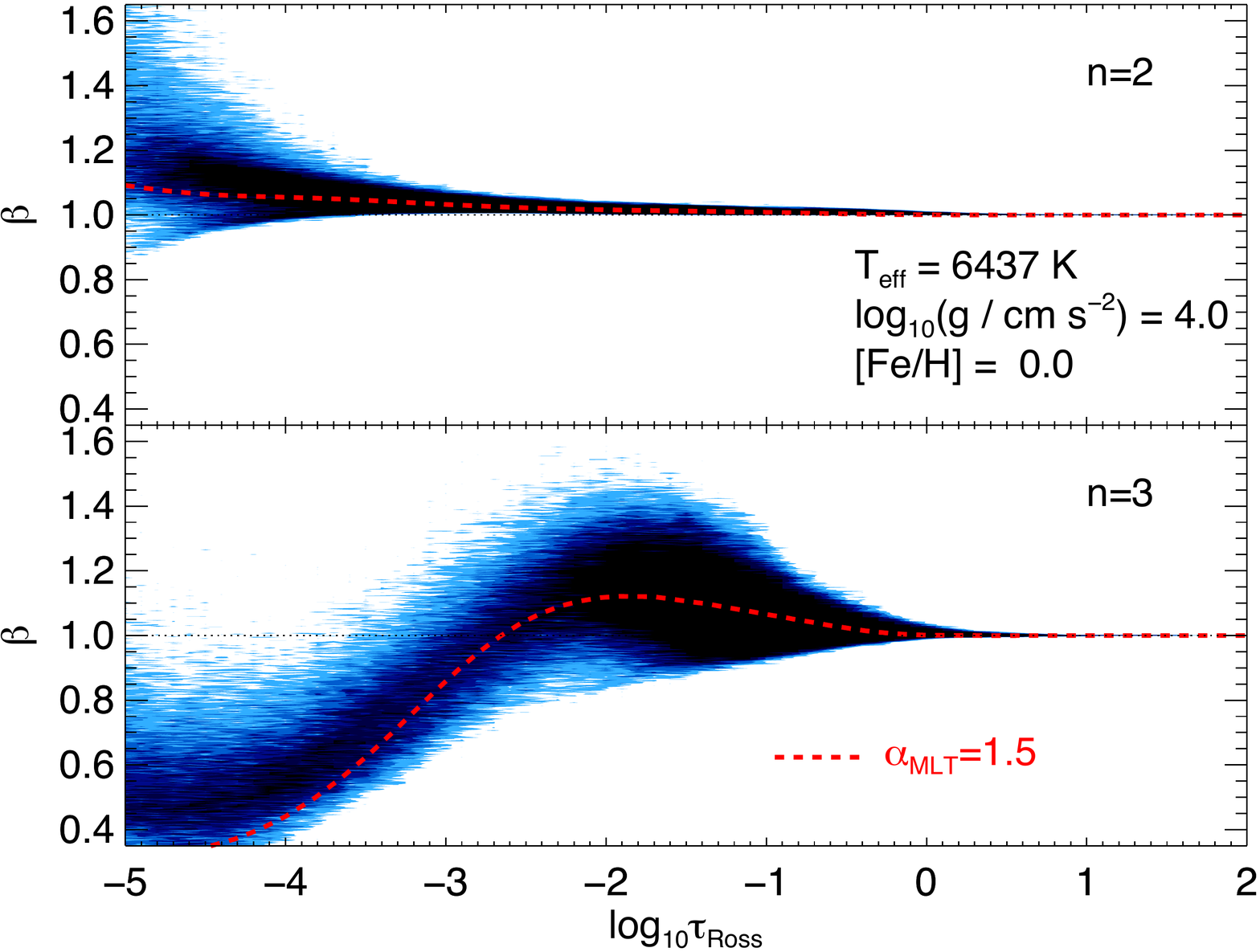}\includegraphics[scale=0.31]{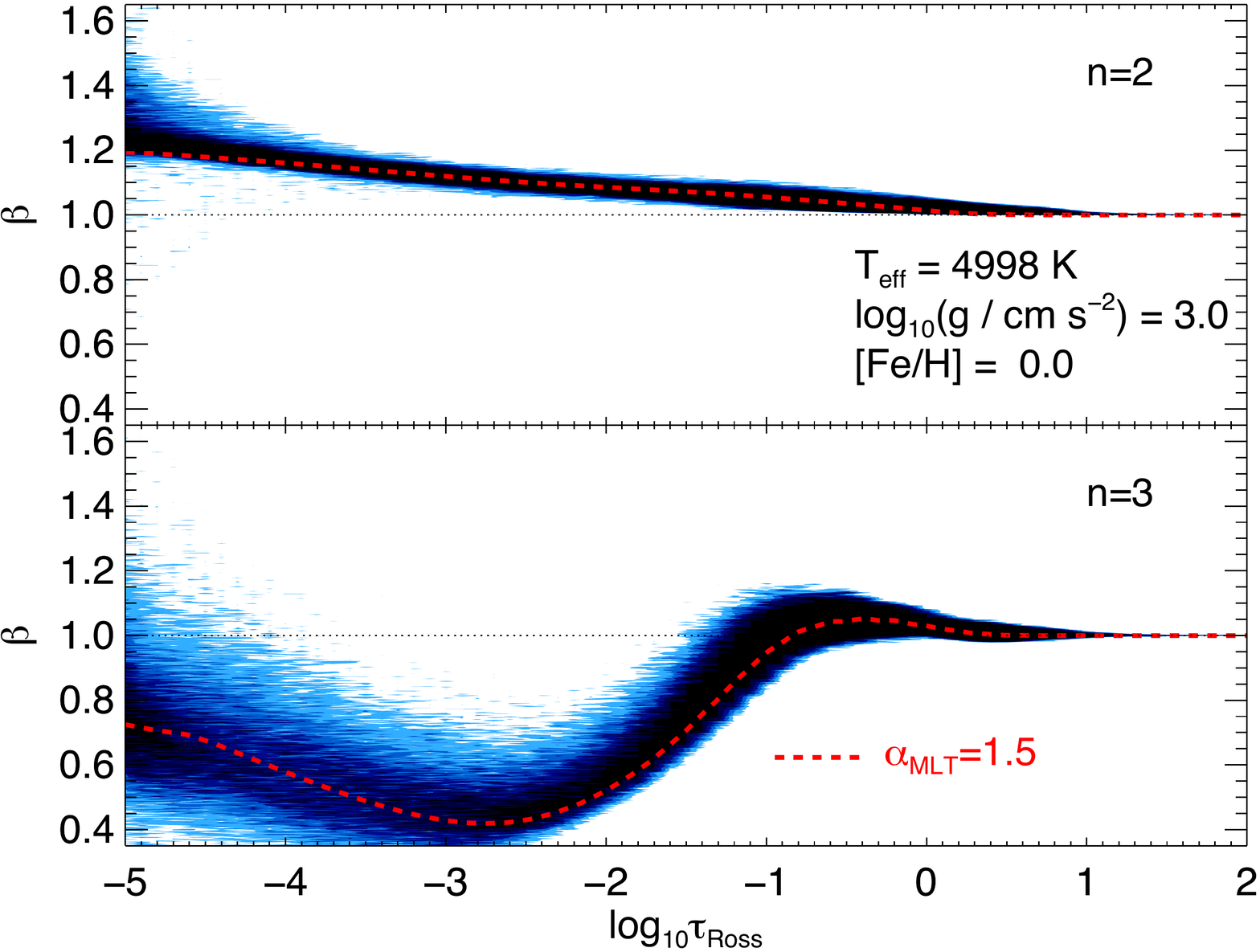}
\includegraphics[scale=0.31]{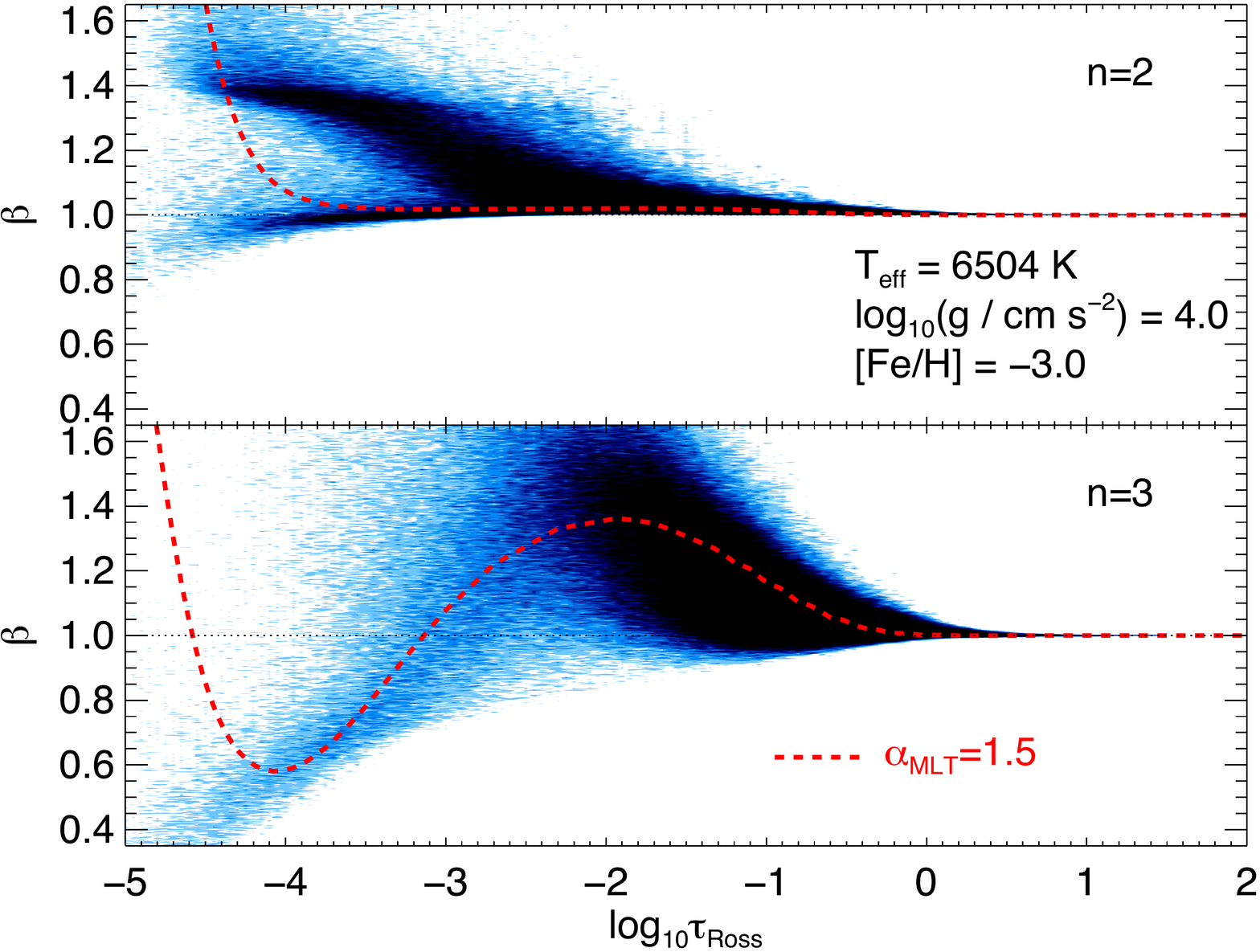}\includegraphics[scale=0.31]{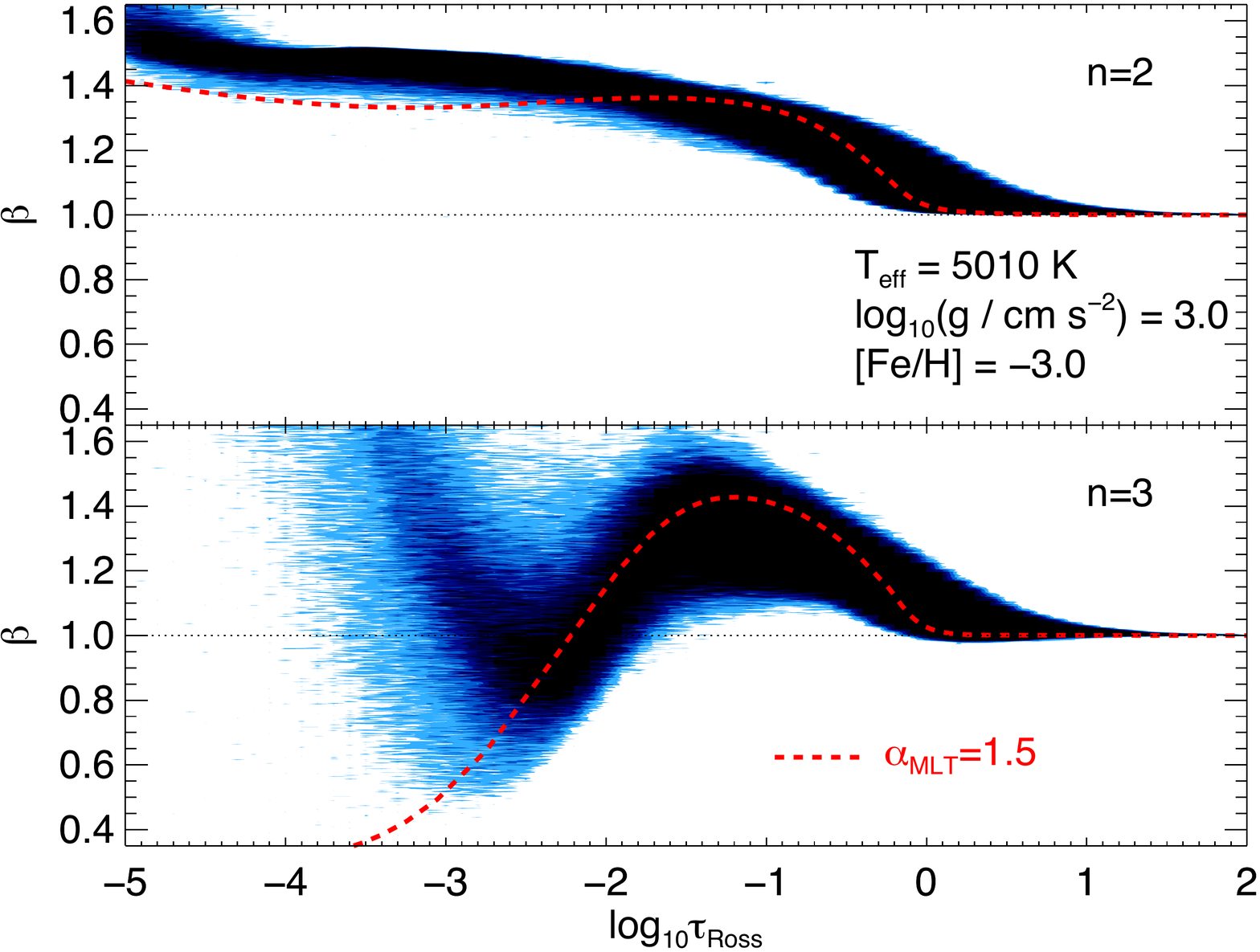}
\caption{Departure coefficient-vertical optical depth
distributions in the 3D hydrodynamic
model atmospheres shown in \fig{figure_temp1}.
Darker shading indicates a larger density of grid-points.
The departure coefficients of the lower
and upper levels of the $\halpha$~line are shown.
The departure coefficients of 
the heavily-populated ground level ($n=1$) always stay close to unity,
while the departure coefficients of the more 
excited levels ($n>3$) and the ionised state (\ion{H}{II}) 
roughly follow those of the $n=3$~state.}
\label{figure_b1}
\end{center}
\end{figure*}

\begin{figure*}
\begin{center}
\includegraphics[scale=0.31]{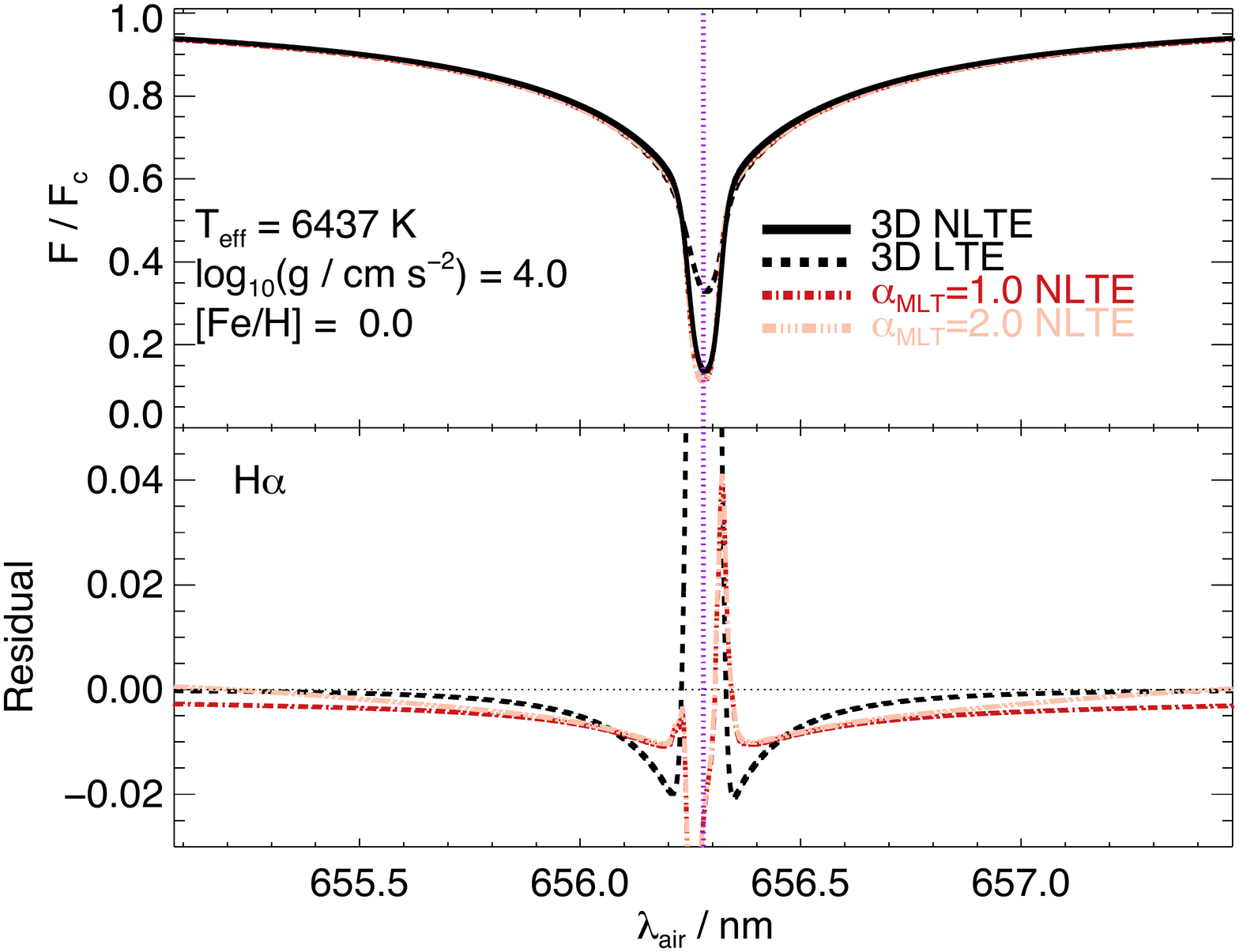}\includegraphics[scale=0.31]{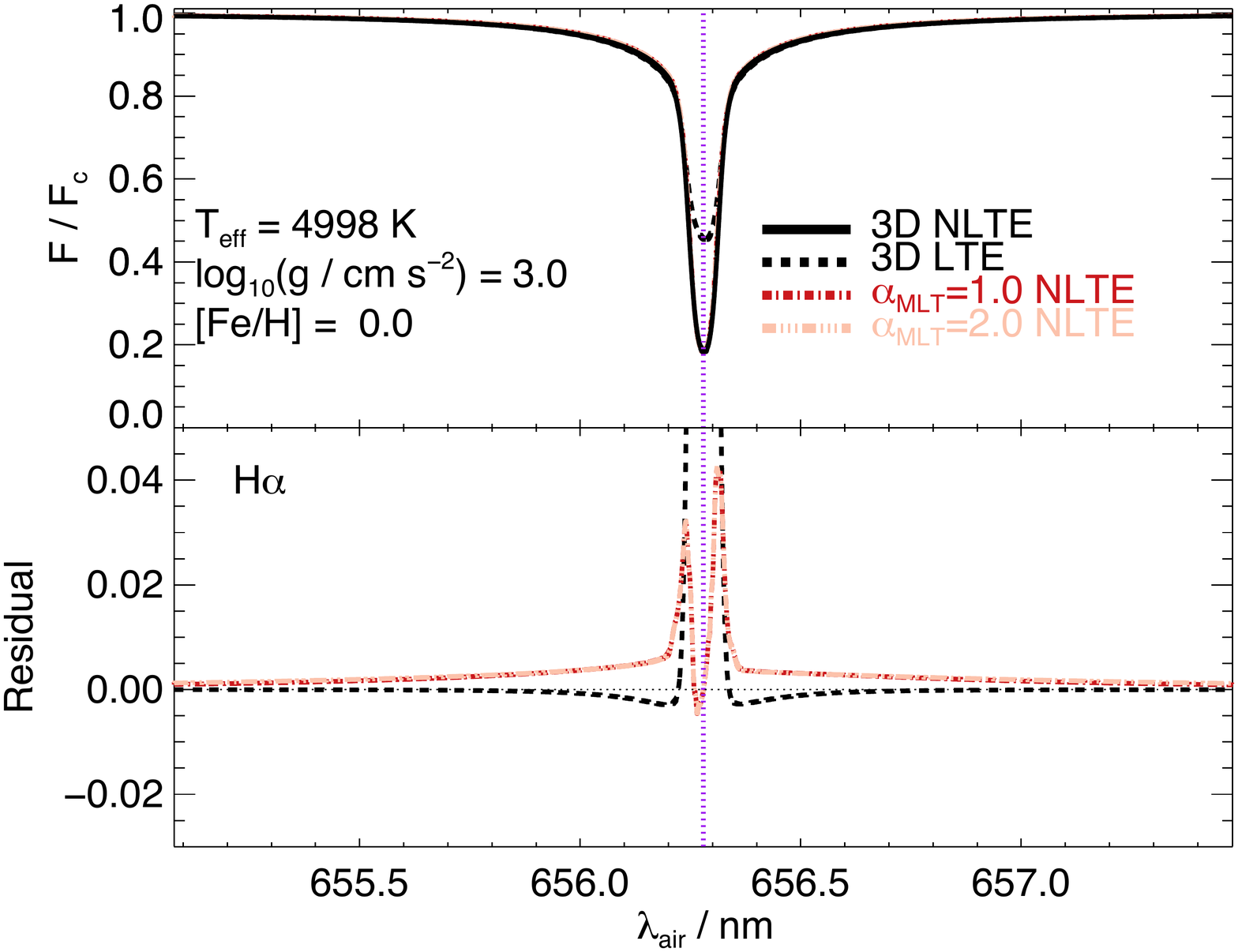}
\includegraphics[scale=0.31]{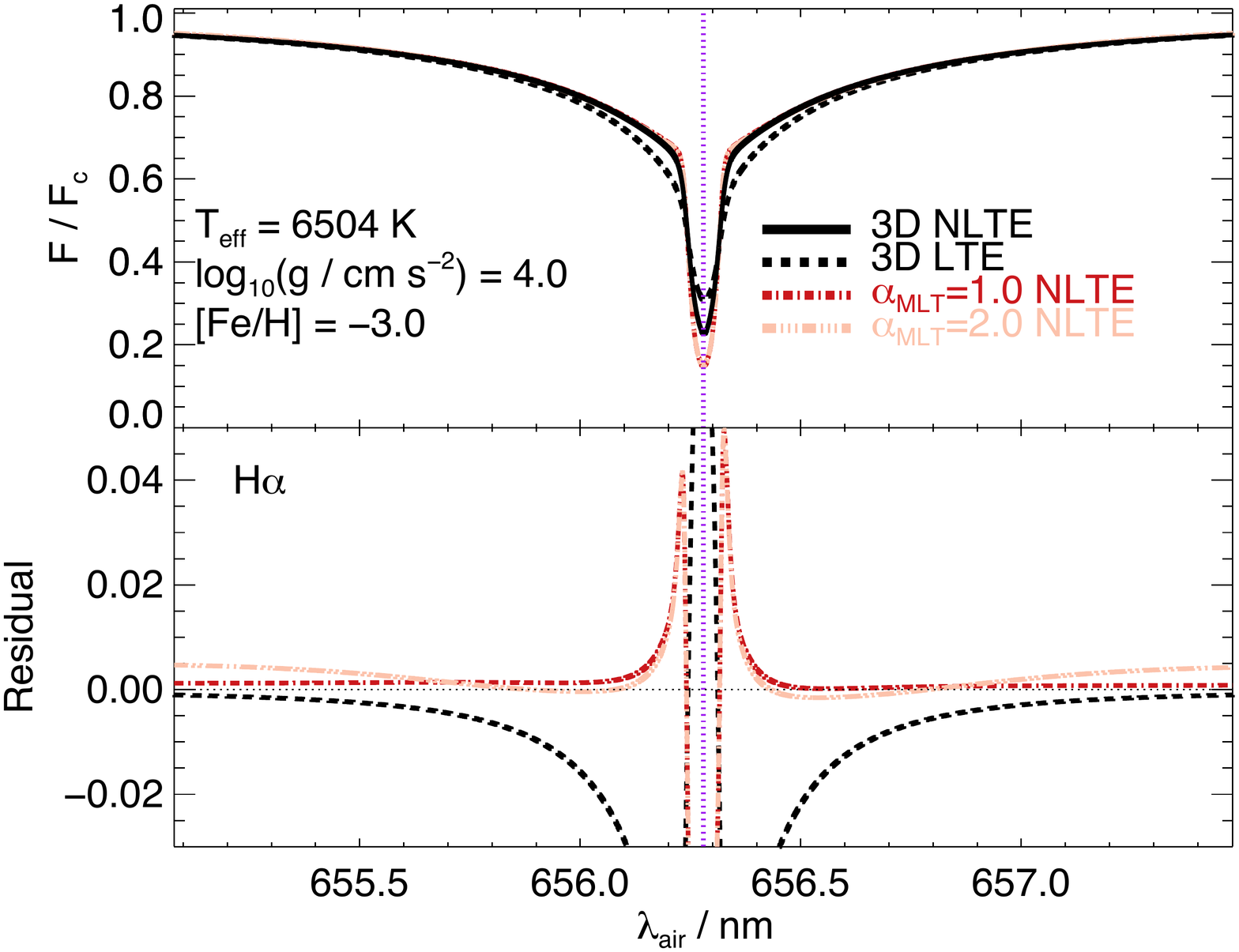}\includegraphics[scale=0.31]{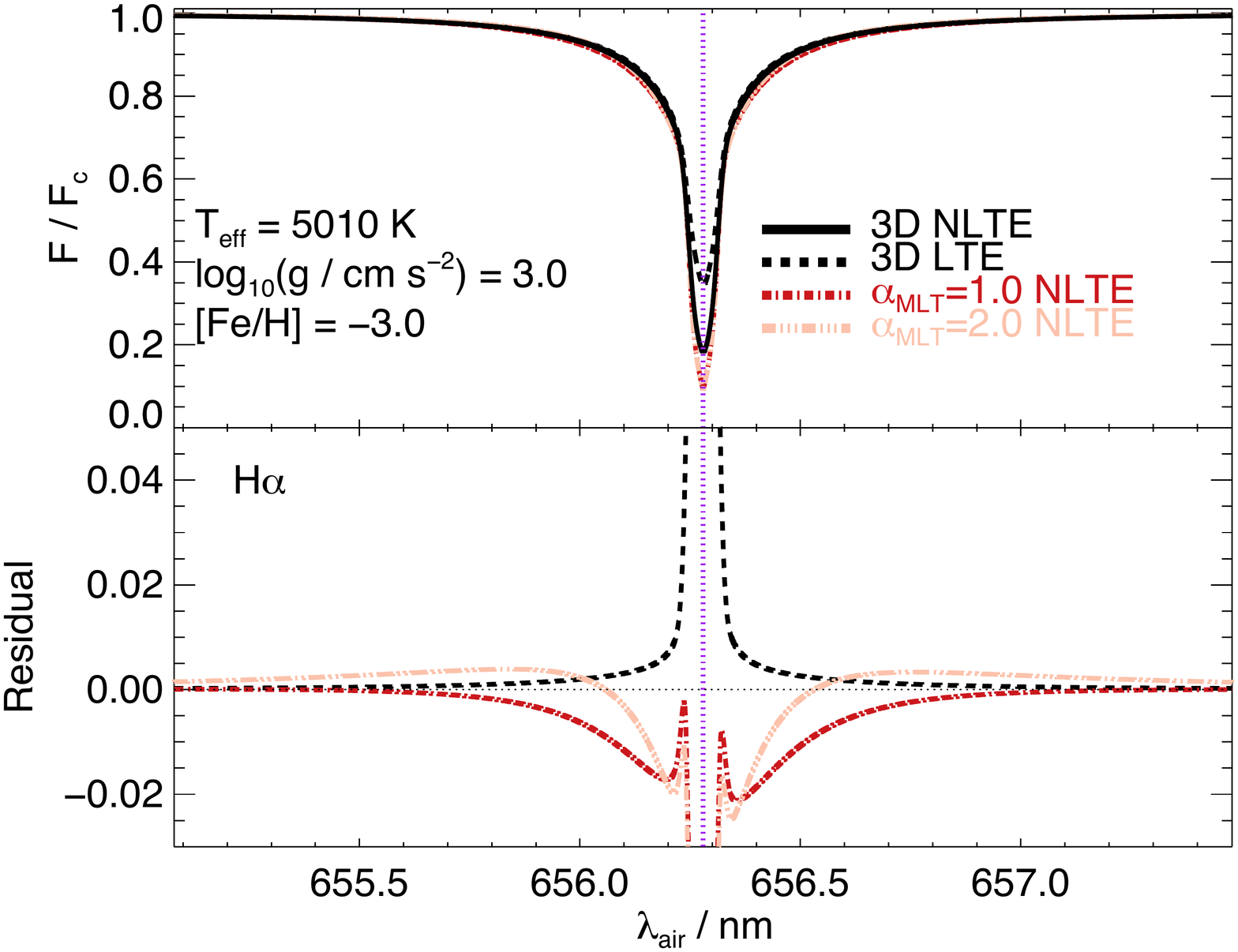}
\caption{$\halpha$~lines emergent from
the model atmospheres shown in \fig{figure_temp1}.
Rotational broadening is neglected.
The residuals are given with respect to the 3D non-LTE model.}
\label{figure_flux1}
\end{center}
\end{figure*}

\begin{figure*}
\begin{center}
\includegraphics[scale=0.31]{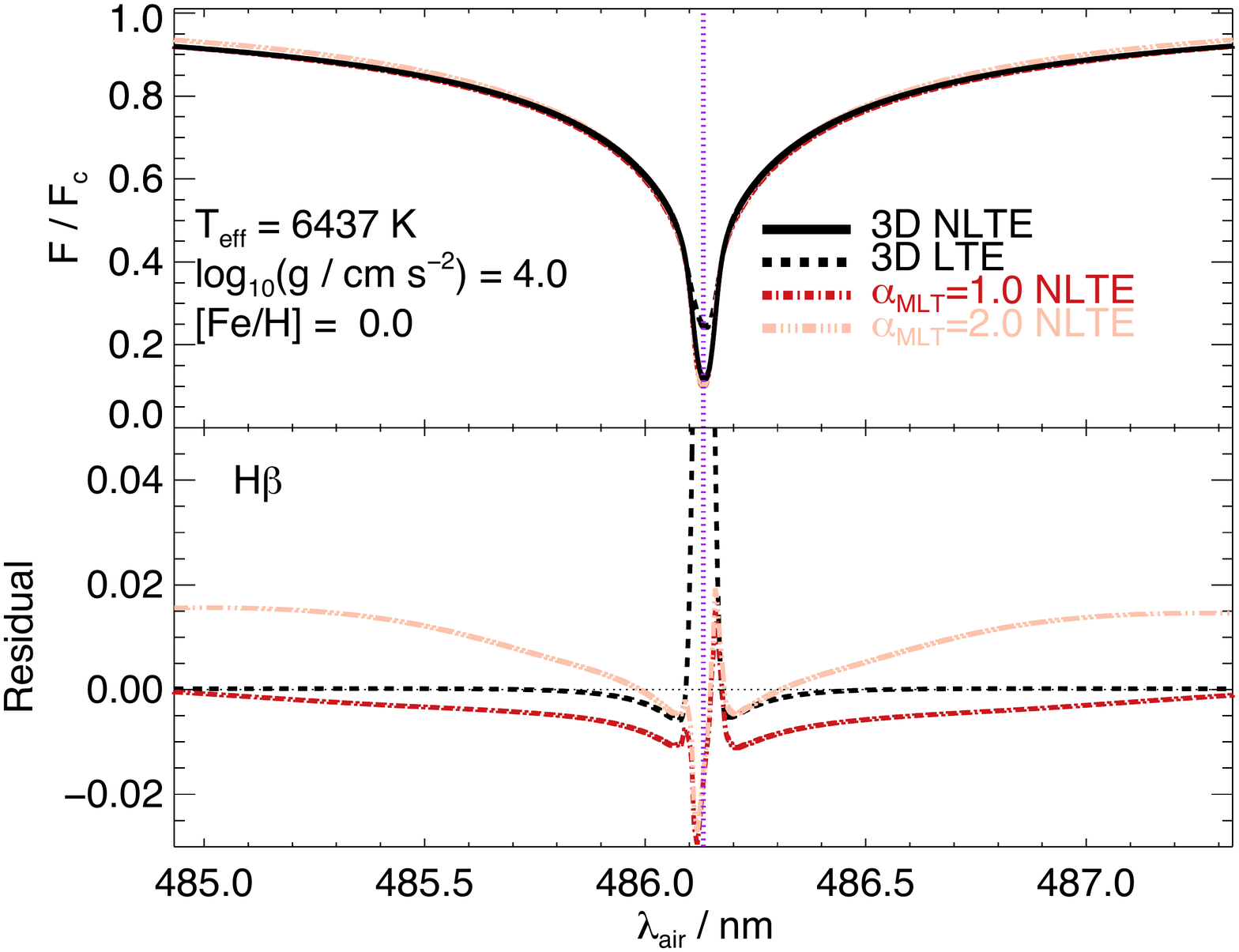}\includegraphics[scale=0.31]{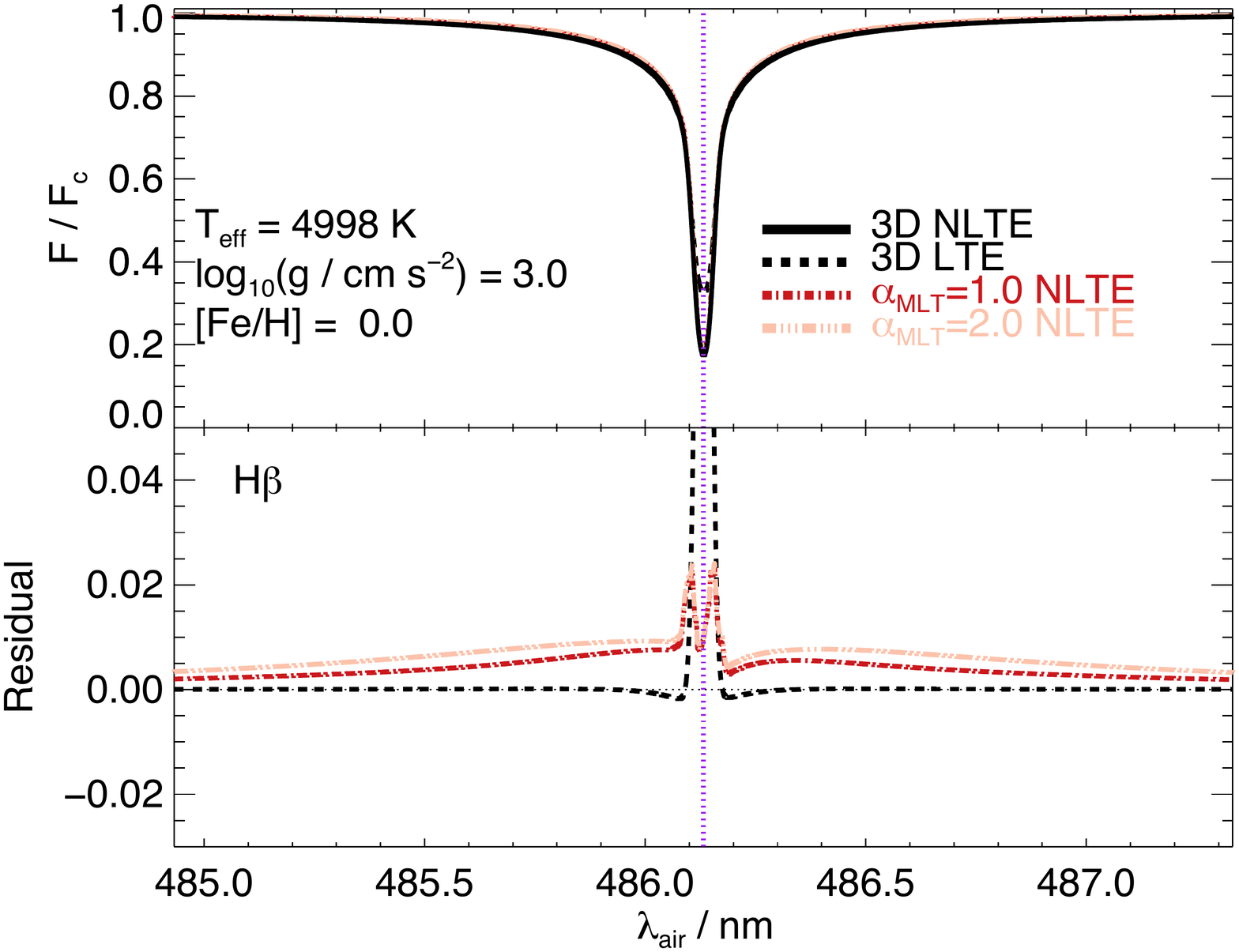}
\includegraphics[scale=0.31]{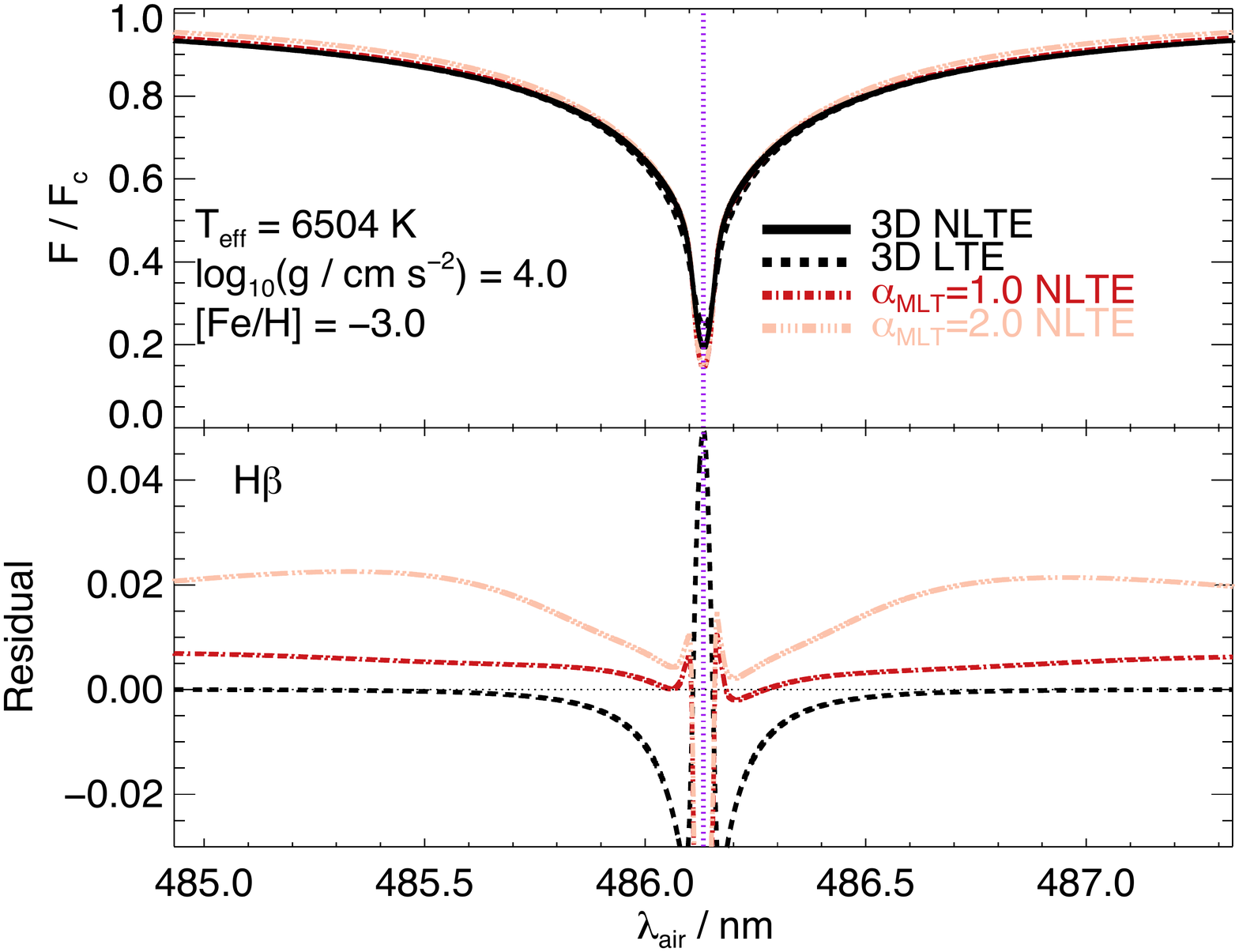}\includegraphics[scale=0.31]{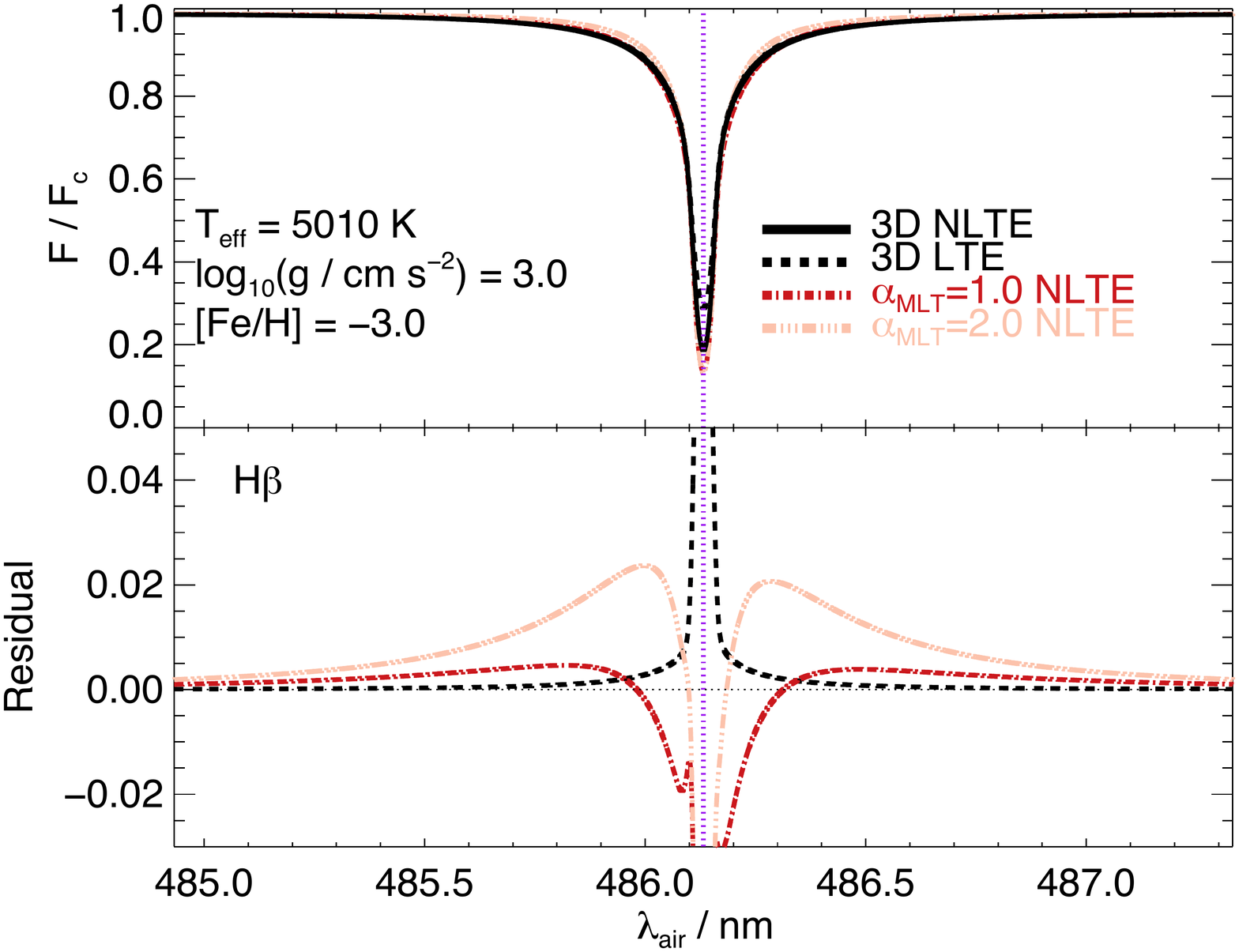}
\caption{$\hbeta$~lines emergent from
the model atmospheres shown in \fig{figure_temp1}.
Rotational broadening is neglected.
The residuals are given with respect to the 3D non-LTE model.}
\label{figure_flux2}
\end{center}
\end{figure*}

\begin{figure*}
\begin{center}
\includegraphics[scale=0.31]{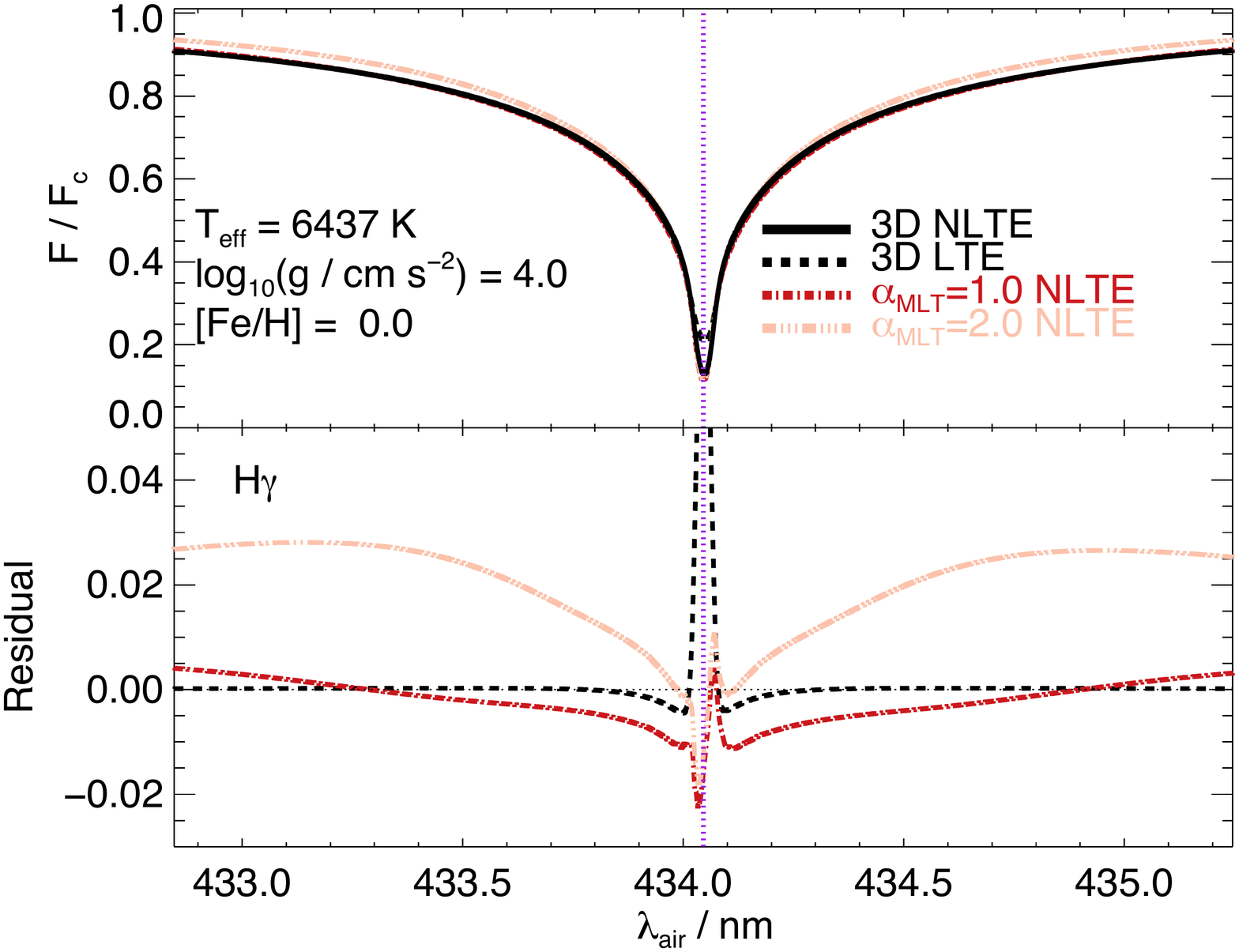}\includegraphics[scale=0.31]{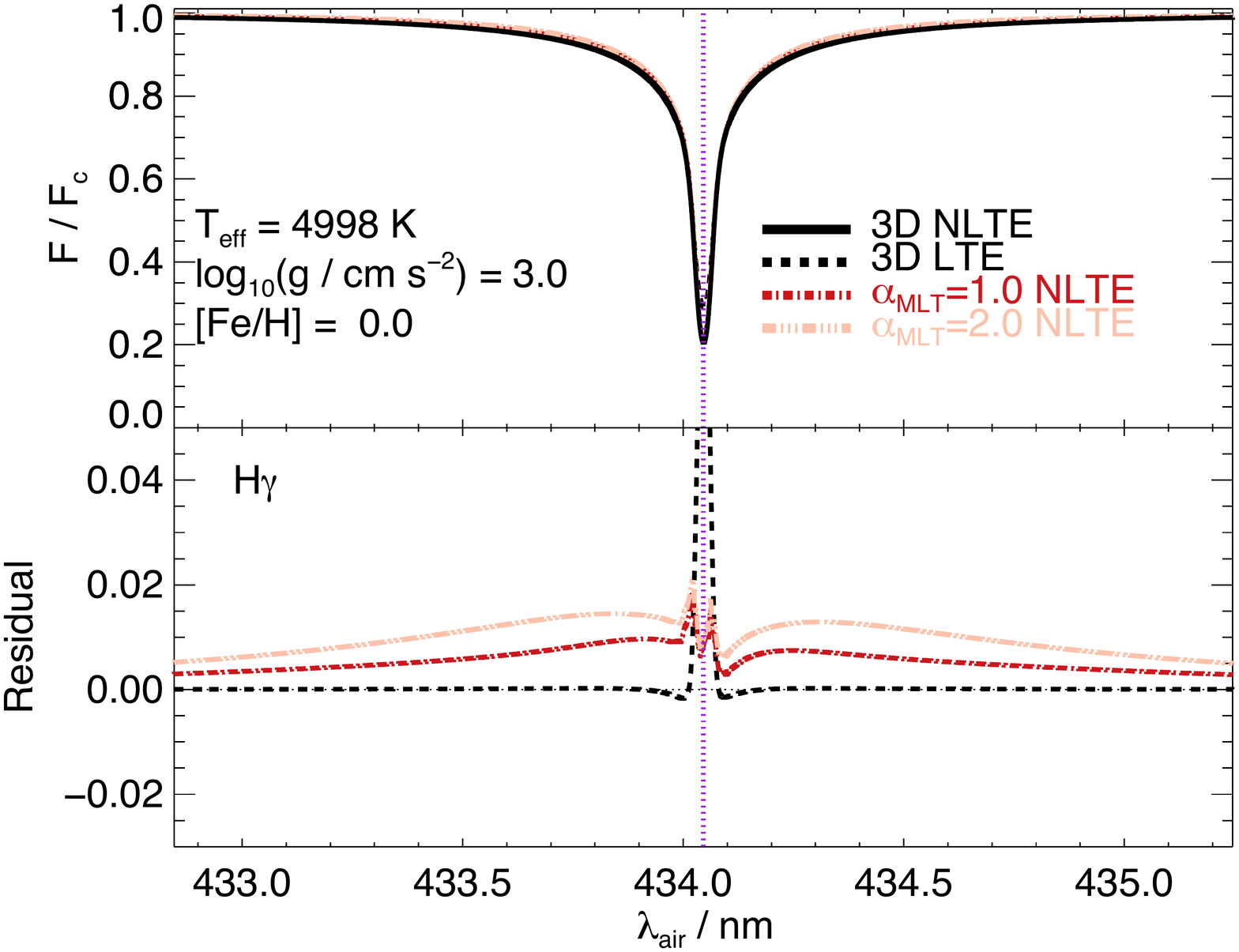}
\includegraphics[scale=0.31]{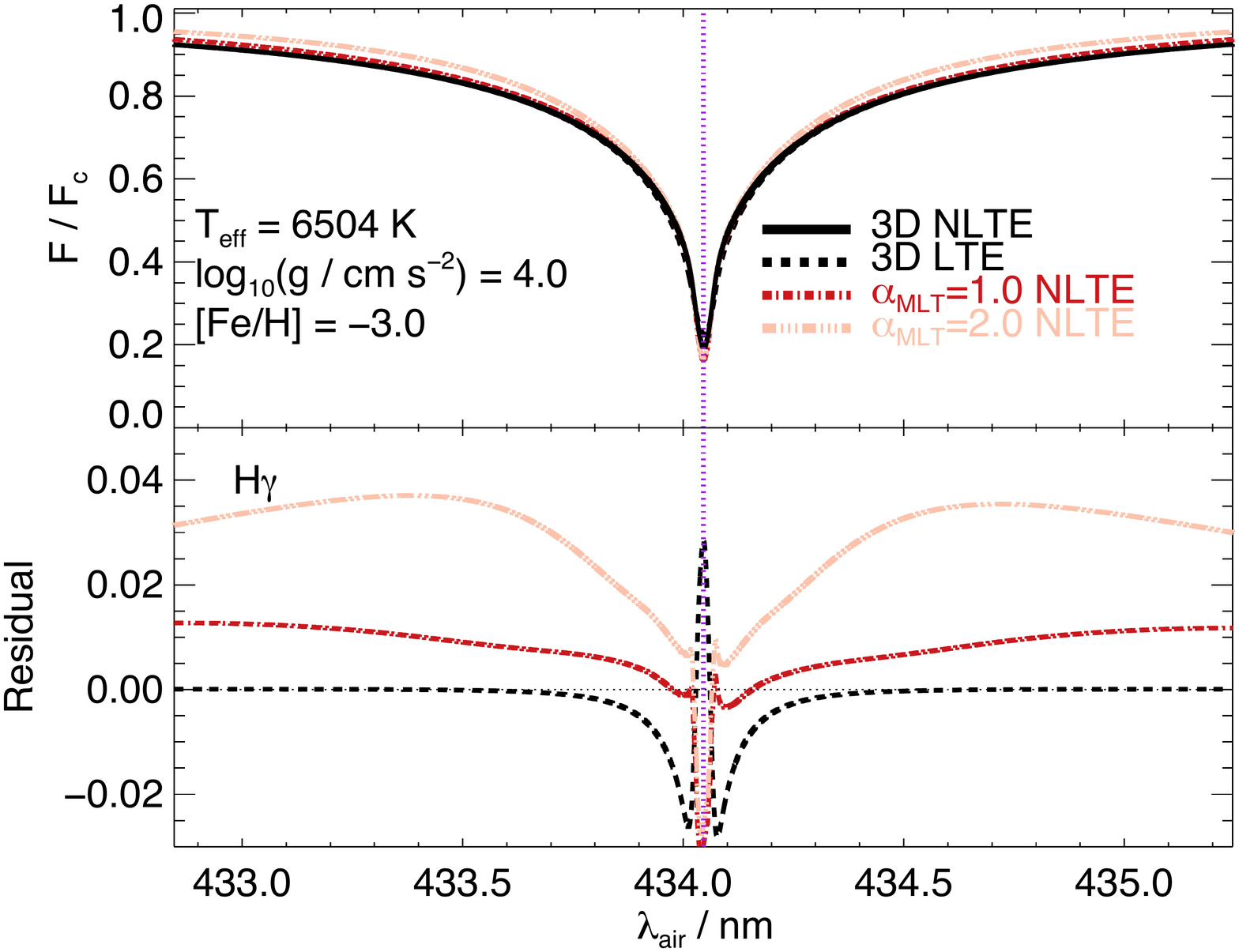}\includegraphics[scale=0.31]{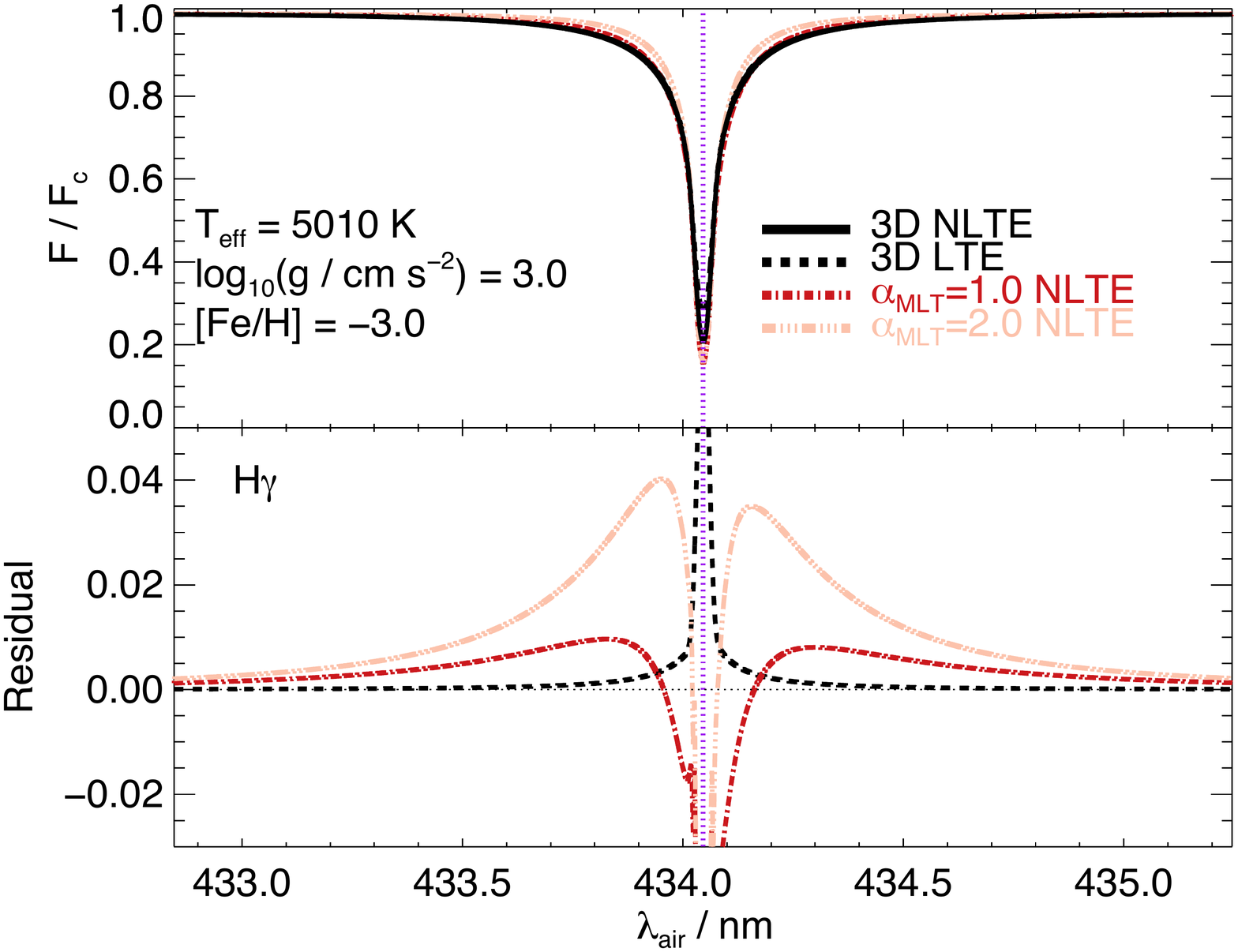}
\caption{$\hgamma$~lines emergent from
the model atmospheres shown in \fig{figure_temp1}.
Rotational broadening is neglected.
The residuals are given with respect to the 3D non-LTE model.}
\label{figure_flux3}
\end{center}
\end{figure*}

\subsection{Formation depths}
\label{results_cf}

It is useful to first consider where in the model
atmospheres that the Balmer lines form. The monochromatic
contribution function to the flux depression 
in the lines, 
$C_{\nu}\left(\bm{r},\,t\right)$~\citep[][Eq.~14]{2015MNRAS.452.1612A},
is useful here. 
For this discussion we
define a monochromatic mean formation depth
\citep[][Sect.~2.4]{2015MNRAS.452.1612A},
\phantomsection\begin{IEEEeqnarray}{rCl}
\label{equation_formationdepth}
q_{\nu}&=&
\frac{\int{\int{\lgr\,C_{\nu}\,\mathrm{d}^{3}r}}\,\mathrm{d}t}
{\int{\int{C_{\nu}}\,\mathrm{d}^{3}r}\,\mathrm{d}t}\, ,
\end{IEEEeqnarray}
where we adopted $\lgr$, the 
logarithmic vertical Rosseland-mean optical depth, 
as the reference depth scale.

In \fig{figure_cf1}~we illustrate
contribution functions for the $\halpha$~and $\hgamma$~lines
in vertical slices of the turn-off model
atmospheres shown in \fig{figure_temp1}.
Line formation generally follow the contours of 
vertical optical depth, rather than geometric height.
In turn, in the deeper regions these contours trace the 
granulation pattern.
It can be seen from the lighter shadings in the plots
for $\hgamma$~how 
the contribution function is sharply peaked
around $\lgr\approx0$; moreover,
the line formation is biased towards the hot convective upflows,
rather than the cool downflows.
For $\halpha$, the contribution function extends
further up into the atmosphere.

Across the grid, the Balmer line wings form mainly 
in the region $-1.0\lesssim\lgr\lesssim0.5$,
with weaker lines forming in the more optically-thick layers
(thus, in cooler stars, or for higher members 
of the Balmer series,
the formation region is pushed to deeper layers).
For the metal-poor turn-off model atmosphere,
the mean formation depths of the wings,
here taken to be $0.2\,\mathrm{nm}$~redward of the line centre,
are $q_{\nu}\approx-0.78$~($\halpha$), $-0.43$~($\hbeta$), 
and $-0.32$~($\hgamma$). 
For the metal-poor 
sub-giant model atmosphere, 
the mean formation depths of the wings 
are $q_{\nu}\approx-0.35$~($\halpha$), $-0.05$~($\hbeta$), 
and $0.06$~($\hgamma$).

Based on this discussion, the 
3D effects on the Balmer line wings are
expected to be more positive for $\hgamma$~than for $\hbeta$,
and for $\hbeta$~than for $\halpha$~(where a positive difference
implies weaker, or less depressed absorption lines in 1D than in 3D),
at least for higher values of mixing-length.
The contribution functions for higher members
of the Balmer series peak in 
deeper regions of the photosphere,
probing regions in the (upflows of the) 3D model atmospheres
that typically have increasingly steeper temperature gradients 
compared to in the corresponding 1D model atmospheres.
It also follows that the 
3D effects are expected to be
more sensitive to the mixing-length for higher members of the
Balmer series, since this parameter has a larger effect
deeper in the photosphere (\fig{figure_temp1}).
However, the signs and absolute magnitudes of these 
1D non-LTE versus 3D non-LTE differences
depend on the atmospheric
parameters, adopted mixing-length, and wavelength (\sect{results_balmer}).

We briefly note that
the distribution of line formation of the Balmer wings
is slightly skewed towards the optically thin regions of the photosphere.
For $\halpha$~in particular, some line formation
is able to occur very high up in the atmosphere, $\lgr\approx-3.0$,
in the regions where the local
gas temperature is higher than its surroundings.
These hot regions are associated with reversed granulation,
and this effect is enhanced in the metal-poor regime
\citep[][Appendix A]{2013A&amp;A...560A...8M},
as can be seen in \fig{figure_cf1}.

For the Sun, Balmer line cores are known to have a 
significant chromospheric contribution
\citep[e.g.][]{2012ApJ...749..136L},
despite the ability of 1D model photospheres to approximately 
reproduce time-averaged observations of the core flux 
\citep[e.g.][]{2004ApJ...610L..61P}. 
For cooler stars, such as the red giant Arcturus, 
1D model photospheres fail to satisfactorily reproduce observations
of the Balmer line cores \citep[e.g.][]{2004ApJ...610L..61P,
2016A&amp;A...594A.120B}.
Modelling the chromospheric contribution to the emergent Balmer line cores
is beyond the scope of the present work;
we just note here that 
our models are not reliable for the emergent Balmer line cores.

\subsection{Departure coefficients}
\label{results_departurecoefficients}

The departure coefficients are defined as the ratio
of non-LTE to LTE level populations:
\phantomsection\begin{IEEEeqnarray}{rCl}
\label{equation_departurecoefficients}
    \beta_{i}&=&\frac{n_{i,\text{non-LTE}}} {n_{i,\text{LTE}}}\, .
\end{IEEEeqnarray}
To first order, the non-LTE line opacity
is enhanced by a factor $\beta_{\text{lo}}$,
while the non-LTE line source function is enhanced by a factor 
$\beta_{\text{up}}/\beta_{\text{lo}}$~\citep[e.g.][Chapter 2]{2003rtsa.book.....R}.

In \fig{figure_b1}~we illustrate departure coefficients in
the model atmospheres shown in \fig{figure_temp1}.
They are plotted against the reference vertical optical depth
($\lgr$); this correlates well with line formation,
with the Balmer line wings mostly forming 
around $-1.0\lesssim\lgr\lesssim0.5$~(\sect{results_cf}).

\fig{figure_b1}~demonstrates broad distributions
for the departure coefficients of the 
$n=2$~and $n>2$~levels, through which the Balmer lines form.
The departure mechanisms (which we briefly describe at the end
of this section) are similar in 1D and in 3D, and consequently 
the departure coefficients in the 1D model atmospheres
follow the distributions in the 3D model atmospheres.
A notable exception is for 
$n=2$~in the metal-poor turn-off star.
Here, the distribution of departure coefficients is bimodal.
The mode corresponding to larger departures~(reaching 
$\beta_{2}\approx1.4$~in the upper layers) corresponds
to line formation in the regions of high temperature in the upper 
atmosphere (see \fig{figure_cf1}, second column).  These reverse granulation
features are obviously not present in the 1D model atmospheres.
As such, the departure coefficients from the 1D model atmospheres
follows the other mode, which stays close to unity.

The departure coefficients have a strong dependence on metallicity.
Generally, at solar metallicity,
the departures in the region $-1.0\lesssim\lgr\lesssim0.5$~are
not very severe.
For the Balmer levels, they are less than around $5\%$.
The departures grow towards lower metallicities.
At $\feh=-3.0$, the departures are
of the order $10$-$20\%$.
This dependence of the departure coefficients
on metallicity translates into more severe 
non-LTE effects on the emergent Balmer line wings
at lower $\feh$, consistent with \citet[][]{2007A&amp;A...466..327B},
and with our own results in \sect{results_balmer}.

The departure coefficients also have a dependence 
on effective temperature.
At higher effective temperatures, there is predominantly a
source function effect
on the Balmer line wings: for the lower Balmer level
$\beta_{2}$~stays close to unity,
while for the upper Balmer levels $\beta_{\text{up}}$~are larger than unity.
The Balmer line wings are thus weaker when departures from LTE are taken
into account. 
On the other hand, towards lower effective temperatures,
an opacity effect becomes increasingly more important:
$\beta_{2}$~becomes larger than unity, and the ratios
$\beta_{2}$~to $\beta_{\text{up}}$~move closer to unity.
The opacity effect strengthens the Balmer line wings,
whereas the source function effect weakens them.
Since these effects are in competition,
the non-LTE effects
on the emergent line profiles
are not necessarily more severe at lower $\teff$,
as we shall 
demonstrate in \sect{results_balmer}.

To aid intuition, we provide a brief, qualitative description
on what causes these opacity and source function effects.
The overpopulation of the $n=2$~level, and thus the opacity effect,
is largely driven by photon pumping through the
$\lyalpha$~line: the suprathermal UV radiation field
leads to a flow from the $n=1$~population reservoir into the $n=2$~level;
this effect is enhanced at lower metallicities
and lower effective temperatures
(\fig{figure_b1}).
Were radiative coupling via the $\lyalpha$~line to be omitted,
the lower Balmer level would satisfy
$\beta_{2}\approx1$, even at the top of the
simulation domain $\lgr\approx-5$.
A careful treatment of $\lyalpha$~and the UV flux
is likely important for accurate
non-LTE modelling of the Balmer lines:
a future work should examine how
relaxing the assumption of complete redistribution
for the $\lyalpha$~wings in the stellar photosphere,
may affect the departure coefficients
of the $n=2$~level and through that the Balmer line wings
\citep[as was done for a solar chromospheric
model by][]{2012ApJ...749..136L}.

The statistical equilibrium of the $n>2$~levels 
is more complicated.
Close collisional coupling within these excited levels,
as well as with \ion{H}{II}, means that at high effective temperatures, 
numerous strong lines and continua significantly affect the statistical
equilibrium. The picture becomes simpler
towards lower effective temperatures, 
where \ion{H}{I} lines are weaker and so
collisional coupling with the $n=2$~level becomes more 
effective at pushing the departure coefficients
of the upper Balmer levels, $\beta_{\text{up}}$~towards that of 
the lower Balmer level, $\beta_{2}$.

\subsection{Emergent Balmer lines}
\label{results_balmer}

In \fig{figure_flux1} for $\halpha$,
\fig{figure_flux2} for $\hbeta$,
and \fig{figure_flux3} for $\hgamma$,
we illustrate the line profiles
emergent from the model atmospheres shown in \fig{figure_temp1}.
These plots directly illustrate the impact of assuming LTE
and of using 1D hydrostatic model atmospheres
on the synthetic Balmer lines.
We focus on the line wings, rather than the cores,
for the reasons given in \sect{results_cf}.

The far outer regions of the Balmer line wings
are typically stronger in 3D non-LTE models than in 1D non-LTE models.
The line formation in the (upflows of the) 3D model atmospheres
is enhanced owing to their steeper temperature gradients
relative to in the corresponding 1D model atmospheres.
As expected from \sect{results_cf},
this effect is more pronounced for higher values of mixing-length
which predict shallower 1D temperature stratifications
in the deepest regions (\fig{figure_temp1}),
and for higher members of the Balmer series
(e.g.~\fig{figure_flux3})~rather than for $\halpha$, because
the latter forms higher up in the atmosphere
where differences in the temperature stratifications 
of the 1D and (upflowing columns of the) 3D model atmospheres are lesser.
On its own, stronger outer wings
implies lower inferred effective temperatures
from the 3D models than from the 1D models
(and this is in fact the case for
e.g.~$\hgamma$~and high $\alpha_{\mathrm{MLT}}$; 
see \sect{discussion_effects}).

On the other hand, the inner regions of the Balmer line wings
can be weaker in 3D non-LTE models than in 1D non-LTE models,
at least for stronger lines that form higher up in the atmosphere 
such as for example $\halpha$~in the solar-metallicity turn-off star
(\fig{figure_flux1}).
This result reflects how the
average 3D temperature stratifications 
(\mtd~in \fig{figure_temp1}) can be shallower than the 1D 
temperature stratifications, in the region
$0.0\lesssim\lgr\lesssim-1.0$.
On its own, weaker inner wings
implies higher inferred effective temperatures
from the 3D models than from the 1D models.

Figs~\ref{figure_flux1},~\ref{figure_flux2}, and
\ref{figure_flux3}~indicate that smaller values of mixing-length
tend to reduce the 
3D effects, particularly in the outer wings. 
This is because low mixing-lengths imply
a 1D stratification below $\lgr=0$~closer
to the high-temperature upflows in the 3D model
(\fig{figure_temp1}),
which is where much of the Balmer line formation occurs (\sect{results_cf}).
This is consistent with the
low mixing-lengths, $\alpha_{\text{MLT}}\approx0.5$, calibrated by
\citet{1993A&amp;A...271..451F,1994A&amp;A...285..585F}
and \citet{2002A&amp;A...385..951B}. 
However, there does not appear to be a value of $\alpha_{\text{MLT}}$~for 
which the 1D model line shapes adequately reproduce the 3D model line shapes,
for any particular set of atmospheric parameters
\citep[e.g.][]{2009A&amp;A...502L...1L},
and inhomogeneities in the 3D model atmospheres have a non-negligible
impact on the Balmer lines
\citep[e.g.][]{2013A&amp;A...554A.118P}.

The non-LTE effects are generally less pronounced than the 
3D effects.
They mainly affect the inner wings of $\halpha$,
and become less severe for higher members 
of the Balmer series.
This is because the stronger $\halpha$~line 
forms in more optically thin regions (\sect{results_cf}), 
where the departure coefficients deviate more
significantly from unity (\sect{results_departurecoefficients}).
For similar reasons, the 
non-LTE effects are 
most severe closer to the cores of the lines, 
whereas they are typically insignificant in the outer wings.

We discussed in \sect{results_departurecoefficients} that
the departure coefficients have a strong dependence on
metallicity. 
Figs~\ref{figure_flux1},~\ref{figure_flux2}, and
\ref{figure_flux3}~show that this
translates into more pronounced 
non-LTE effects
on the inner wings (of all Balmer lines) towards lower metallicities.
However, even at low metallicities,
it is mainly $\halpha$~that is influenced by departures from LTE.

We also discussed in \sect{results_departurecoefficients} that the
departure coefficients have a complicated
dependence on effective temperature.
A source function effect dominates at higher effective temperatures:
it serves to weaken the Balmer line wings with respect to LTE.
This is demonstrated in \fig{figure_flux1} and in \fig{figure_flux2},
in the inner wings of 
$\halpha$~and $\hbeta$, in the turn-off stars at either metallicity.
The competing opacity effect becomes more important
at lower effective temperatures:
it serves to strengthen the Balmer line wings with respect to LTE.
This is also demonstrated in these figures, in the inner wings of 
$\halpha$~and $\hbeta$, this time in the metal-poor 
sub-giant.
Although the departures are more severe at lower effective temperature,
this does not necessarily translate into
more pronounced non-LTE effects
towards lower effective temperatures,
because the opacity and source function effects are in competition.

As a result of how the 
3D non-LTE effects depend
on distance from the line core of 
the Balmer line in question,
it is not possible to define a definitive, quantitative
3D non-LTE versus 1D LTE (for example) effective temperature correction:
the value inevitably depends on the adopted method of
analysis. Nevertheless, some intuition for the errors incurred
by a 1D LTE analysis can be drawn from our analysis of 
a sample of benchmark stars in the next section,
\sect{benchmark}.

%-------------------------------------------------------------------------------
\section{Fits to benchmark stars}
\label{benchmark}

\begin{figure*}
\begin{center}
\includegraphics[scale=0.55]{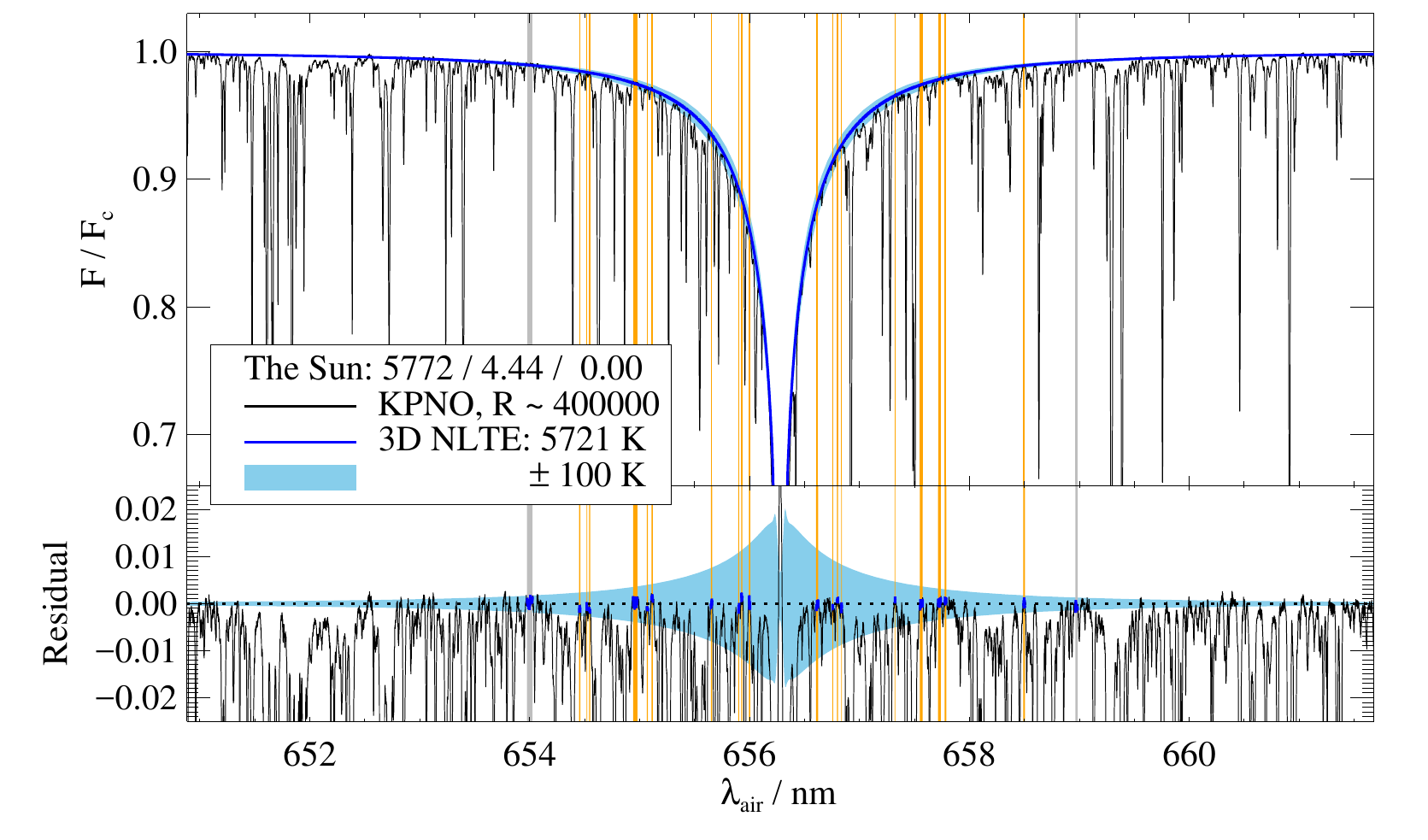}\includegraphics[scale=0.55]{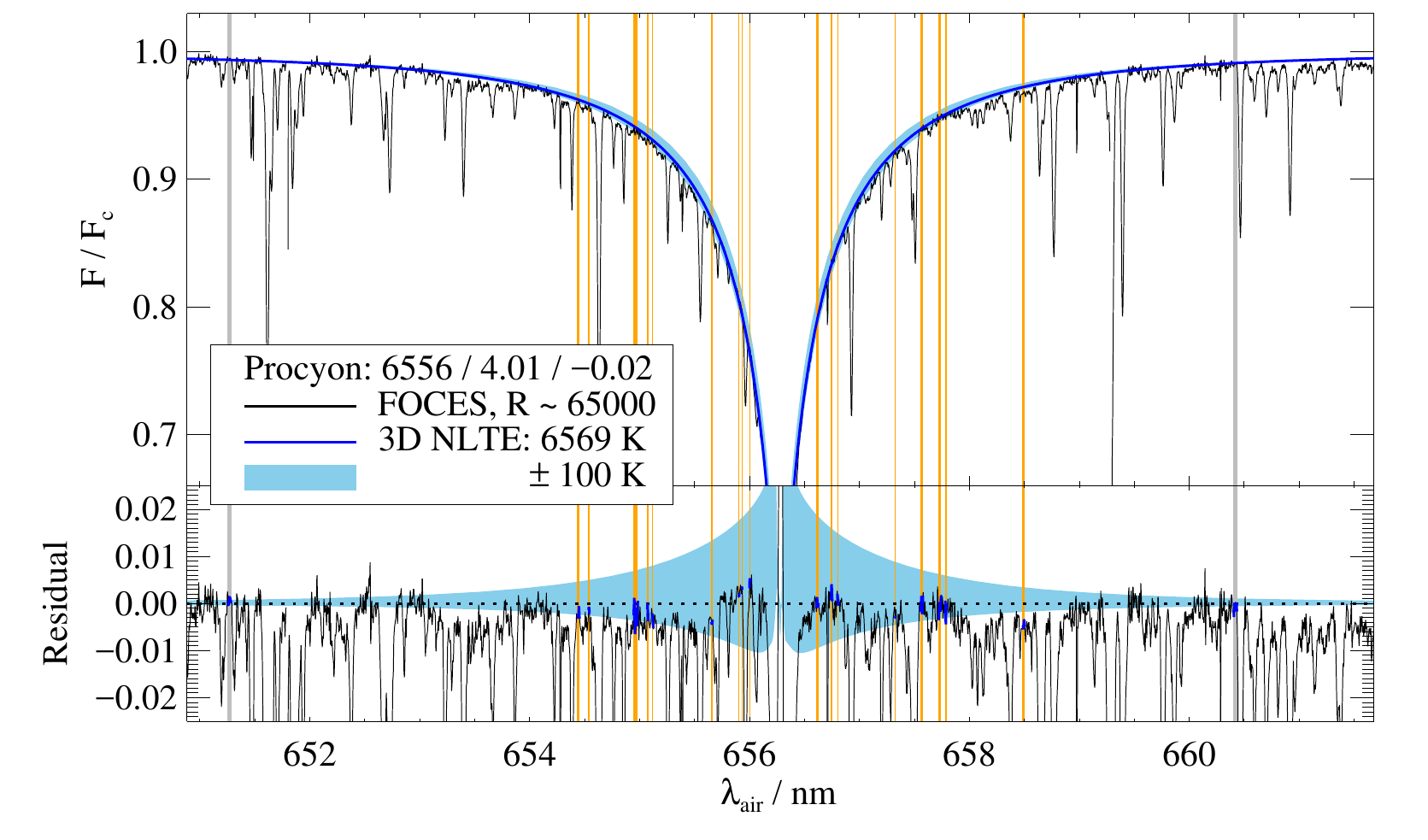}
\includegraphics[scale=0.55]{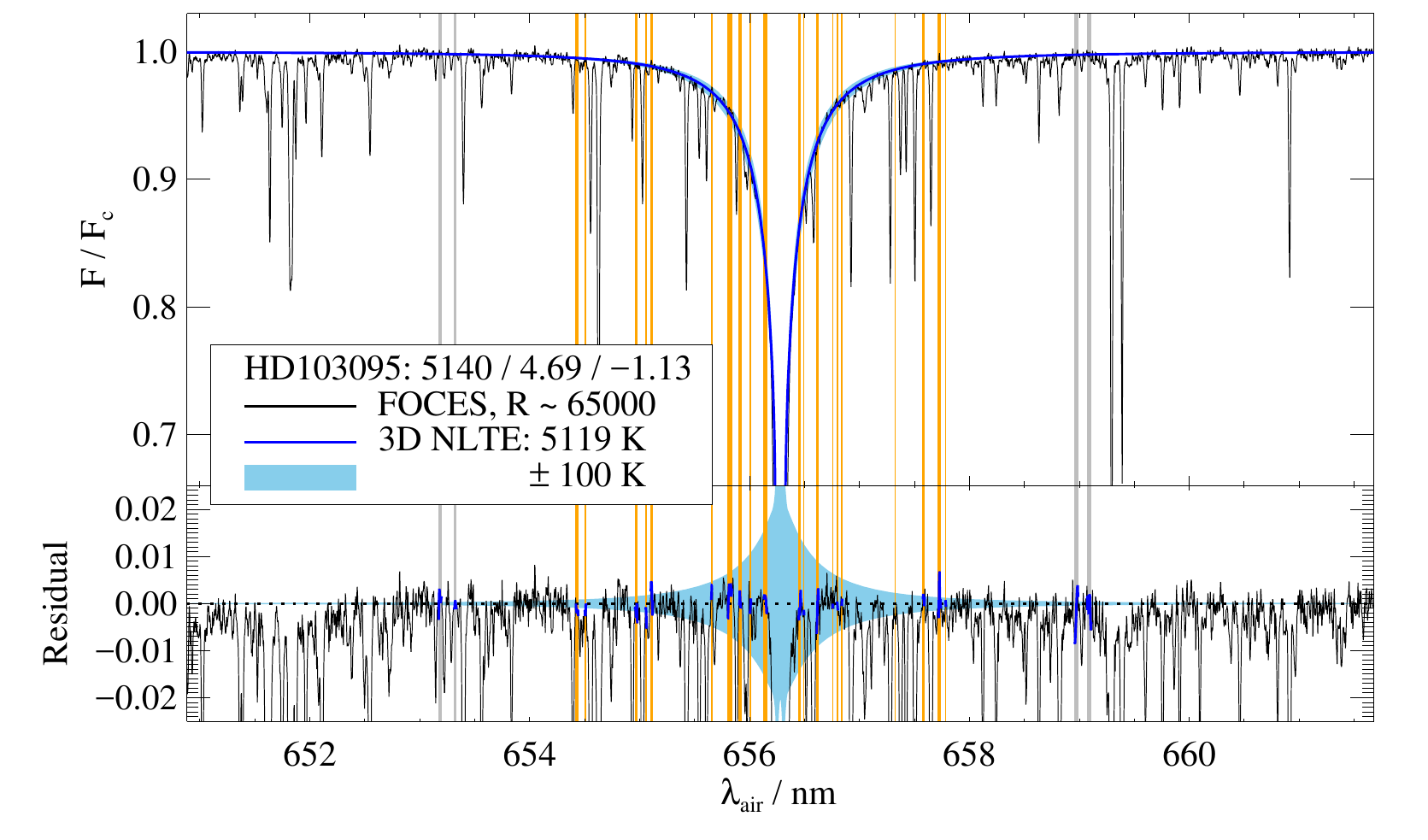}\includegraphics[scale=0.55]{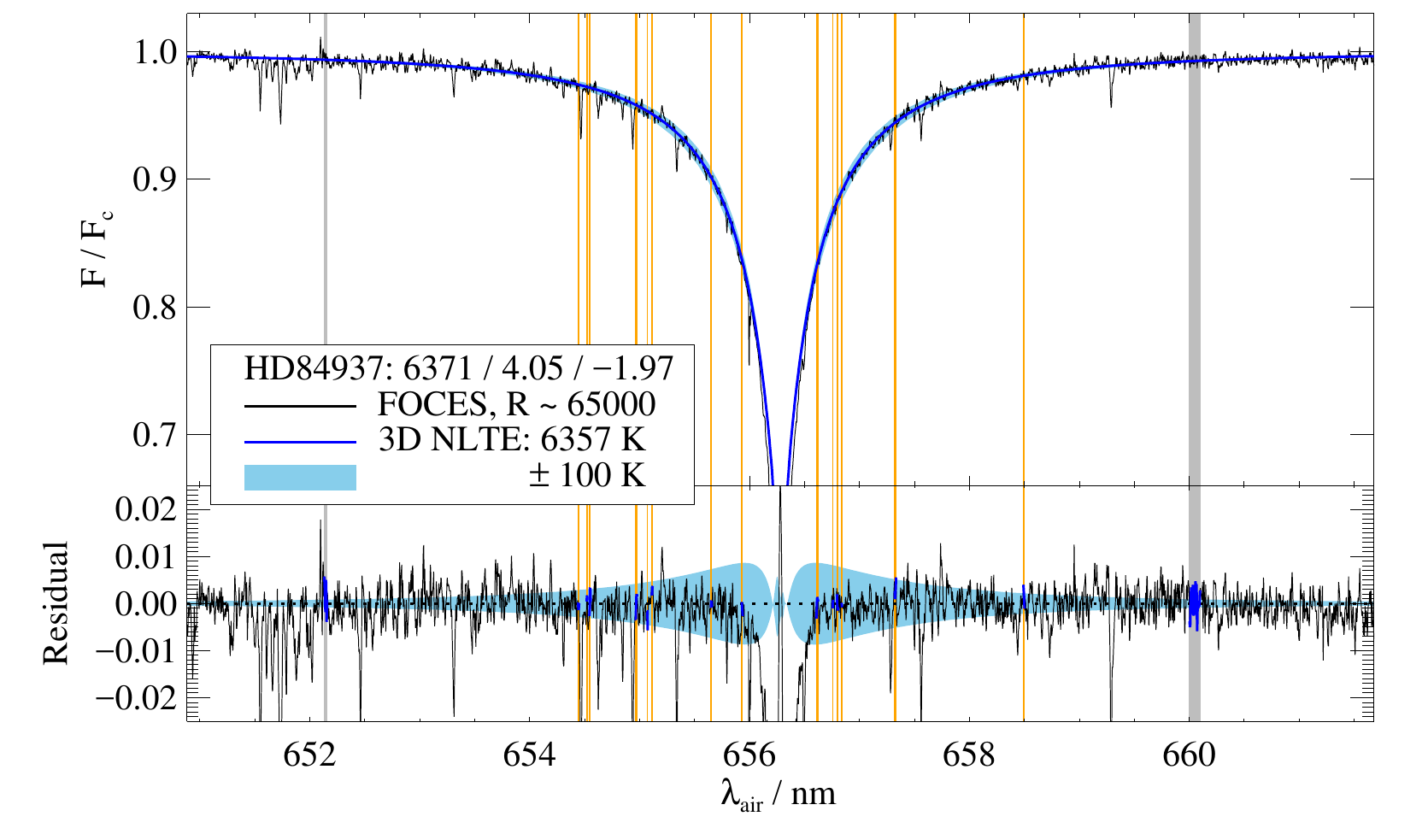}
\includegraphics[scale=0.55]{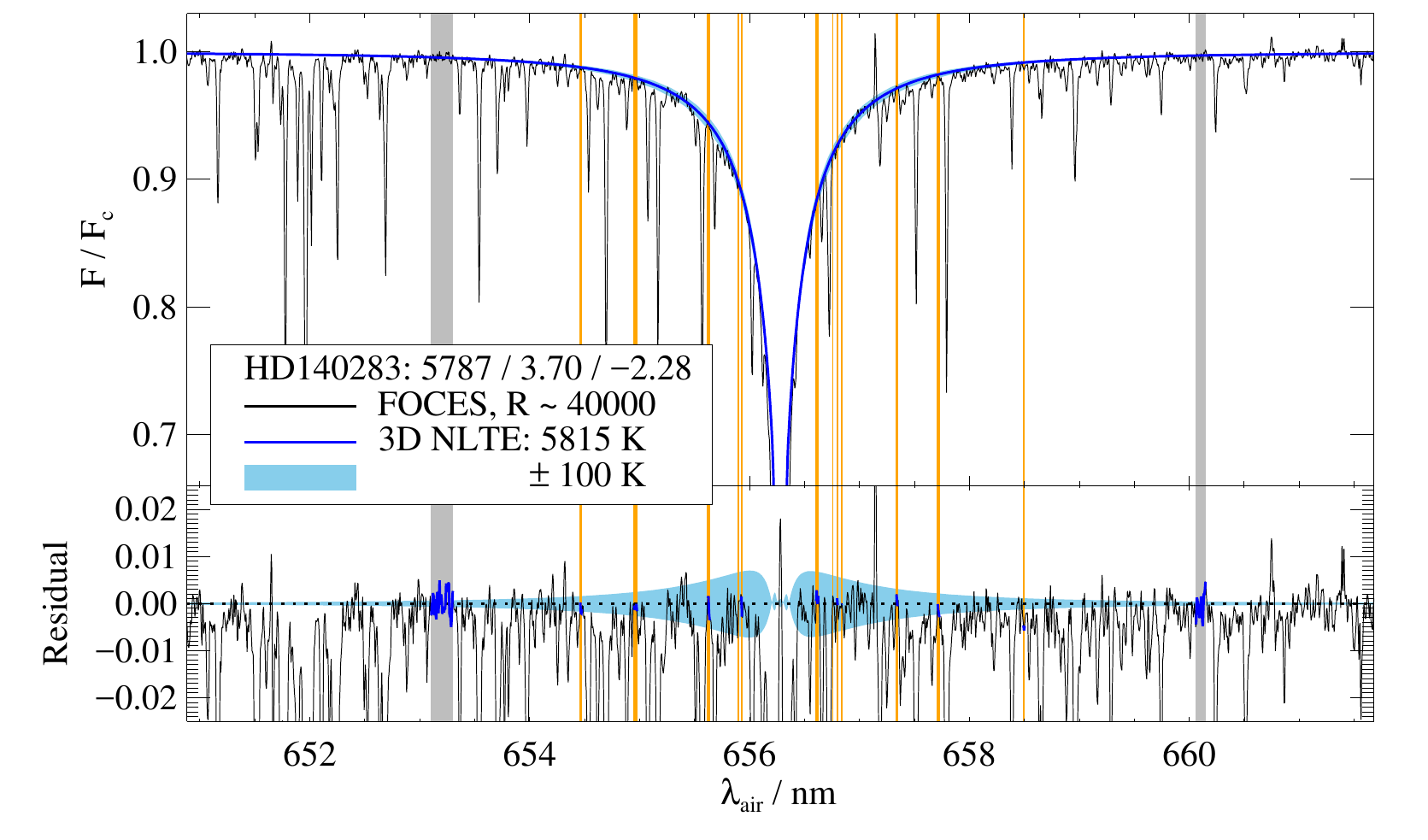}\includegraphics[scale=0.55]{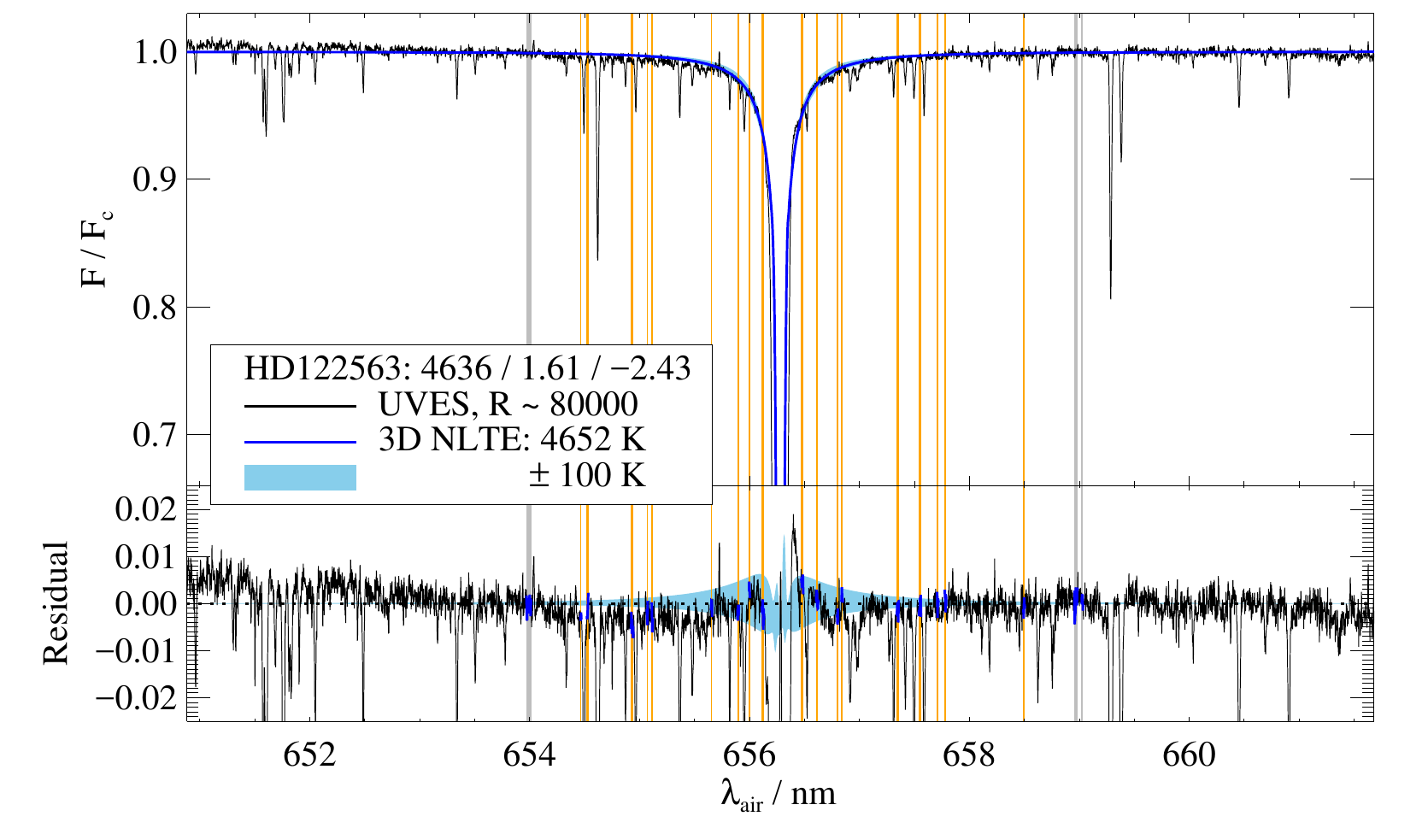}
\caption{$\halpha$ line profiles observed in benchmark stars, 
compared to the best-fitting
3D non-LTE model when effective temperature is taken as a free parameter. 
The reference parameters $\teff$/$\lgg$/$\feh$~of each star are given in the
legends. The continuum and line masks are shown as 
dark and light vertical bands, respectively.
The light shaded region indicates the effect
of adjusting the effective temperature by $\pm 100\,\mathrm{K}$, 
where lower effective temperatures result in a
weaker line and thus a higher normalised flux.
Residuals between the 3D non-LTE model and the observations 
are shown in the lower panel.
Residuals inside the masks 
are highlighted using thick lines; only these pixels 
have any influence on the fitting procedure.}
\label{figure_fit1}
\end{center}
\end{figure*}

\begin{figure*}
\begin{center}
\includegraphics[scale=0.55]{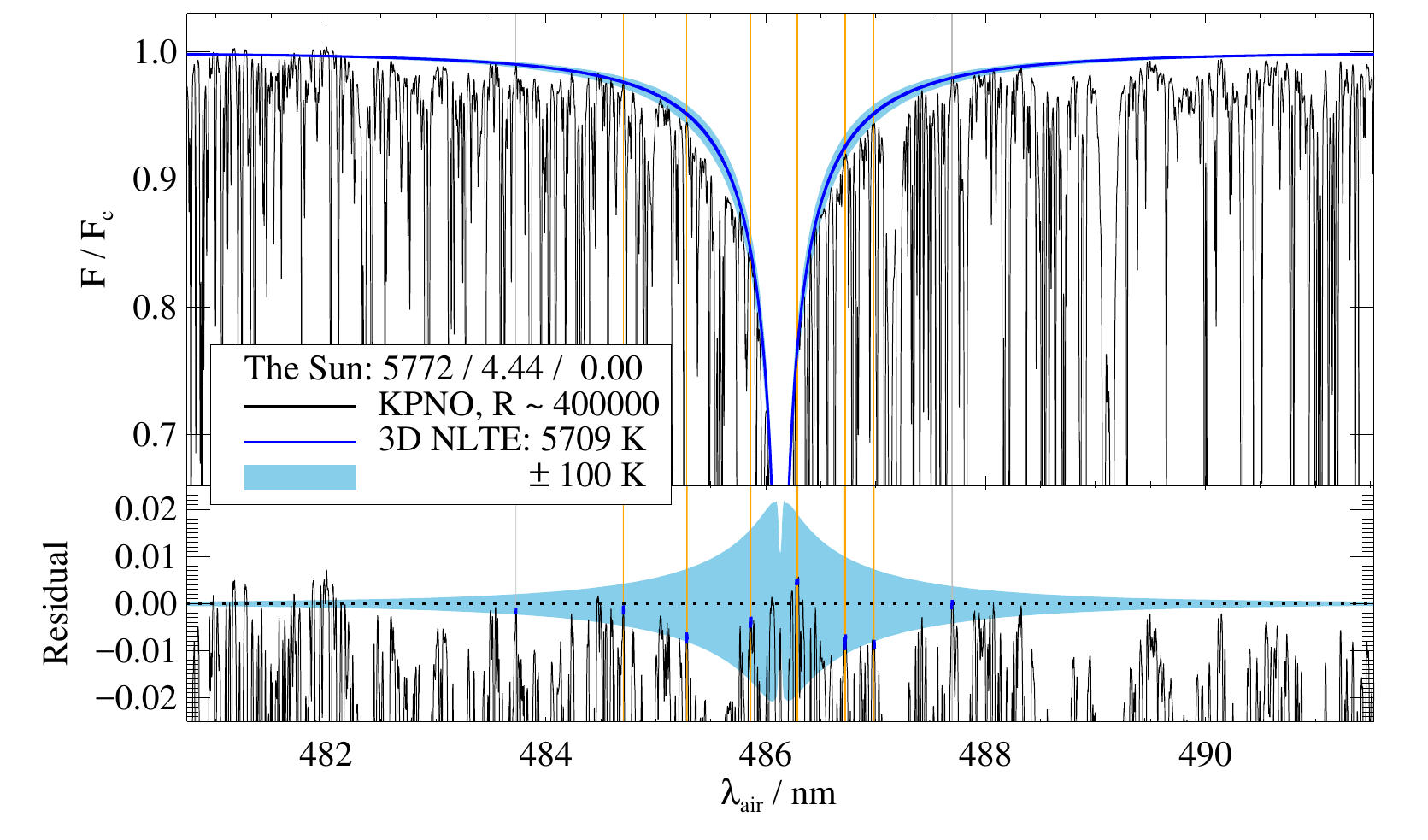}\includegraphics[scale=0.55]{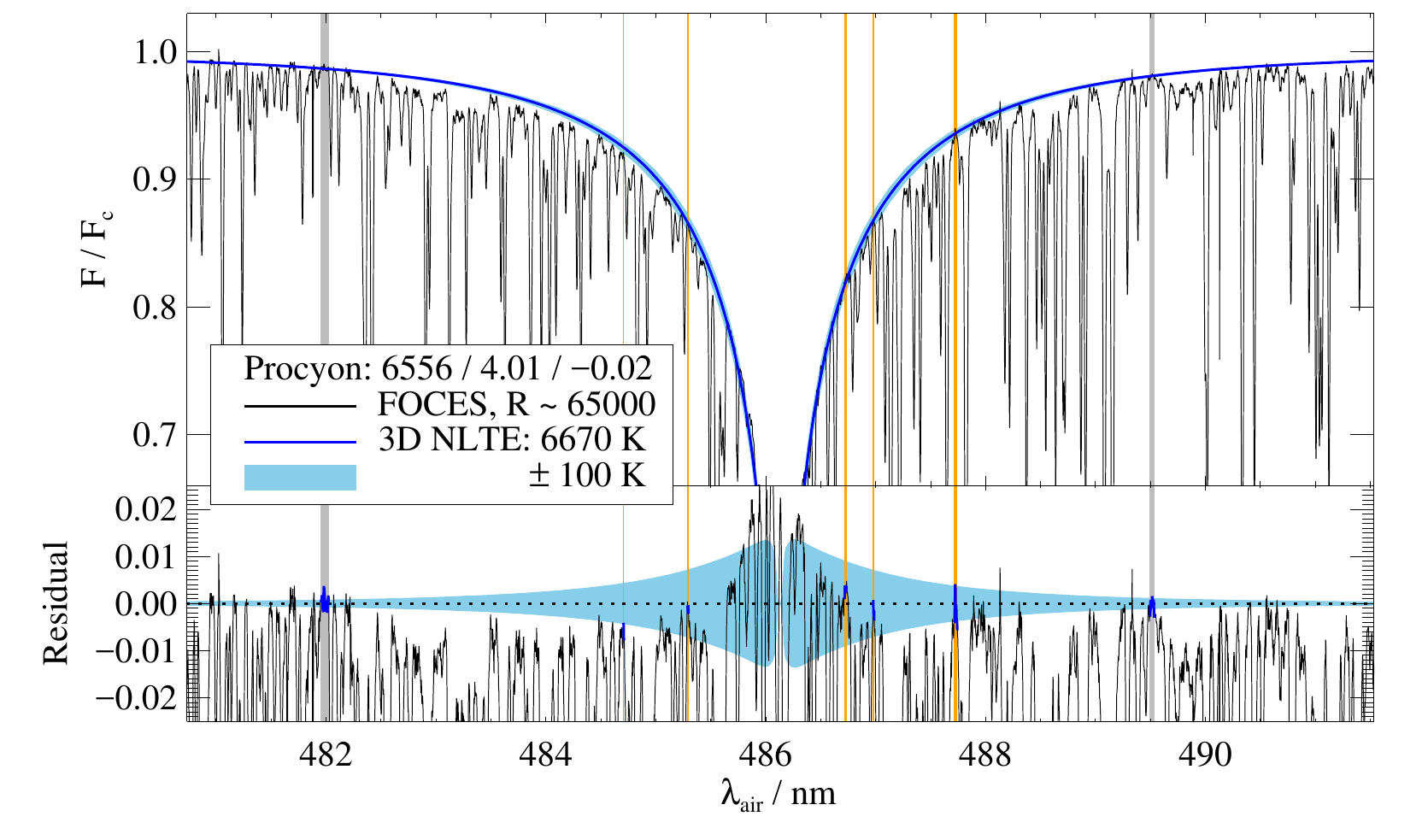}
\includegraphics[scale=0.55]{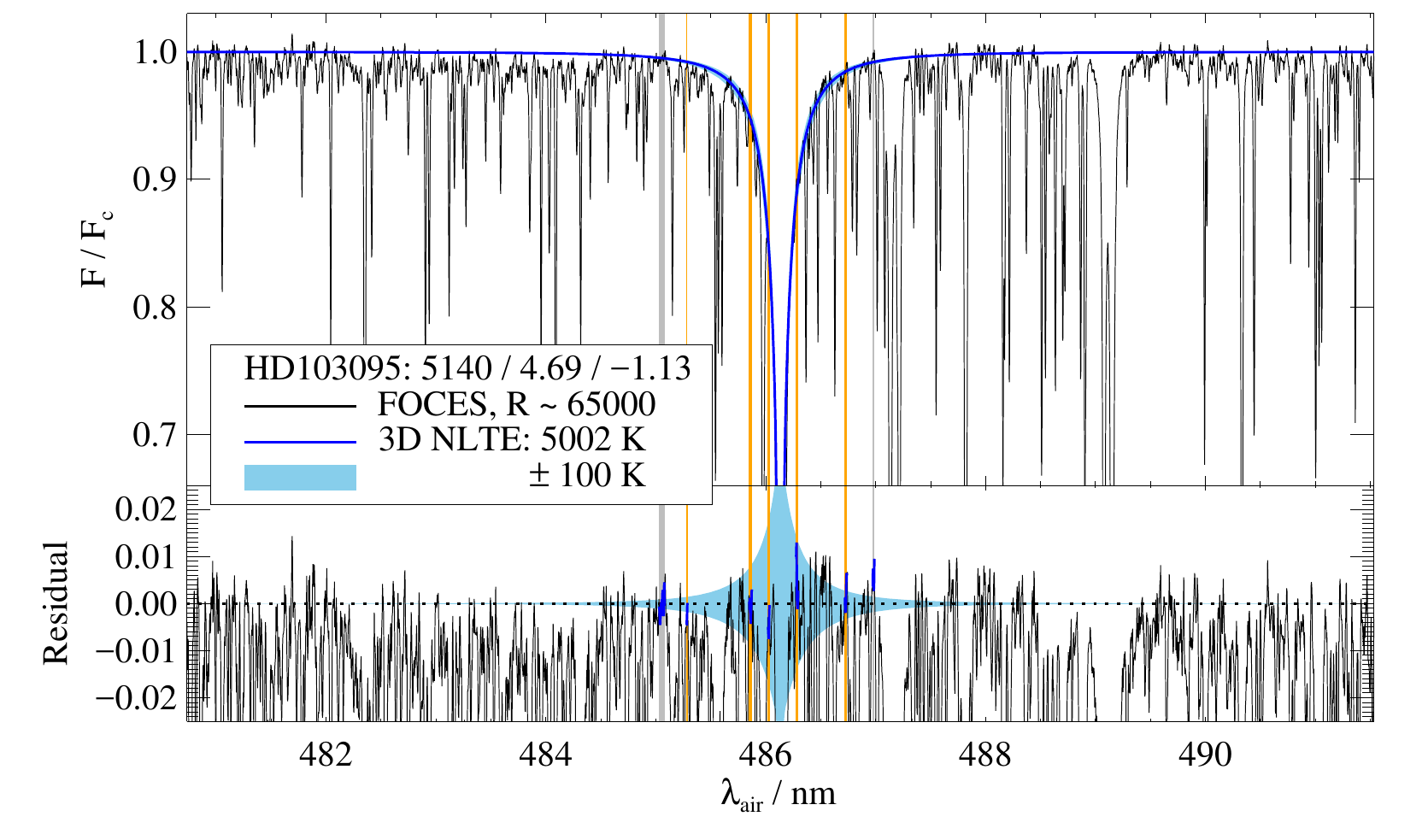}\includegraphics[scale=0.55]{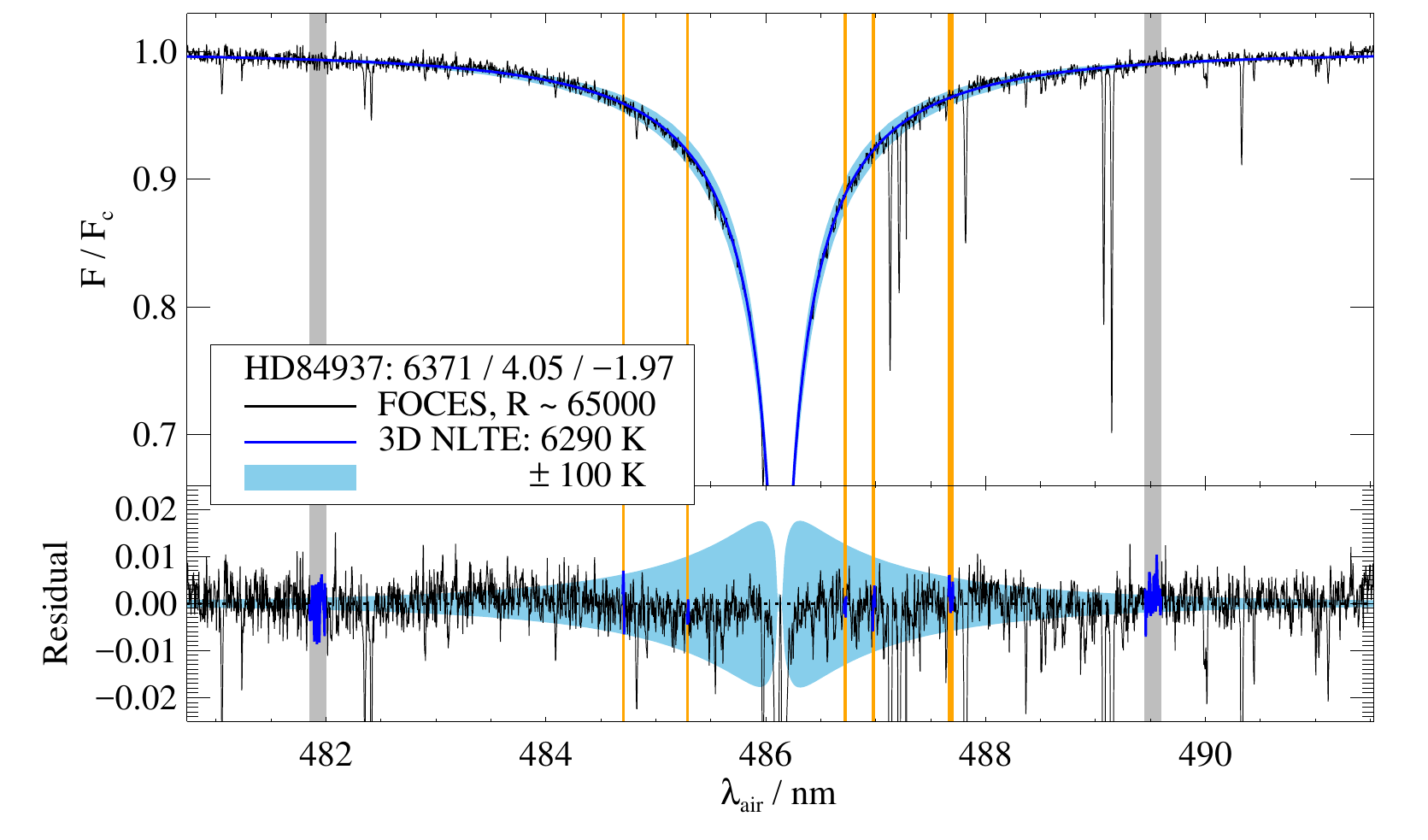}
\includegraphics[scale=0.55]{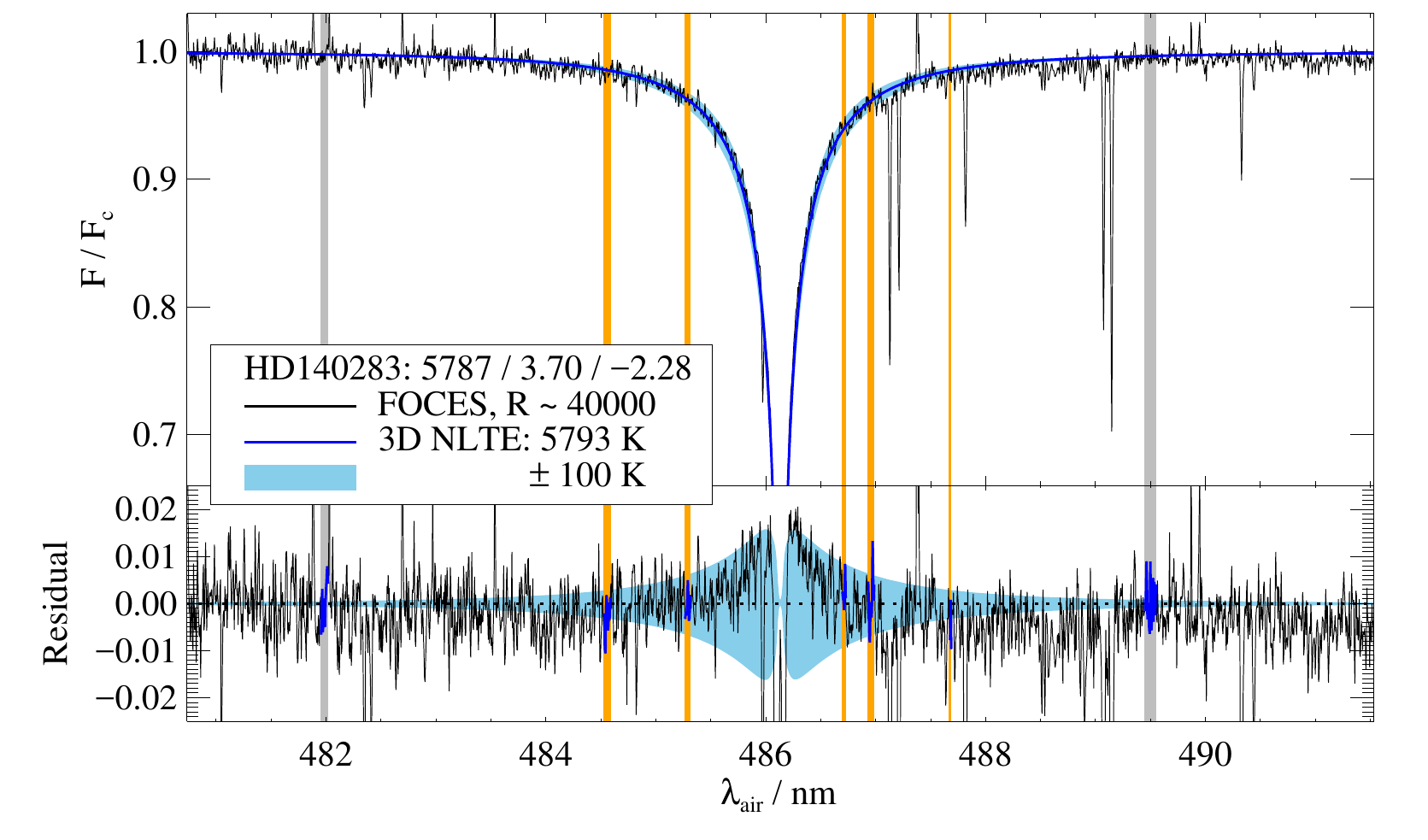}\includegraphics[scale=0.55]{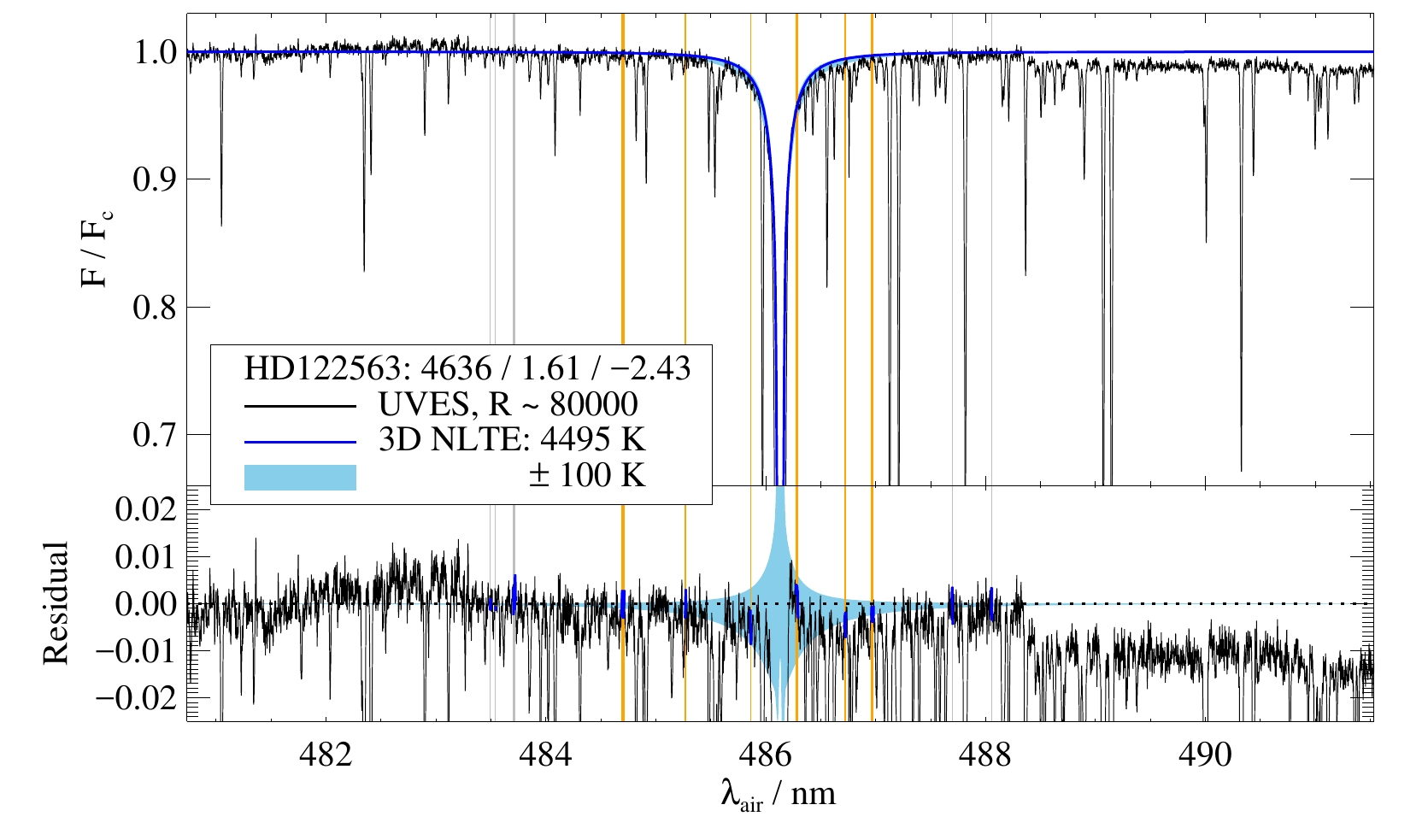}
\caption{$\hbeta$ line profiles observed in benchmark stars, 
compared to the best-fitting
3D non-LTE model when effective temperature is taken as a free parameter. 
The reference parameters $\teff$/$\lgg$/$\feh$~of each star are given in the
legends. The continuum and line masks are shown as 
dark and light vertical bands, respectively.
The light shaded region indicates the effect
of adjusting the effective temperature by $\pm 100\,\mathrm{K}$, 
where lower effective temperatures result in a
weaker line and thus a higher normalised flux.
Residuals between the 3D non-LTE model and the observations 
are shown in the lower panel.
Residuals inside the masks 
are highlighted using thick lines; only these pixels 
have any influence on the fitting procedure.}
\label{figure_fit2}
\end{center}
\end{figure*}

\begin{figure*}
\begin{center}
\includegraphics[scale=0.55]{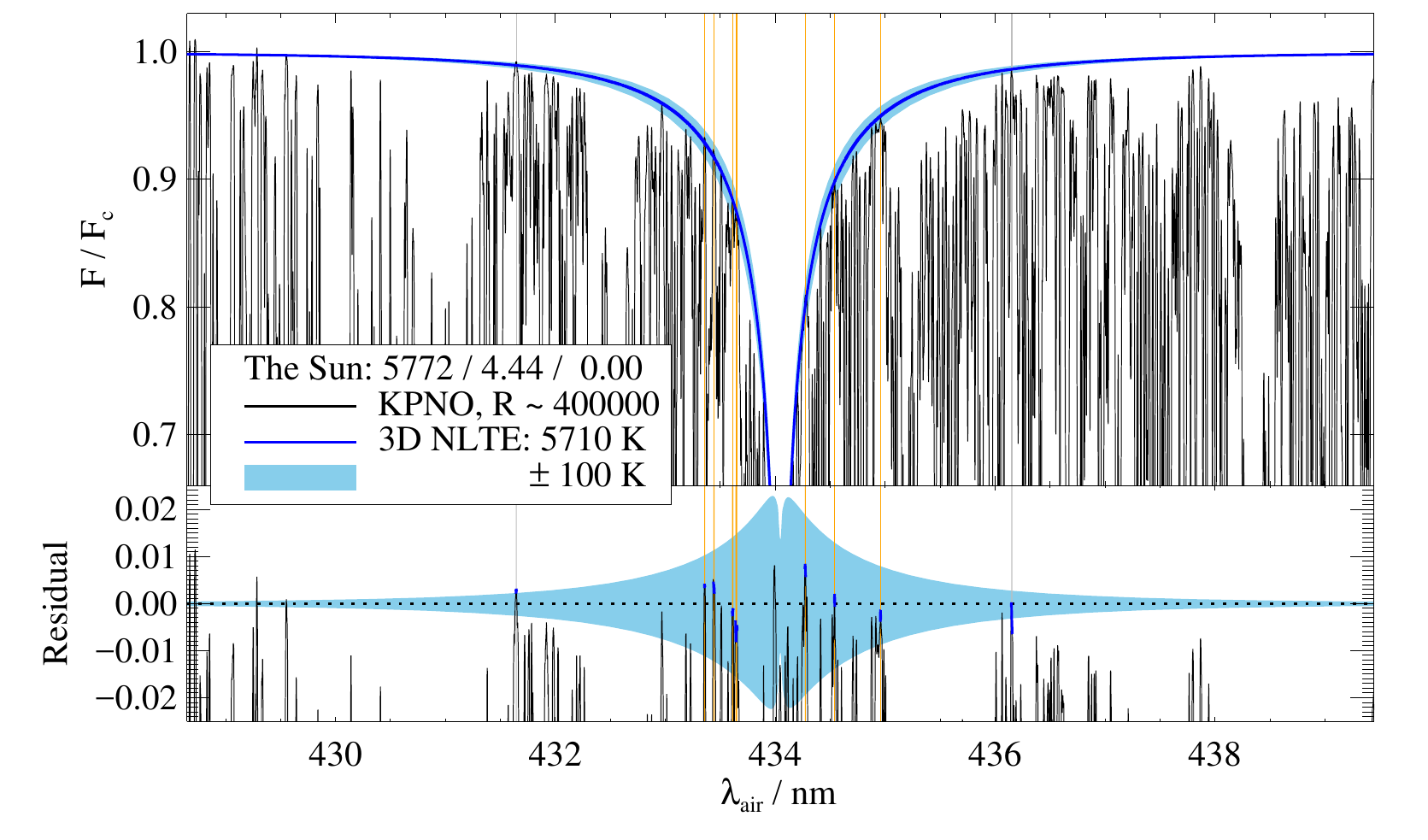}\includegraphics[scale=0.55]{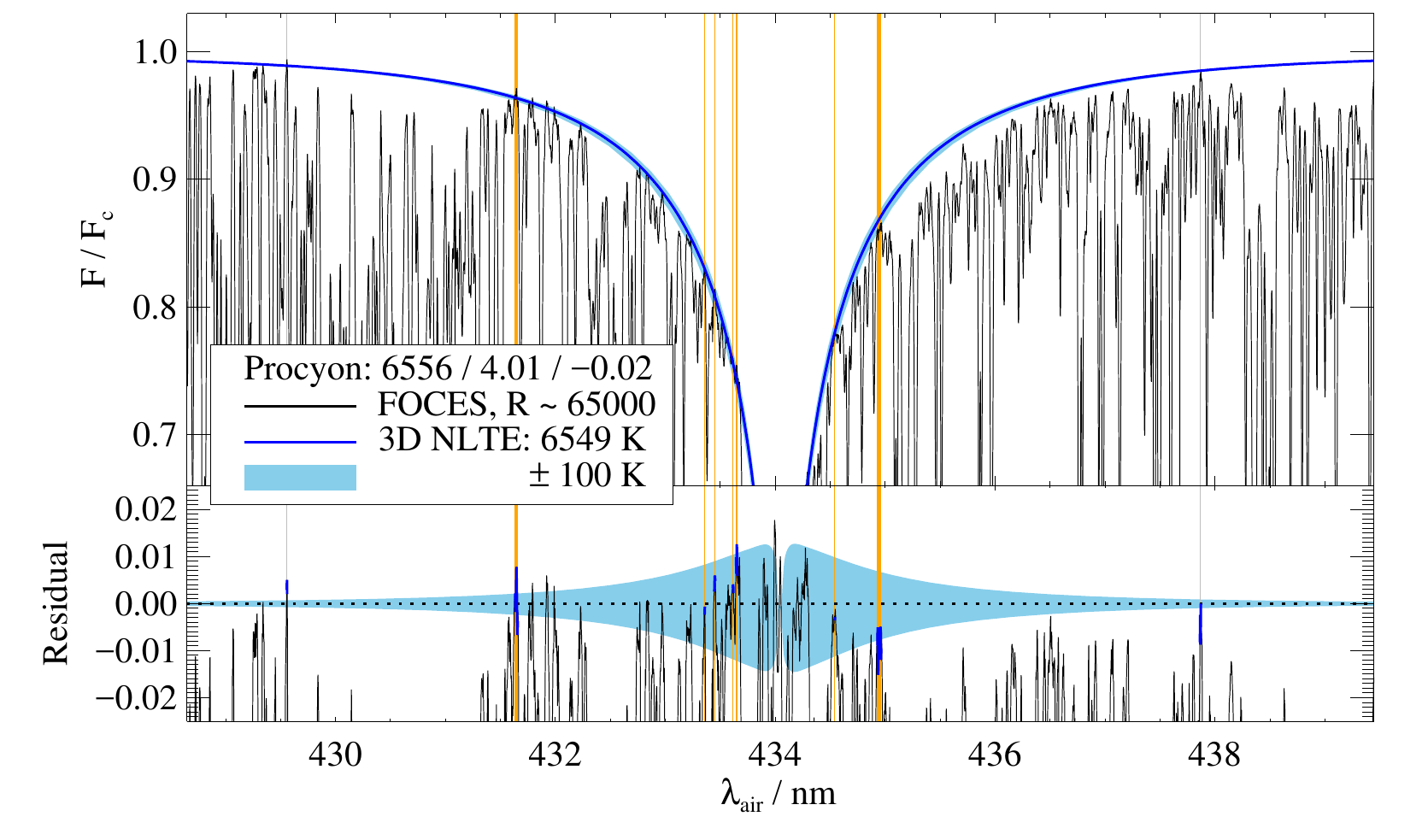}
\includegraphics[scale=0.55]{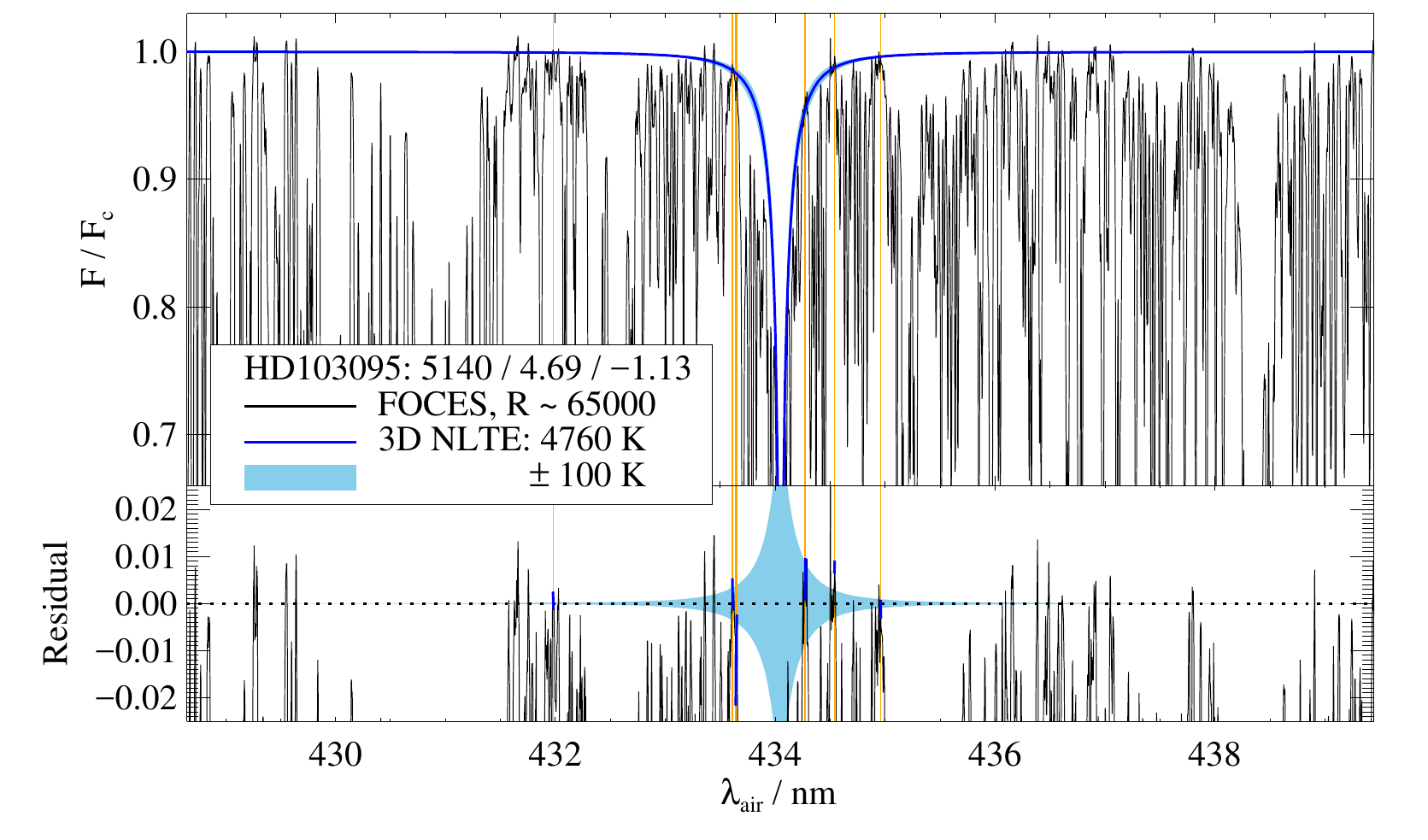}\includegraphics[scale=0.55]{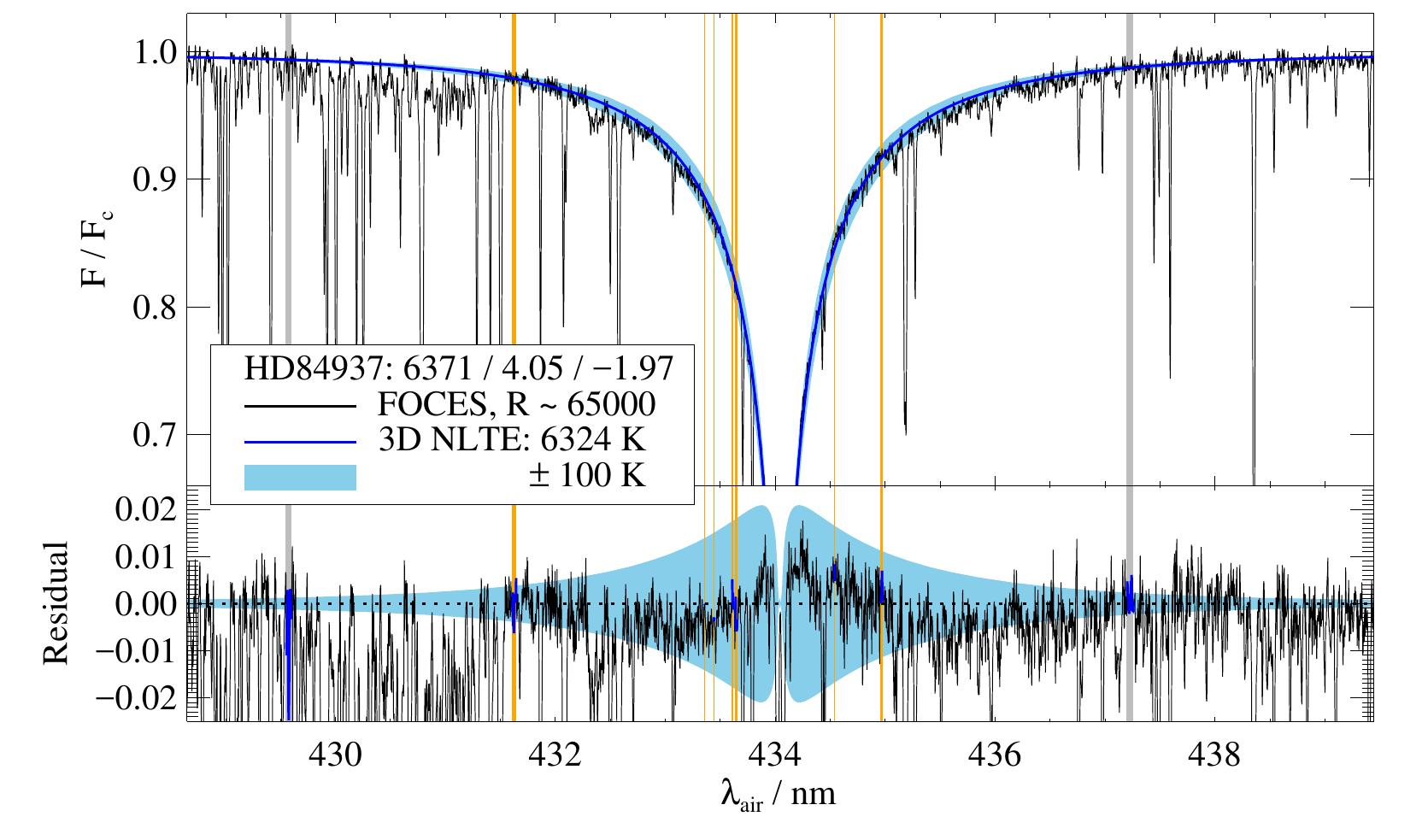}
\includegraphics[scale=0.55]{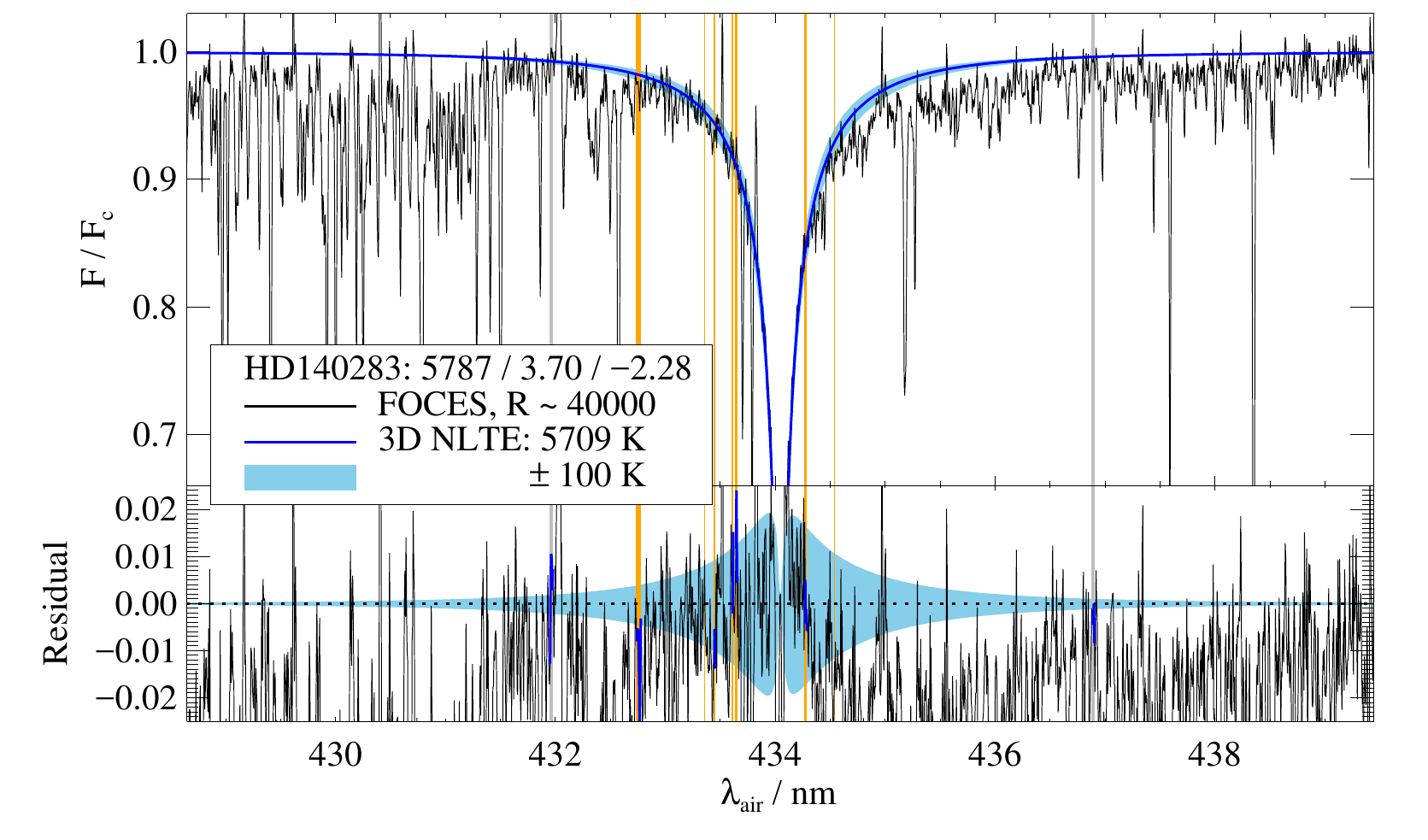}\includegraphics[scale=0.55]{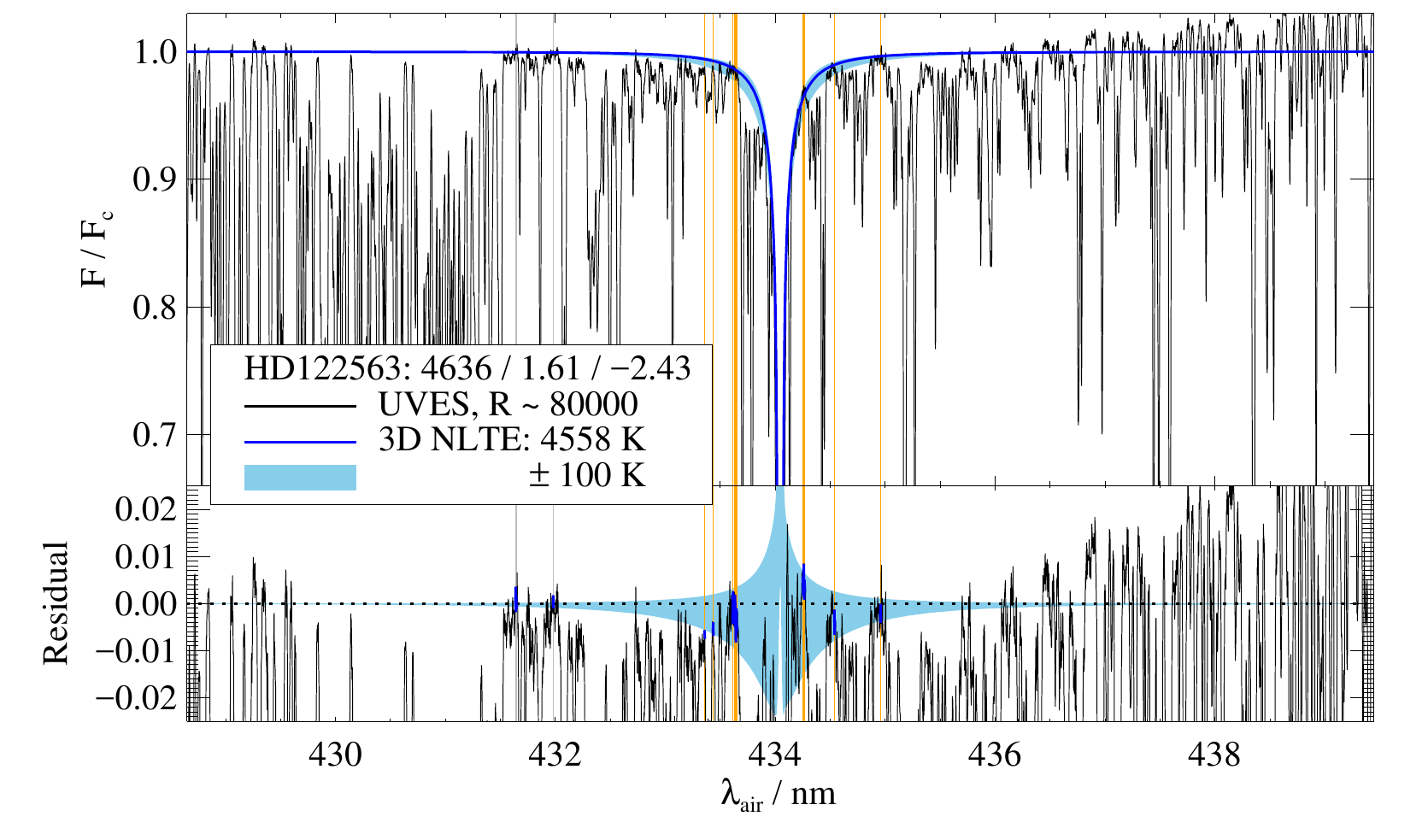}
\caption{$\hgamma$ line profiles observed in benchmark stars, 
compared to the best-fitting
3D non-LTE model when effective temperature is taken as a free parameter. 
The reference parameters $\teff$/$\lgg$/$\feh$~of each star are given in the
legends. The continuum and line masks are shown as 
dark and light vertical bands, respectively.
The light shaded region indicates the effect
of adjusting the effective temperature by $\pm 100\,\mathrm{K}$, 
where lower effective temperatures result in a
weaker line and thus a higher normalised flux.
Residuals between the 3D non-LTE model and the observations 
are shown in the lower panel.
Residuals inside the masks 
are highlighted using thick lines; only these pixels 
have any influence on the fitting procedure.}
\label{figure_fit3}
\end{center}
\end{figure*}

\begin{figure*}
\begin{center}
\includegraphics[scale=0.31]{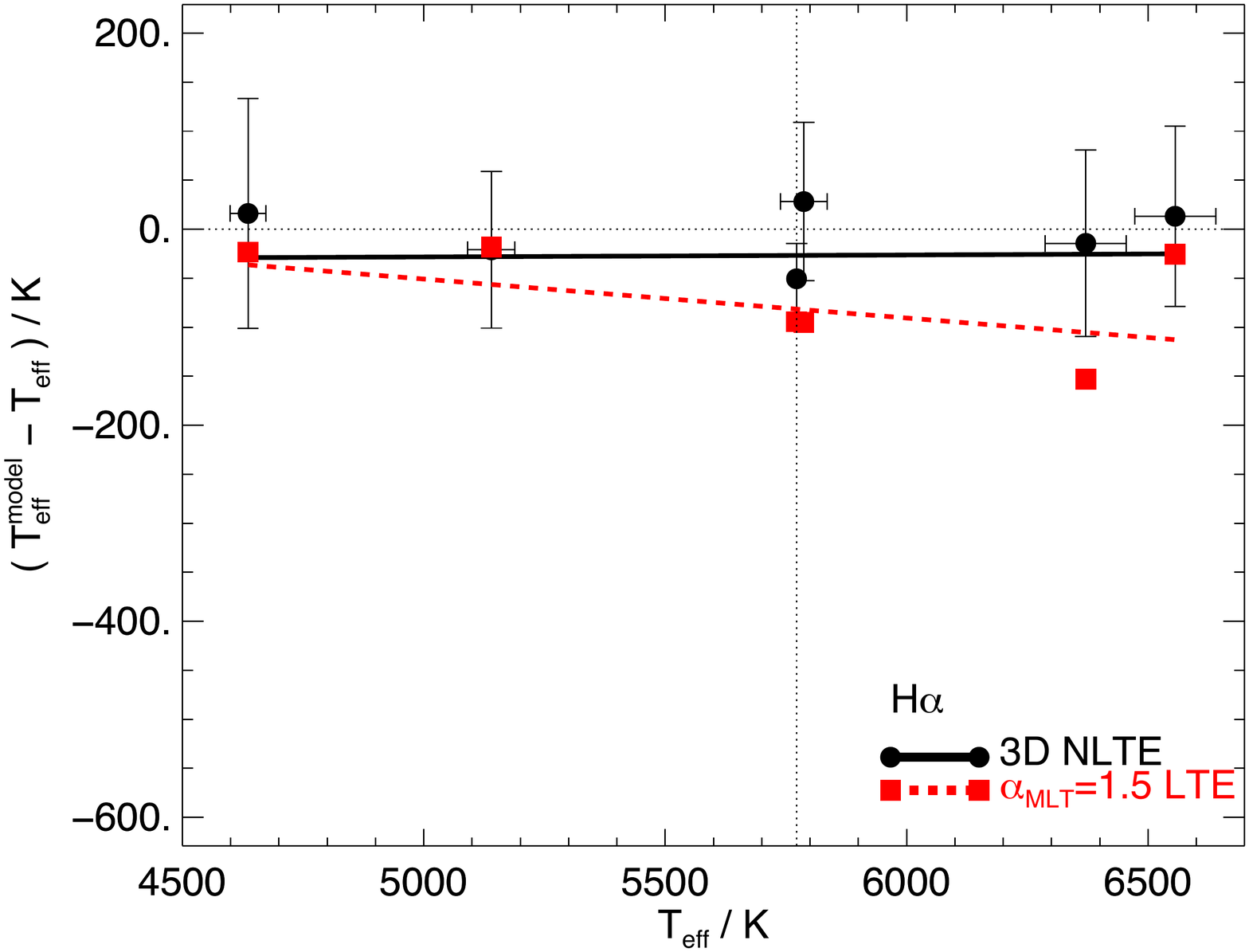}\includegraphics[scale=0.31]{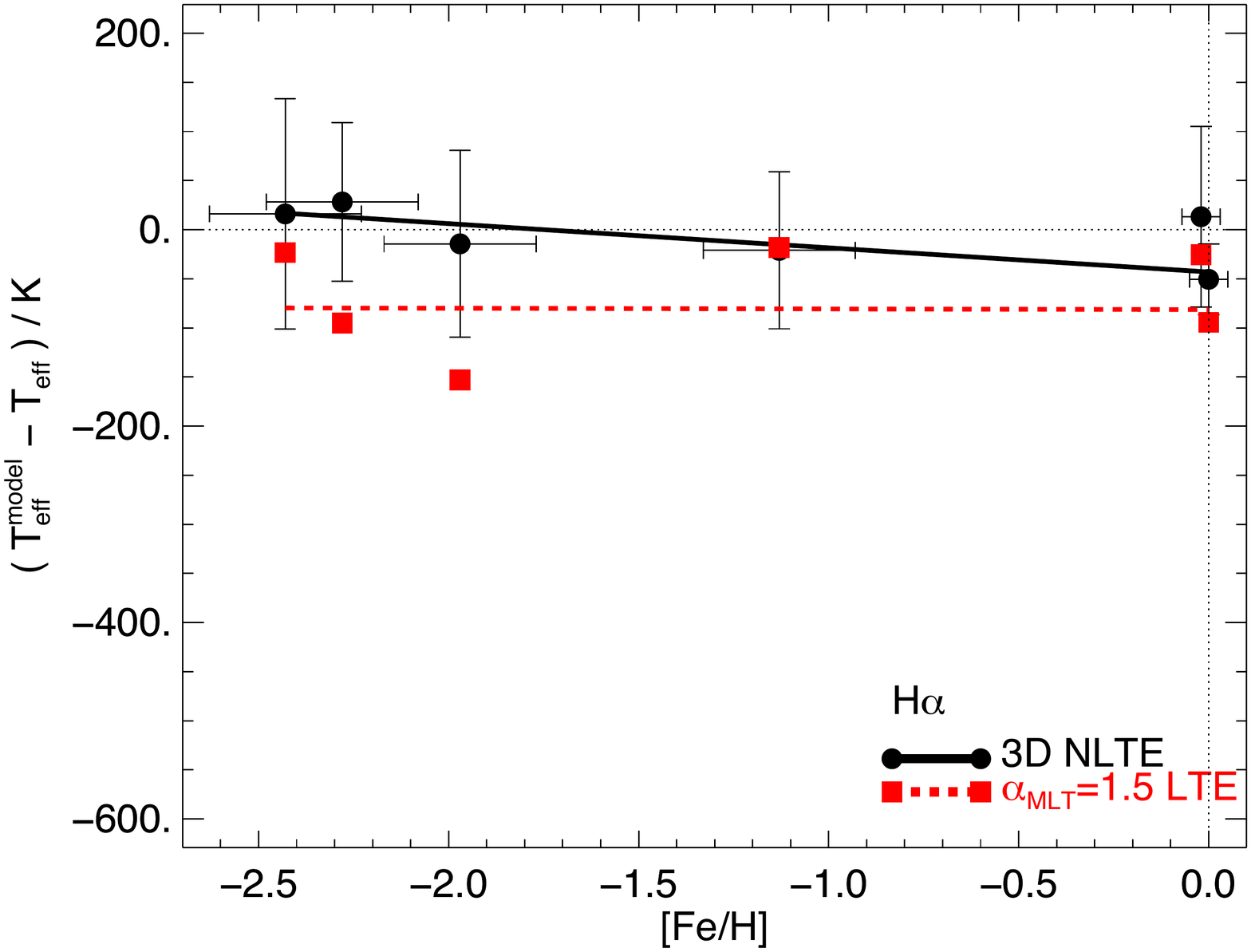}
\includegraphics[scale=0.31]{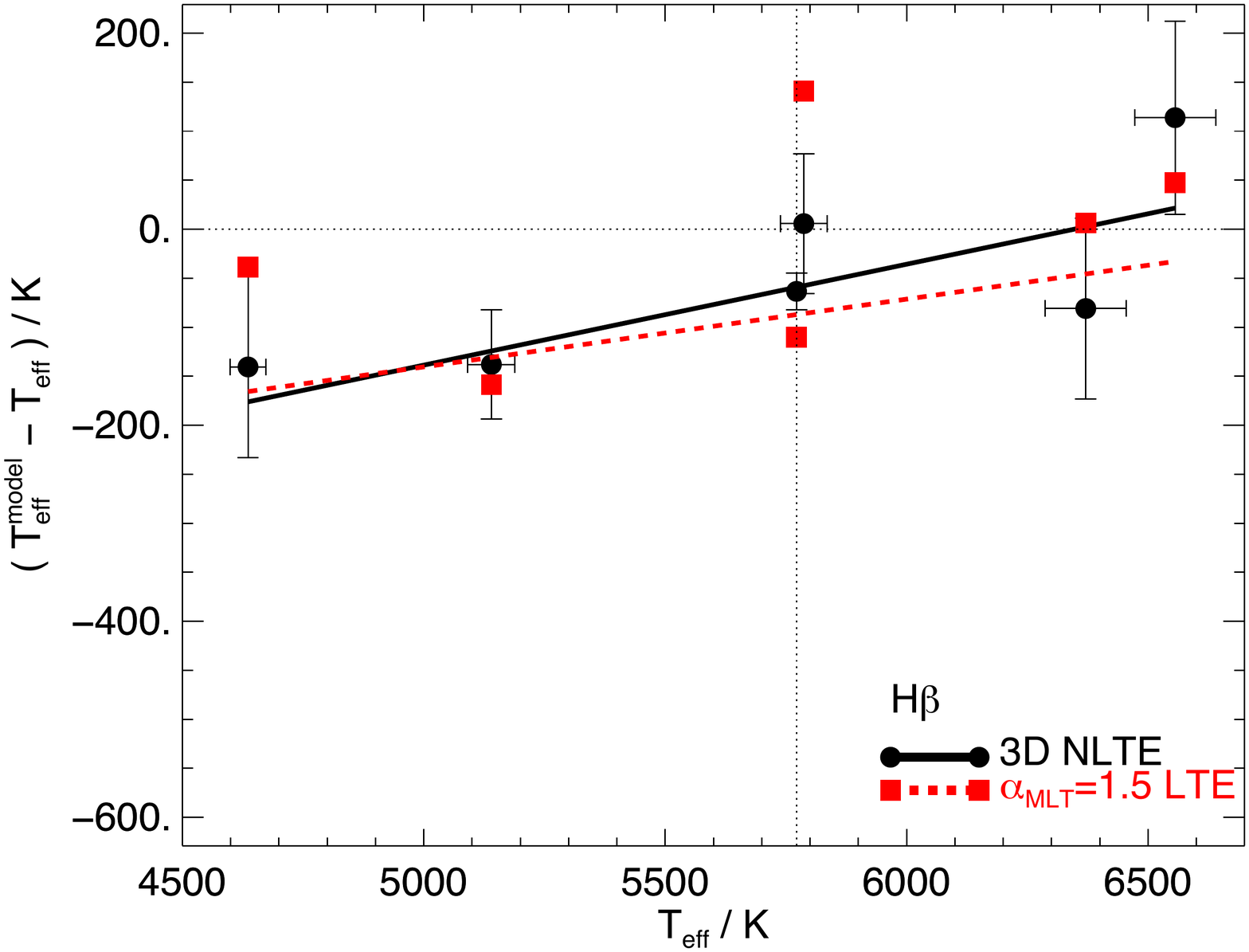}\includegraphics[scale=0.31]{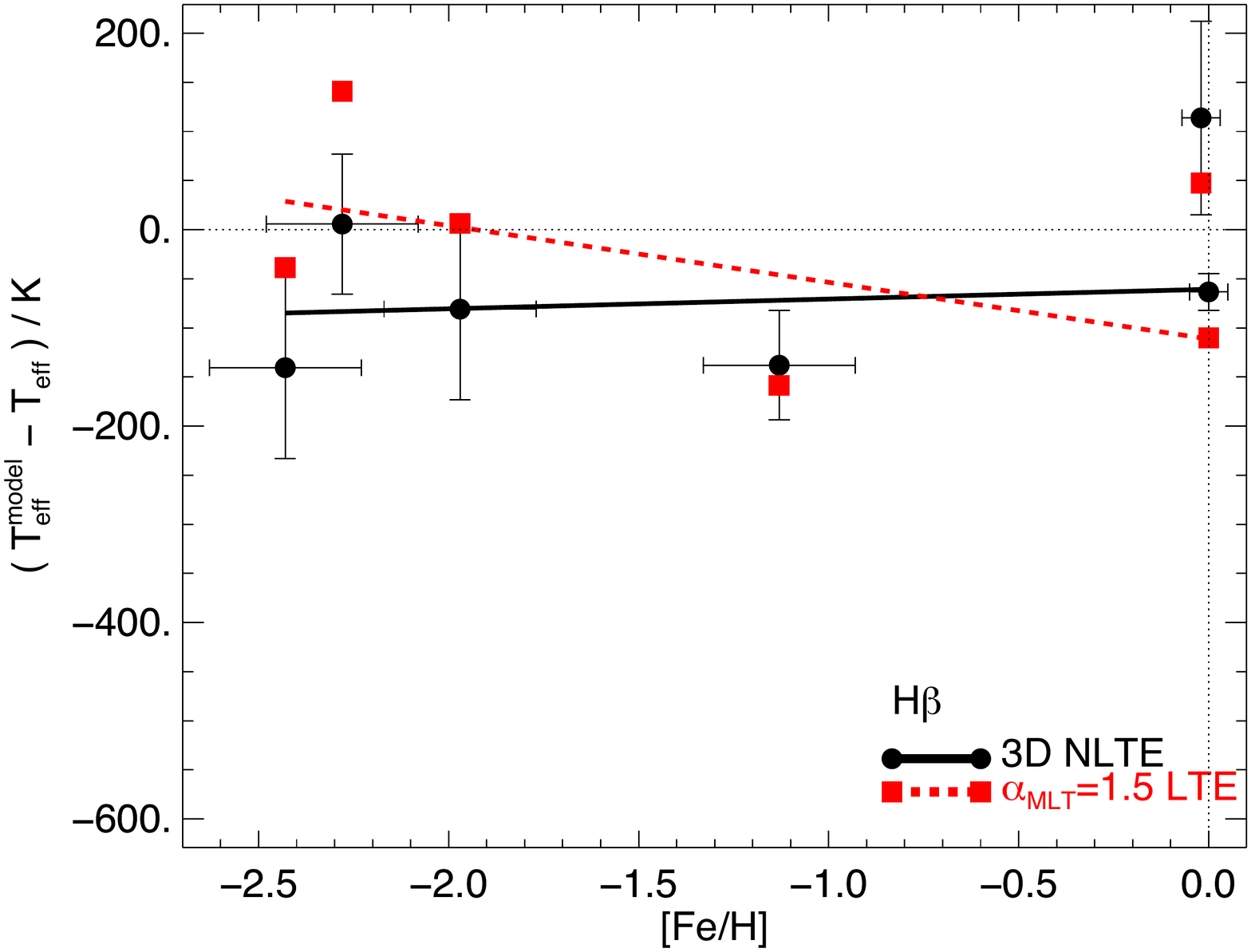}
\includegraphics[scale=0.31]{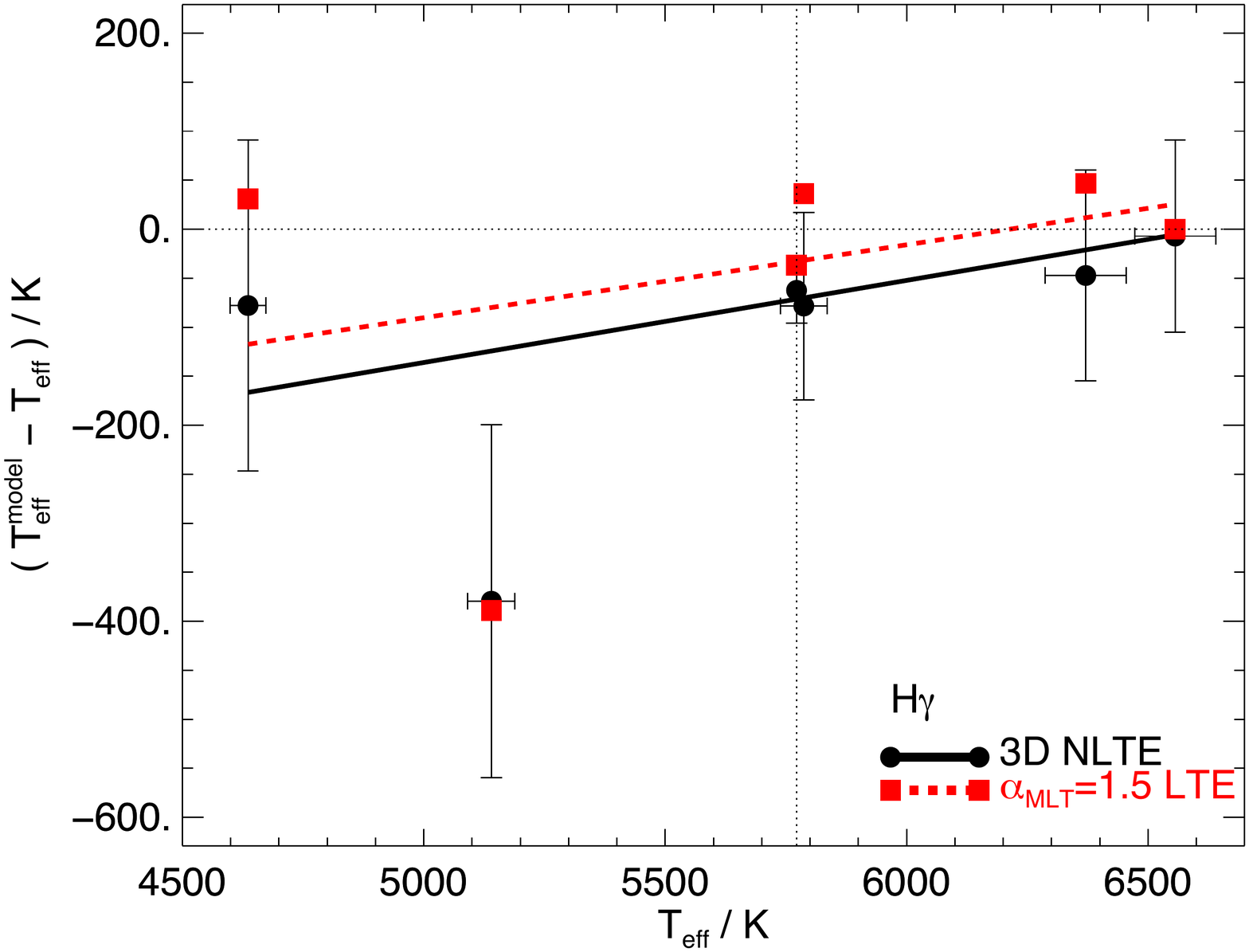}\includegraphics[scale=0.31]{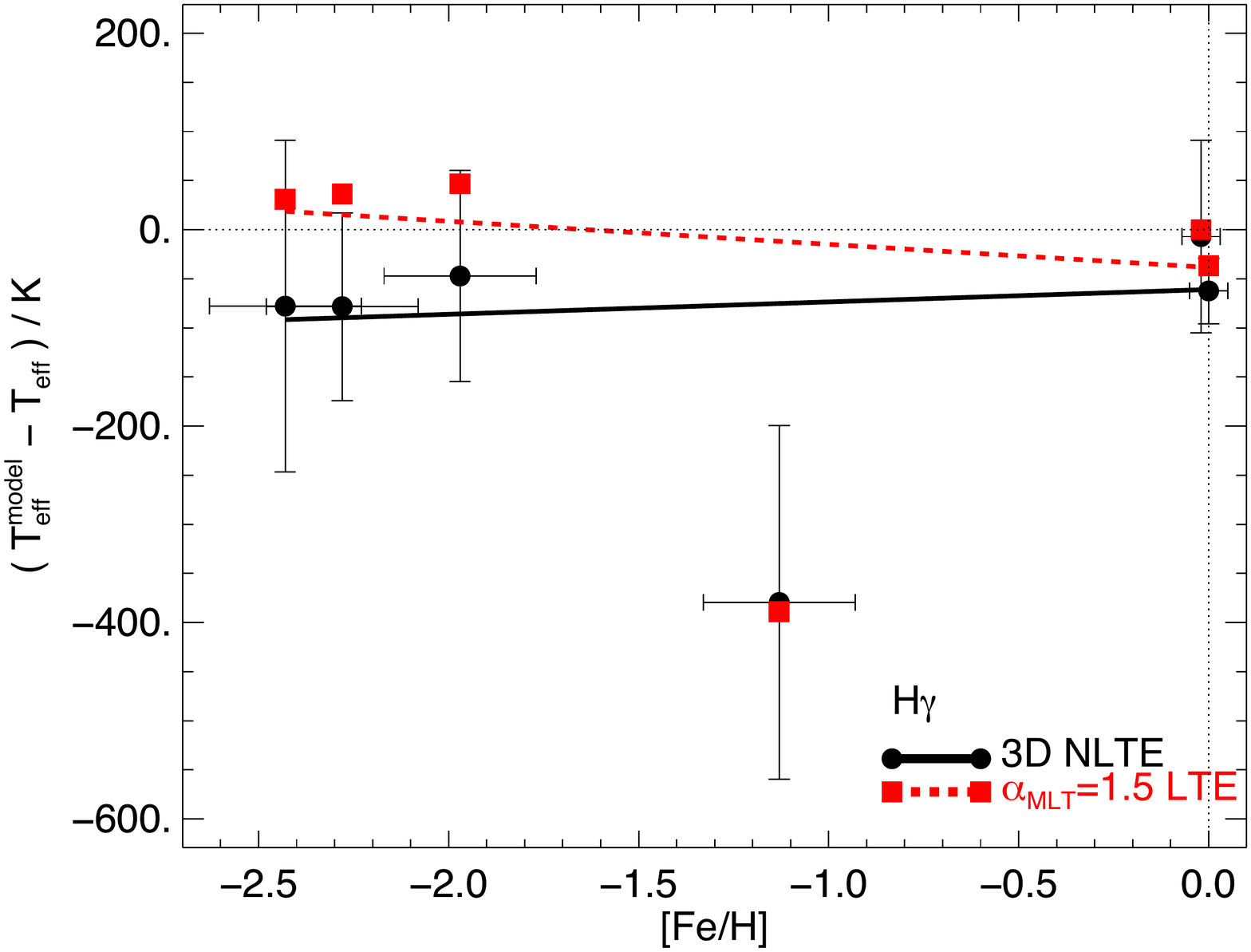}
\caption{Differences between effective temperatures
of benchmark stars inferred 
from 3D non-LTE or 1D LTE model spectra
(the latter calculated on 
$\alpha_{\mathrm{MLT}}=1.5$~model atmospheres),
and independent literature values,
as functions of literature effective temperatures (left)
and metallicities (right).
Error bars are shown for the 3D non-LTE models,
and omitted for the 1D LTE models for clarity. 
Also shown are lines of best fit, that take 
uncertainties into account.}
\label{figure_results}
\end{center}
\end{figure*}

\subsection{Sample}
\label{benchmark_sample}

\begin{table*}
\begin{center}
\caption{Benchmark stars, their literature atmospheric parameters, sources for their observed spectra, nominal spectral resolving power $R=\lambda/\Delta\lambda$, and the assumed $1\sigma$~uncertainties in the atmospheric parameters and in the placement of the continuum.}
\label{table_param1}
\begin{tabular}{l | cc | cc | cc | cc ccc}
\hline
\multirow{2}{*}{Star} &
\multirow{2}{*}{$\teff/\mathrm{K}$} &
\multirow{2}{*}{$\sigma_{\teff}/\mathrm{K}$} &
\multirow{2}{*}{$\lgg$} &
\multirow{2}{*}{$\sigma_{\lgg}$ }& 
\multirow{2}{*}{$\feh$ }&
\multirow{2}{*}{$\sigma_{\feh}$ }&
\multirow{2}{*}{Obs.} &
\multirow{2}{*}{$R/10^{5}$} &
\multicolumn{3}{c}{$\sigma_{\text{cont.}}/\%$} \\
 & & & & & & & & & 
$\halpha$ & $\hbeta$ & $\hgamma$ \\
\hline
\hline
\multirow{1}{*}{Sun} & 
\multirow{1}{*}{$5772^{\text{a}}$ } & 
\multirow{1}{*}{ } & 
\multirow{1}{*}{$4.44^{\text{a}}$ } & 
\multirow{1}{*}{ } & 
\multirow{1}{*}{$ 0.00^{\text{h}}$ } & 
\multirow{1}{*}{$0.05$ } & 
KPNO$^{\text{l}}$ & 
$$>4$$  & 
$0.3$  & $0.3$ & $0.5$  \\ 
\multirow{1}{*}{Procyon} & 
\multirow{1}{*}{$6556^{\text{b}}$ } & 
\multirow{1}{*}{$84$ } & 
\multirow{1}{*}{$4.01^{\text{b}}$ } & 
\multirow{1}{*}{$0.03$ } & 
\multirow{1}{*}{$-0.02^{\text{i}}$ } & 
\multirow{1}{*}{$0.05$ } & 
FOCES$^{\text{m}}$ & 
$$0.65$$  & 
$0.3$  & $0.3$ & $1.0$  \\ 
\multirow{1}{*}{HD 103095} & 
\multirow{1}{*}{$5140^{\text{c}}$ } & 
\multirow{1}{*}{$49$ } & 
\multirow{1}{*}{$4.69^{\text{e}}$ } & 
\multirow{1}{*}{$0.10$ } & 
\multirow{1}{*}{$-1.13^{\text{j}}$ } & 
\multirow{1}{*}{$0.20$ } & 
FOCES$^{\text{m}}$ & 
$$0.65$$  & 
$0.3$  & $0.3$ & $1.0$  \\ 
\multirow{1}{*}{HD 84937} & 
\multirow{1}{*}{$6371^{\text{d}}$ } & 
\multirow{1}{*}{$84$ } & 
\multirow{1}{*}{$4.05^{\text{f}}$ } & 
\multirow{1}{*}{$0.03$ } & 
\multirow{1}{*}{$-1.97^{\text{k}}$ } & 
\multirow{1}{*}{$0.20$ } & 
FOCES$^{\text{m}}$ & 
$$0.65$$  & 
$0.3$  & $0.3$ & $1.0$  \\ 
\multirow{1}{*}{HD 140283} & 
\multirow{1}{*}{$5787^{\text{c}}$ } & 
\multirow{1}{*}{$48$ } & 
\multirow{1}{*}{$3.70^{\text{f}}$ } & 
\multirow{1}{*}{$0.03$ } & 
\multirow{1}{*}{$-2.28^{\text{k}}$ } & 
\multirow{1}{*}{$0.20$ } & 
FOCES$^{\text{m}}$ & 
$$0.4$$  & 
$0.3$  & $0.3$ & $1.0$  \\ 
\multirow{1}{*}{HD 122563} & 
\multirow{1}{*}{$4636^{\text{c}}$ } & 
\multirow{1}{*}{$37$ } & 
\multirow{1}{*}{$1.61^{\text{g}}$ } & 
\multirow{1}{*}{$0.07$ } & 
\multirow{1}{*}{$-2.43^{\text{k}}$ } & 
\multirow{1}{*}{$0.20$ } & 
UVES$^{\text{n}}$ & 
$$0.8$$  & 
$0.5$  & $0.5$ & $1.0$  \\ 
\hline
\hline
\end{tabular}
\end{center}
\tablebib{(a) Reference value from \citet{2016AJ....152...41P}; (b) Fundamental value from \citet{2012A&amp;A...540A...5C}; (c) Fundamental value from \citet{2018MNRAS.tmpL..10K}; (d) IRFM value from \citet{2011A&amp;A...530A.138C}; (e) Fundamental value from \citet{2008A&amp;A...492..823B}; (f) Fundamental value from \citet{2014ApJ...792..110V}; (g) Fundamental value from \citet{2015A&amp;A...582A..49H}; (h) \citet{2009ARA&amp;A..47..481A}; (i) \mtd~non-LTE \ion{Fe}{II} value from \citet{2012MNRAS.427...27B}; (j) 1D LTE \ion{Fe}{II} value from \citet{2013ApJ...764...78R} with \mtd~non-LTE corrections from \citet{2016MNRAS.463.1518A}; (k) 3D non-LTE \ion{Fe}{II} value from \citet{2016MNRAS.463.1518A}; (l) \citet{2011ApJS..195....6W}; (m) \citet{2003A&amp;A...407..691K}; (n) \citet{2003Msngr.114...10B}. }
\end{table*}

The models that we presented in the previous section, 
\sect{results}, were compared to 
high-quality observations of well-studied benchmark stars:
the Sun, Procyon (HD 61421), 
HD 103095, HD 84937, HD 140283, and HD 122563.
The effective temperatures as well as the
surface gravities of these benchmark stars
are well-constrained by photometric, interferometric,
or astrometric 
measurements that are independent of spectrum analysis methods.
Since rotational broadening is unimportant for fitting the
Balmer line wings, the $\vsini$~parameter was neglected here.
We summarise the sample in \tab{table_param1}.

For the Sun, the Fourier Transform Spectrometer
(FTS) solar flux atlas of \citet{2011ApJS..195....6W}
was used,
based on high-resolution, high signal-to-noise ratio observations
taken at the Kitt Peak National Observatory (KPNO).
For the other benchmark stars,
high-quality spectra \citep{2003A&amp;A...407..691K}
taken with the FOCES
spectrograph \citep{1998A&amp;AS..130..381P} were used.
These have been proven to accurately reproduce the
intrinsic shapes of broad lines thanks to
the fibre-fed design of the instrument
\citep{2002sdef.conf..199K}.

High-resolution spectra from the UVES-POP catalogue
\citep{2003Msngr.114...10B} are also available for some
of the benchmark stars considered here.
These data are generally unreliable for Balmer line analyses
because of percent-level residuals from
the flat-fielding process occurring over scales of 
$1.5$-$2.0\,\mathrm{nm}$.
This can be seen in the spectrum of HD 122563,
which is the only star in the sample for which UVES data were used.
As the line wings of this star are not very extended, 
our analysis is not significantly affected 
by these shortcomings, in comparison to the relatively
large influence of blending lines.
Percent-level residuals 
in the $\halpha$ profile are apparent in most data sets
based on cross-dispersed echelle spectrographs 
\citep[e.g.][Fig.~2]{2014A&amp;A...566A..98B};
special care must be taken in the data reduction of such observations
\citep[e.g.][Sect.~2]{2002A&amp;A...385..951B}.

\subsection{Fitting procedure}
\label{benchmark_method}

Effective temperatures were determined from individual
Balmer line wings by 
profile fitting of
the continuum normalised model spectra.
These were performed
by $\chi^{2}$-minimisation, using the IDL routine
\texttt{MPFIT} \citep{2009ASPC..411..251M}.
We discuss the line and continuum masks
in \sect{benchmark_method_masks},
the continuum normalisation 
in \sect{benchmark_method_normalisation},
the interpolation procedure
in \sect{benchmark_method_interpolation},
and the error analysis
in \sect{benchmark_method_errors}.
The only free parameters in the fits were
{$\teff$}, and the continuum placement as we describe below.

\subsubsection{Fitting masks}
\label{benchmark_method_masks}

Following \citet{2002A&amp;A...385..951B}, 
wavelength masks were used to 
isolate unblended regions;
the line cores were also avoided,
for the reasons given in \sect{results_cf}.
Continuum masks were used to fit the observed continuum,
while line masks were used to fit the Balmer lines
and infer the effective temperature. 
We illustrate the masks 
in Figs~\ref{figure_fit1},~\ref{figure_fit2}, and
\ref{figure_fit3}, and describe how these masks were constructed below.

Basic wavelength masks were constructed first, from
which continuum and line masks were derived. For a particular benchmark star, 
the basic masks were constructed by comparing two sets of
1D LTE model spectra: one set containing all known blends
in the vicinity of the Balmer lines, and one set without them.
These model spectra were constructed on the adopted
atmospheric parameters of the benchmark star
(\tab{table_param1}); for the metal-poor
HD 84937 and HD 140283,
the masks were based on the warm solar-metallicity
benchmark star Procyon with only minor adjustments, instead.
Clean wavelength regions were then identified using the criterion
that the difference between the blended and unblended
1D LTE model spectra corresponds to an effect
of less than $30\,\mathrm{K}$.
Following that, these masks were refined
by using the observed, high-resolution KPNO
spectrum of the Sun to identify and screen residual 
missing blends and telluric lines.

From these basic masks, 
continuum and line masks were constructed 
based on the sensitivity of the 1D non-LTE and 3D non-LTE
model spectra to the effective temperature.
The continuum masks were chosen so as to consider wavelength regions
where neither the 1D nor the 3D model spectra had significant 
sensitivity to the effective temperature. 
Conversely, the line masks were chosen so as to consider
wavelength regions
where both the 1D and the 3D model spectra had significant 
sensitivity to the effective temperature. 
Owing to the severe line blending in the region surrounding $\hgamma$,
especially in the nearby CH G-band, 
the continuum masks had to be placed closer to the line core.
Consequently for $\hgamma$,
residual unresolved blends inside the continuum and line
mask regions can affect the results at
the level of $100\,\mathrm{K}$.

\subsubsection{Continuum normalisation}
\label{benchmark_method_normalisation}

The observed spectra were continuum normalised 
during every step of the $\chi^{2}$-minimisation
using the median ratio between
synthesis and observations, within the continuum mask windows;
that is, the
continuum and line fitting were performed simultaneously.
For the continuum placement of the UVES spectra (HD 122563), 
small but significant residual slopes were identified;
therefore for these data a robust slope fitting method
\citep{sen1968estimates,theil1992rank} was adopted instead.

\subsubsection{Interpolation procedure}
\label{benchmark_method_interpolation}

\begin{table}
\begin{center}
\caption{Interpolation/extrapolation errors, as estimated by the difference in effective temperatures inferred from a coarser and incomplete grid of model spectra, relative to those inferred from a finer and complete grid of model spectra, in both cases using 1D non-LTE calculations on \marcs~model atmospheres (see \sect{benchmark_method_interpolation}). A positive value indicates that a higher effective temperature is inferred from the coarser and incomplete grid than from the finer and complete grid.}
\label{table_results_interp}
\begin{tabular}{l | rrr}
\hline
\multirow{2}{*}{Star} & 
\multicolumn{3}{c}{$\delta\teff^{\text{interp.}}/\mathrm{K}$} \\ 
&
\multicolumn{1}{c}{$\halpha$} &
\multicolumn{1}{c}{$\hbeta$} &
\multicolumn{1}{c}{$\hgamma$} \\
\hline
\hline
Sun & 
$0$ 
& 
$-1$ 
& 
$+1$ 
\\ 
Procyon & 
$+41$ 
& 
$+23$ 
& 
$+10$ 
\\ 
HD 103095 & 
$+7$ 
& 
$+6$ 
& 
$+4$ 
\\ 
HD 84937 & 
$-11$ 
& 
$-3$ 
& 
$+1$ 
\\ 
HD 140283 & 
$-17$ 
& 
$-3$ 
& 
$0$ 
\\ 
HD 122563 & 
$+60$ 
& 
$-47$ 
& 
$-136$ 
\\ 
\hline
\hline
\end{tabular}
\end{center}
\end{table}

It was necessary to interpolate 
(and moderately extrapolate) the model spectra
onto arbitrary sets of atmospheric parameters 
when performing the fits.
This interpolation was performed in several steps.
Since the original $\teff$~nodes are not regularly spaced
but the $\lgg$~and $\feh$~nodes are, 
in the first step the continuum and total
model fluxes at each wavelength and on
each $\lgg$~and $\feh$~node were
interpolated onto a regular $\teff$~grid
spanning~$4000\leq\teff/\mathrm{K}\leq6750$ in steps 
of $50\,\mathrm{K}$. Linear extrapolation with respect to $\teff$~was 
permitted up to $25\,\mathrm{K}$, at this stage.

In the second step, the resulting grid was interpolated
onto a finer mesh in 
$\lgg$, spanning~$1.5\leq\log\left(g / \mathrm{cm\,s^{-2}}\right)\leq5.0$~in
steps of $0.1\,\dex$.
In the third step, this was repeated for 
$\feh$, spanning $-4.0\leq\feh\leq0.5$~in 
steps of $0.1\,\dex$.
The interpolations of the continuum and total model
fluxes in these first three steps
were done using cubic splines, or linearly when too few points
were available.  Only in the linear regime are the second and third
steps commutable.  In the fourth step,
where necessary, linear extrapolations of
the continuum and total model fluxes were performed with respect
to $\lgg$, $\feh$, or $\teff$, in order of preference.

Finally, during the $\chi^{2}$-minimisation,
the model fluxes
on this fine, regular and complete grid 
were normalised, and were interpolated tri-linearly with respect
to $\teff$, $\lgg$, and $\feh$~onto the desired parameters.
The normalised model fluxes were interpolated onto 
a uniform wavelength basis, 
convolved with Gaussian instrumental profiles,
then interpolated onto the observed wavelengths,
using cubic splines.

Interpolation and extrapolation errors
were estimated by considering two different sets of model spectra.
The first set of model spectra constitutes 
1D non-LTE calculations across the entire, finely-spaced \marcs-grid
(1100 model atmospheres; see \sect{method_atmospheres_marcs}).
The second set is based on interpolating the first set
of model spectra onto the stellar parameters of the
coarsely-spaced \stagger-grid using cubic splines.
Both sets of model spectra were then
interpolated and extrapolated into regular grids
in the way that we described above,
and used to analyse the benchmark stars.
Then, under the assumption that 
errors incurred when interpolating
across the finer grid of model spectra are much smaller
than those incurred when interpolating
across the coarser grid of model spectra,
we used the differences in the inferred effective
temperatures from the coarse and fine grids
of model spectra to roughly quantify 
interpolation and extrapolation errors on the coarser grid.
These errors in turn are indicative of
interpolation and extrapolation errors on the grid
of 3D non-LTE model spectra.

In \tab{table_results_interp} we illustrate
the interpolation and extrapolation errors,
estimated as described above.
For Procyon and HD 122563, the errors are really 
extrapolation errors, because these stars lie slightly outside of 
the \stagger-grid \citep[see][Fig.~1]{2013A&amp;A...557A..26M}.
The extrapolation errors can be large 
(nearly $150\,\mathrm{K}$~for $\hgamma$ in HD 122563),
and we caution that our results for
these stars are consequently less robust
than our results for the other stars in our sample.
For the Sun, HD 103095, HD 84937, and HD 140283,
the errors mainly reflect interpolation errors
on a coarser $\teff$~and $\feh$~grid,
as we discussed in \sect{method_atmospheres_marcs}.
These errors are most severe for HD 140283,
where their magnitudes are nevertheless still small
(at most around $15\,\mathrm{K}$).
We note, however, that these values do not
reflect interpolation errors in $\lgg$, for which the step-size
is the same in the \marcs-grid as in the \stagger-grid ($0.5\,\dex$);
such errors are nevertheless expected to be small,
since the normalised Balmer lines are not very sensitive
to $\lgg$~(\sect{introduction}).

\subsubsection{Error analysis}
\label{benchmark_method_errors}

\begin{table*}
\begin{center}
\caption{Sensitivity of the effective temperatures inferred from the 1D non-LTE Balmer lines (using \marcs~model atmospheres) to changes in the atmospheric parameters $\lgg$~or $\feh$~with respect to the adopted literature value, or to a change in the placement of the continuum with respect to the best-fitting value. A positive value indicates that a higher effective temperature is inferred upon performing the stated perturbation.}
\label{table_results_sens}
\begin{tabular}{l | c@{$/$}c c@{$/$}c c@{$/$}c | c@{$/$}c c@{$/$}c c@{$/$}c | c@{$/$}c c@{$/$}c c@{$/$}c}
\hline
\multirow{2}{*}{Star} & 
\multicolumn{6}{c|}{$\delta\teff/\mathrm{K}$\,\,$\left(\lgg+\text{/}-0.3\right)$}  &
\multicolumn{6}{c|}{$\delta\teff/\mathrm{K}$\,\,$\left(\feh+\text{/}-0.3\right)$}  &
\multicolumn{6}{c}{$\delta\teff/\mathrm{K}$\,\,$\left(\text{cont.}+\text{/}-0.3\%\right)$}  \\
& 
\multicolumn{2}{c}{$\halpha$} &
\multicolumn{2}{c}{$\hbeta$} &
\multicolumn{2}{c|}{$\hgamma$} &
\multicolumn{2}{c}{$\halpha$} &
\multicolumn{2}{c}{$\hbeta$} &
\multicolumn{2}{c|}{$\hgamma$} &
\multicolumn{2}{c}{$\halpha$} &
\multicolumn{2}{c}{$\hbeta$} &
\multicolumn{2}{c}{$\hgamma$} \\
\hline
\hline
Sun & 
$-47$ & $42$ & 
$-28$ & $19$ & 
$-3$ & $-2$ & 
$9$ & $-10$ & 
$-46$ & $38$ & 
$-97$ & $86$ & 
$-35$ & $37$ & 
$-12$ & $24$ & 
$-19$ & $22$ \\ 
Procyon & 
$-25$ & $0$ & 
$23$ & $-35$ & 
$29$ & $-41$ & 
$-67$ & $64$ & 
$-82$ & $63$ & 
$-87$ & $65$ & 
$-29$ & $31$ & 
$-23$ & $26$ & 
$-14$ & $18$ \\ 
HD 103095 & 
$-42$ & $51$ & 
$-24$ & $29$ & 
$-17$ & $27$ & 
$82$ & $-96$ & 
$6$ & $-11$ & 
$-35$ & $36$ & 
$-23$ & $23$ & 
$-22$ & $22$ & 
$-56$ & $57$ \\ 
HD 84937 & 
$-103$ & $55$ & 
$28$ & $-32$ & 
$36$ & $-38$ & 
$-18$ & $14$ & 
$-17$ & $12$ & 
$-16$ & $12$ & 
$-43$ & $42$ & 
$-32$ & $32$ & 
$-15$ & $20$ \\ 
HD 140283 & 
$-159$ & $107$ & 
$25$ & $-28$ & 
$25$ & $-26$ & 
$-9$ & $8$ & 
$-15$ & $10$ & 
$-16$ & $11$ & 
$-65$ & $73$ & 
$-41$ & $42$ & 
$-18$ & $24$ \\ 
HD 122563 & 
$-93$ & $95$ & 
$-2$ & $-2$ & 
$26$ & $-27$ & 
$31$ & $-28$ & 
$-17$ & $13$ & 
$-25$ & $15$ & 
$-73$ & $80$ & 
$-47$ & $48$ & 
$-49$ & $46$ \\ 
\hline
\hline
\end{tabular}
\end{center}
\end{table*}

To estimate the residual systematic errors in the 3D non-LTE models,
it was first necessary to 
quantify all other contributing sources of uncertainty.
In \tab{table_param1} we list the assumed $1\sigma$~uncertainties
in the adopted atmospheric parameters 
($\lgg$~and $\feh$), as well as in the continuum placement
(arising from the finite signal-to-noise ratio 
of the observations, and imperfections in the 
continuum tracing and residual blends).
The uncertainty in the continuum placement
is only an estimate, since the influence of missing blends may be
significantly larger in the case of $\hgamma$.

These $1\sigma$~uncertainties
were translated into effective temperature errors:
$\sigma_{\teff;\,\lgg}$,
$\sigma_{\teff;\,\feh}$, and
$\sigma_{\teff;\,\text{cont.}}$.
This was done by repeating the 3D non-LTE analysis
(using the interpolated regular 3D non-LTE grid 
we described in \sect{benchmark_method_interpolation}),
but shifting $\lgg$~or $\feh$~relative to their
reference values (\tab{table_param1}),
or shifting the continuum level relative to 
its best-fit value, by the stated $1\sigma$~uncertainties.
To aid intuition on the relative magnitude of these three sources of error,
we illustrate how sensitive the inferred effective temperatures
are to perturbations of $\pm0.3\,\dex$~in
$\lgg$~and $\feh$, and to 
perturbations of $\pm0.3\,\%$~in the continuum level
in \tab{table_results_sens},
based on the 1D non-LTE 
calculations across the entire, finely-spaced \marcs-grid
(\sect{method_atmospheres_marcs}).

Under the assumption that all of these uncertainties are independent, 
they were combined in quadrature, together with the 
formal fitting error $\sigma_{\teff;\,\text{fit.}}$,
to produce the final $1\sigma$~uncertainties
given in \tab{table_results_fits}:
\phantomsection\begin{IEEEeqnarray}{rCl}
\label{sigeq2}
    \sigma_{\teff}^2 &=& 
    \sigma_{\teff;\,\lgg}^2 +
    \sigma_{\teff;\,\feh}^2 +
    \sigma_{\teff;\,\text{cont.}}^2 +
    \sigma_{\teff;\,\text{fit.}}^2
\end{IEEEeqnarray}
Interpolation errors 
(\sect{benchmark_method_interpolation}) were not
folded into these uncertainties.

\subsection{Inferred effective temperatures}
\label{benchmark_fits}

\begin{table*}
\begin{center}
\caption{Differences between the effective temperatures of the benchmark stars inferred from various models compared to those inferred from the 3D non-LTE model (see Figs.~\ref{figure_fit1}, \ref{figure_fit2}, and \ref{figure_fit3} for illustrations of the 3D non-LTE fits). A positive value indicates that a higher effective temperature is inferred from the various models, than from the 3D non-LTE models.}
\label{table_results_diff}
\begin{tabular}{l | ccc | ccc | ccc | ccc | ccc}
\hline
\multirow{2}{*}{Star} & 
\multicolumn{3}{c|}{$\Delta\teff^{\text{3D LTE}}/\mathrm{K}$} & 
\multicolumn{3}{c|}{$\Delta\teff^{\text{1D NLTE;}\,\alpha_{\text{MLT}}=1.0}/\mathrm{K}$} & 
\multicolumn{3}{c|}{$\Delta\teff^{\text{1D NLTE;}\,\alpha_{\text{MLT}}=2.0}/\mathrm{K}$} & 
\multicolumn{3}{c|}{$\Delta\teff^{\text{1D LTE;}\,\alpha_{\text{MLT}}=1.0}/\mathrm{K}$} & 
\multicolumn{3}{c}{$\Delta\teff^{\text{1D LTE;}\,\alpha_{\text{MLT}}=2.0}/\mathrm{K}$} \\ 
&
\multicolumn{1}{c}{$\halpha$} &
\multicolumn{1}{c}{$\hbeta$} &
\multicolumn{1}{c|}{$\hgamma$} &
\multicolumn{1}{c}{$\halpha$} &
\multicolumn{1}{c}{$\hbeta$} &
\multicolumn{1}{c|}{$\hgamma$} &
\multicolumn{1}{c}{$\halpha$} &
\multicolumn{1}{c}{$\hbeta$} &
\multicolumn{1}{c|}{$\hgamma$} &
\multicolumn{1}{c}{$\halpha$} &
\multicolumn{1}{c}{$\hbeta$} &
\multicolumn{1}{c|}{$\hgamma$} &
\multicolumn{1}{c}{$\halpha$} &
\multicolumn{1}{c}{$\hbeta$} &
\multicolumn{1}{c}{$\hgamma$} \\
\hline
\hline
Sun & 
$-5$ 
& 
$-1$ 
& 
$1$ 
& 
$-31$ 
& 
$-53$ 
& 
$-30$ 
& 
$-33$ 
& 
$-26$ 
& 
$94$ 
& 
$-40$ 
& 
$-57$ 
& 
$-31$ 
& 
$-43$ 
& 
$-31$ 
& 
$94$ 
\\ 
Procyon & 
$-21$ 
& 
$4$ 
& 
$0$ 
& 
$-29$ 
& 
$-120$ 
& 
$-9$ 
& 
$-29$ 
& 
$3$ 
& 
$86$ 
& 
$-38$ 
& 
$-120$ 
& 
$-10$ 
& 
$-37$ 
& 
$1$ 
& 
$85$ 
\\ 
HD 103095 & 
$55$ 
& 
$33$ 
& 
$19$ 
& 
$-45$ 
& 
$-107$ 
& 
$-125$ 
& 
$-18$ 
& 
$-1$ 
& 
$34$ 
& 
$-10$ 
& 
$-80$ 
& 
$-106$ 
& 
$21$ 
& 
$30$ 
& 
$58$ 
\\ 
HD 84937 & 
$-77$ 
& 
$-2$ 
& 
$-1$ 
& 
$-20$ 
& 
$25$ 
& 
$26$ 
& 
$-23$ 
& 
$157$ 
& 
$189$ 
& 
$-133$ 
& 
$21$ 
& 
$23$ 
& 
$-129$ 
& 
$153$ 
& 
$187$ 
\\ 
HD 140283 & 
$-61$ 
& 
$0$ 
& 
$-1$ 
& 
$-51$ 
& 
$56$ 
& 
$34$ 
& 
$-5$ 
& 
$213$ 
& 
$188$ 
& 
$-150$ 
& 
$53$ 
& 
$30$ 
& 
$-81$ 
& 
$212$ 
& 
$185$ 
\\ 
HD 122563 & 
$-51$ 
& 
$23$ 
& 
$4$ 
& 
$102$ 
& 
$88$ 
& 
$70$ 
& 
$93$ 
& 
$132$ 
& 
$154$ 
& 
$-45$ 
& 
$81$ 
& 
$68$ 
& 
$-29$ 
& 
$127$ 
& 
$154$ 
\\ 
\hline
\hline
\end{tabular}
\end{center}
\end{table*}

\begin{table*}
\begin{center}
\caption{Inferred effective temperatures of the benchmark stars from the 3D non-LTE model (see Figs.~\ref{figure_fit1}, \ref{figure_fit2}, and \ref{figure_fit3} for illustrations of the 3D non-LTE fits). The $1\sigma$~uncertainties in the 3D non-LTE models take into account the uncertainties in the adopted surface gravities and metallicities, and uncertainties in placing the continuum, given in \tab{table_param1}, as well as the formal fitting error. The uncertainties in the continuum placement dominate, except for $\halpha$~in HD 103095, where the uncertainty in $\feh$~dominates. The last three columns show the differences between the models and the reference values adopted from the literature (Col.~2, also \tab{table_param1}), with the uncertainties combined in quadrature (see \fig{figure_results}~for an illustration of these differences). }
\label{table_results_fits}
\begin{tabular}{l | c | c |r@{$\pm$}r r@{$\pm$}r r@{$\pm$}r | r@{$\pm$}r r@{$\pm$}r r@{$\pm$}r}
\hline
\multirow{2}{*}{Star} & 
\multirow{2}{*}{$\teff/\mathrm{K}$} & 
\multirow{2}{*}{Spectrum} & 
\multicolumn{6}{c|}{$\teff^{\text{3D NLTE}}/\mathrm{K}$} & 
\multicolumn{6}{c}{$\Delta\teff/\mathrm{K}$} \\ 
& 
& 
& 
\multicolumn{2}{c}{$\halpha$} &
\multicolumn{2}{c}{$\hbeta$} &
\multicolumn{2}{c|}{$\hgamma$} &
\multicolumn{2}{c}{$\halpha$} &
\multicolumn{2}{c}{$\hbeta$} &
\multicolumn{2}{c}{$\hgamma$} \\
\hline
\hline
\multirow{1}{*}{Sun} & 
\multirow{1}{*}{$5772\pm1$ } & 
KPNO & 
$5721$ & $36$ & 
$5709$ & $19$ & 
$5710$ & $33$ & 
$-51$ & $36$ &
$-63$ & $19$ &
$-62$ & $33$ \\
\multirow{1}{*}{Procyon} & 
\multirow{1}{*}{$6556\pm84$ } & 
FOCES & 
$6569$ & $37$ & 
$6670$ & $52$ & 
$6549$ & $50$ & 
$13$ & $92$ &
$114$ & $99$ &
$-7$ & $98$ \\
\multirow{1}{*}{HD 103095} & 
\multirow{1}{*}{$5140\pm49$ } & 
FOCES & 
$5119$ & $63$ & 
$5002$ & $26$ & 
$4760$ & $173$ & 
$-21$ & $80$ &
$-138$ & $56$ &
$-380$ & $180$ \\
\multirow{1}{*}{HD 84937} & 
\multirow{1}{*}{$6371\pm84$ } & 
FOCES & 
$6357$ & $45$ & 
$6290$ & $38$ & 
$6324$ & $67$ & 
$-14$ & $95$ &
$-81$ & $92$ &
$-47$ & $108$ \\
\multirow{1}{*}{HD 140283} & 
\multirow{1}{*}{$5787\pm48$ } & 
FOCES & 
$5815$ & $65$ & 
$5793$ & $53$ & 
$5709$ & $83$ & 
$28$ & $81$ &
$6$ & $71$ &
$-78$ & $96$ \\
\multirow{1}{*}{HD 122563} & 
\multirow{1}{*}{$4636\pm37$ } & 
UVES & 
$4652$ & $111$ & 
$4495$ & $85$ & 
$4558$ & $165$ & 
$16$ & $117$ &
$-141$ & $92$ &
$-78$ & $169$ \\
\hline
\hline
\end{tabular}
\end{center}
\end{table*}

We illustrate the 3D non-LTE model fits to the benchmark
spectra in \fig{figure_fit1} for $\halpha$,
\fig{figure_fit2} for $\hbeta$,
and \fig{figure_fit3} for $\hgamma$.
The quality of the fits for the line wings
remains satisfactory, in particular for 
$\halpha$, in the reduced $\chi^{2}$~sense,
for the whole sample.
This is despite the 3D non-LTE models 
lacking any leverage in the form of free parameters.

In \tab{table_results_diff}
we tabulate the differences in the effective temperatures 
inferred, for a given benchmark star and Balmer line,
from the 3D LTE models, and from the
1D non-LTE models, compared to those inferred
from the 3D non-LTE models.
We use these differences to quantify the non-LTE effects
and the 3D effects, respectively.
We also show the differences
between the 1D LTE models and
the 3D non-LTE models, that reflect
the errors that may be present in standard 
1D LTE Balmer line analyses.
We tabulate the actual effective temperatures inferred 
from the 3D non-LTE models in \tab{table_results_fits}.
In \fig{figure_results} we plot the effective temperature
errors as inferred from different models, as functions
of literature atmospheric parameters.
We discuss these results in the following section, 
\sect{discussion}.

%-------------------------------------------------------------------------------
\section{Discussion}
\label{discussion}

\subsection{Quantifying the 3D non-LTE effects}
\label{discussion_effects}

\tab{table_results_diff}~shows that the 
3D effects, as quantified
by the 1D non-LTE versus 3D non-LTE differences, are 
significant for all of the Balmer lines.
Their magnitudes depend on the line and atmospheric parameters. 
For $\halpha$, the absolute differences range from
negligible to around $100\,\mathrm{K}$,
and a typical absolute value is around $50\,\mathrm{K}$.
The 3D effects
tend to grow in magnitude for higher 
members of the Balmer series,
and for $\hbeta$~and $\hgamma$~the differences are also sensitive to
the adopted mixing-length (\sect{results_cf}):
typical absolute values are around 
$50-150\,\mathrm{K}$~for
$\hbeta$~and $\hgamma$,
depending on the star and on the adopted mixing-length.

\tab{table_results_diff}~shows that
for $\halpha$, the 3D effects
tend to be negative. 
This is because the inner wings of $\halpha$~tend to be 
weaker in the 3D non-LTE models 
than in the 1D non-LTE models (\sect{results_balmer}). 
For higher 
members of the Balmer series
($\hbeta$~and $\hgamma$), the 
3D effects
tend to be more positive, 
at least for
larger values of mixing-length (\sect{results_cf});
in particular, the differences are all positive
for $\hgamma$~with $\alpha_{\mathrm{MLT}}=2.0$.
\citet{2009A&amp;A...502L...1L} also found 
that the differences get more positive for the higher members,
based on a differential 1D LTE versus 3D LTE comparison
(see their Table 2, noting the opposite sign convention).
Lowering $\alpha_{\text{MLT}}$~generally 
acts to reduce the absolute differences
in effective temperatures inferred from the 1D and 3D 
model spectra (as expected from the theoretical fluxes,
in \sect{results_balmer}).

In contrast to the 
3D effects,
\tab{table_results_diff}~shows that the 
non-LTE effects, as quantified by the
3D LTE versus 3D non-LTE differences,
become less severe for higher 
members of the Balmer series.
This is expected, since the higher 
members
form deeper in the atmosphere, and the departure coefficients
are much closer to unity there (\sect{results_departurecoefficients}).
In warmer stars, the absolute differences 
are usually only significant for $\halpha$;
however in cooler stars, 
such as HD 103095, the 
non-LTE effects can remain significant even for $\hgamma$.
For $\halpha$, the absolute non-LTE effects
range from negligible to nearly $100\,\mathrm{K}$
and a typical absolute value is around $50\,\mathrm{K}$.

We showed in \sect{results_balmer}~that 
the non-LTE effects are sensitive to the effective temperature. 
At high effective temperatures
a source function effect dominates, weakening the Balmer lines,
whereas towards lower effective temperatures 
a competing opacity effect becomes increasingly important
(\sect{results_departurecoefficients}).
This is somewhat apparent in \tab{table_results_diff}, 
with the 
non-LTE effects being
positive for HD 103095
and, at least for $\hbeta$~and 
$\hgamma$, for HD 122563, but negative
(or negligible) for the other benchmark stars.
In those sections we also explained that the non-LTE effects
become more severe towards lower metallicities;
accordingly, \tab{table_results_diff} shows larger
absolute 
non-LTE effects for 
HD 84937 than for Procyon, two turn-off stars with
similar effective temperatures and surface gravities
but very different metallicities.

For $\halpha$~at high effective temperature, the 
non-LTE effects
and the
3D effects
tend to go in the same direction: 
they are usually both negative.
Since the non-LTE effects
are more severe towards lower metallicities,
3D non-LTE modelling of $\halpha$~is especially important
for warmer metal-poor stars.
And indeed, \tab{table_results_diff} shows that some of the largest 
1D LTE versus 3D non-LTE differences 
for $\halpha$~are in HD 84937 and HD 140283.

\subsection{Solar effective temperature}
\label{discussion_sun}

\tab{table_results_fits} shows that
the solar effective temperature inferred from the 3D non-LTE 
models is too low. The effective 
temperature was inferred to be 
around $5710$-$5720\,\mathrm{K}$,
from the different Balmer lines.
This corresponds to an error of around 
$50$-$65\,\mathrm{K}$~in effective temperature,
slightly larger than the $1\sigma$~uncertainty of 
around $20$-$40\,\mathrm{K}$~in placing the continuum.
Other observations of the solar flux
\citep{brault1987spectral,2003A&amp;A...407..691K}
tend to give similar results,
as do observations of the solar disk-centre intensity
\citep{1984SoPh...90..205N}.
This solar effective temperature error is also consistent with the
analysis of \citet{2013A&amp;A...554A.118P}.

This systematic error is small ($50\,\mathrm{K}$~in effective temperature
corresponds to around $0.5\%$~in the emergent flux),
which makes it difficult to pin down exactly what is missing
from the models.  Uncertainties in the Stark-broadening 
and self-broadening theories may separately impart errors of the order 
$10$-$15\,\mathrm{K}$~\citep[][Table 4]{2002A&amp;A...385..951B}.
Furthermore, considering these two mechanisms as independent of
each other may also lead to a significant error on its own.
On the other hand, tests on different tailored model solar atmospheres,
with the chemical composition of \citet{2009ARA&amp;A..47..481A}
as well as the more metal-poor composition of \citet{2005ASPC..336...25A},
indicated a sensitivity of the results on the microphysics adopted in the
model atmosphere, of the order $20$-$30\,\mathrm{K}$.
Interpolation errors are apparently 
unimportant in this region of parameter space,
at least according to \tab{table_results_interp}.
In practical applications, however, 
errors of the order $50$-$65\,\mathrm{K}$~are usually smaller
than the other uncertainties intrinsic to the method,
in particular, in placing the continuum.

\subsection{Use as effective temperature diagnostics}
\label{discussion_diagnostics}

\tab{table_results_fits} and \fig{figure_results} show that
the 3D non-LTE models 
are generally able to
reproduce the effective temperatures of the benchmark stars
to within the combined $1\sigma$~uncertainties;
the only exceptions are the Sun
(which we discussed in \sect{discussion_sun}),
$\hbeta$~and $\hgamma$~in HD 103095 (which we discuss below),
and $\hbeta$~in
Procyon and HD 122563 (both of which 
are somewhat influenced by extrapolation errors).
These $1\sigma$~uncertainties are 
mainly influenced by 
uncertainties in placing the continuum;
this is very significant for 
$\hgamma$, because the spectral region is very blended,
as can be seen in \fig{figure_fit3}.
Very blended stellar spectra
combined with a finite spectral resolution can lead to
the observed continuum being systematically underestimated
by our fitting procedure,
as can be seen by comparing the
results for $\halpha$~with those 
for $\hgamma$~in
\tab{table_results_fits} and \fig{figure_results}.

All of the effective temperatures inferred 
from $\halpha$~agree with the corresponding reference values to within
$50\,\mathrm{K}$.  For the Sun, HD 84937, and HD 140283,
the effective temperatures inferred from 
$\halpha$~and $\hbeta$~agree to better than $70\,\mathrm{K}$.
These results suggest that our 3D non-LTE
$\halpha$~and $\hbeta$~models can be used
for reliable effective temperature determinations.

The results in \tab{table_results_fits} do however indicate 
some failure in the analysis for HD 103095. 
While $\halpha$~gives an effective temperature that
is consistent with the reference value to well within
the $1\sigma$~uncertainties, $\hbeta$~and $\hgamma$~do not.
The 1D LTE models of HD 103095, perform similarly badly, 
if not slightly worse, than the 3D non-LTE models (\fig{figure_results}).
We note that HD 103095 has a non-standard chemical composition,
having a lower $\upalpha$-enhancement than the
value of $0.4\,\dex$~adopted here 
\citep[][Fig.~8]{2016ApJ...833..225Z},
and we suspect that this is the origin of the errors
in the effective temperatures inferred from $\hbeta$~and $\hgamma$.
Test calculations on \marcs~model atmospheres 
that were both enhanced and not enhanced in $\upalpha$-element abundances
\citep{2008A&amp;A...486..951G} support this hypothesis.
These tests
indicated significant differences that go in the 
opposite directions for $\halpha$,
as for $\hbeta$~and $\hgamma$, owing to 
the temperature structure in the $\upalpha$-poor model atmospheres
being shallower in the region $0.0\lesssim\lgr\lesssim1.0$~(resulting
in weaker $\hbeta$~and $\hgamma$~line wings),
and steeper in the region $-0.5\lesssim\lgr\lesssim0.0$~(resulting
in stronger $\halpha$~line wings),
compared to that in the standard-composition model atmospheres.
Moreover, the residuals in Figs~\ref{figure_fit2}~and
\ref{figure_fit3}~illustrate that small changes
to the model $\hbeta$~and $\hgamma$~fluxes have
a large influence on the inferred effective temperature,
at least compared to $\halpha$~in \fig{figure_fit1}.
To confirm this hypothesis quantitatively,
one would need to recalculate the 3D model atmospheres
using a custom chemical composition, and repeat the analysis;
this is beyond the scope of the present work.

With the majority of analyses today still based on 
1D LTE models, it is interesting to 
briefly consider how our 1D LTE $\halpha$~models
fare in comparison to our 3D non-LTE models.
For $\halpha$, \fig{figure_results} illustrates that
the 1D LTE models tend to underestimate the effective temperatures.
This is most apparent for the warmer benchmark stars.
As we discussed in \sect{discussion_effects}, the 
3D effects
and the
non-LTE effects
go in the same direction
at higher effective temperatures and lower metallicities;
we thus find that the 1D LTE models perform worst
for the metal-poor turn-off HD 84937, 
underestimating 
the effective temperature by around
$150\,\mathrm{K}$~compared
to the IRFM method \citep{2011A&amp;A...530A.138C}.
This highlights the importance of adopting the full 3D non-LTE approach
in this regime.

%-------------------------------------------------------------------------------
\section{Accessing the grid}
\label{access}

We make publicly available the 3D non-LTE model spectra 
calculated on the \stagger-grid nodes
for the astronomy community to use
for effective temperature determinations. 
The data can be accessed at an online
repository\footnote{\url{https://zenodo.org/record/1288078}},
or by contacting one of the authors directly.
We also make our interpolation routines 
(\sect{benchmark_method_interpolation}) available.
The grids span $\halpha$~through to $\hgamma$.
However, analyses of $\hgamma$~are usually prone
to uncertainties caused by neighbouring blends
(\sect{discussion_diagnostics}), 
so we recommend giving most weight to
$\halpha$~and $\hbeta$~in practice.

%-------------------------------------------------------------------------------
\section{Conclusion}
\label{conclusion}

We carried out 3D non-LTE radiative transfer calculations
for \ion{H}{I}/\ion{H}{II}~on the extensive \stagger~grid of
3D hydrodynamic model atmospheres.
We used these calculations to study Balmer line formation in the context 
of effective temperature determinations of late-type stars.
We summarise our main findings below.

The absolute 3D effects,
as quantified by the 1D non-LTE versus 3D non-LTE differences,
are typically around $50\,\mathrm{K}$~for $\halpha$,
and can reach around $100\,\mathrm{K}$.
The differences tend to be negative: the inner wings of $\halpha$~are
significantly weaker in the 3D models compared to in the 1D models.
The 3D effects
tend to become more severe and 
more positive for higher 
members of the Balmer series
and for higher values of mixing-length;
they can reach around $+200\,\mathrm{K}$~for $\hgamma$~when
$\alpha_{\mathrm{MLT}}=2.0$.

The absolute 
non-LTE effects, 
as quantified by the 3D LTE versus 3D non-LTE differences,
are typically around $50\,\mathrm{K}$~for $\halpha$,
and can also reach around $100\,\mathrm{K}$.
The non-LTE effects become more severe towards lower metallicities.
The signs of the non-LTE effects
are sensitive to the effective temperature.
At higher effective temperatures the 
non-LTE effects
tend to be negative;
the LTE models of $\halpha$~usually underestimate the effective temperature.
The non-LTE effects
become less significant for higher 
members of the Balmer series.

At higher effective temperatures and lower metallicities
the 3D effects and 
non-LTE effects go in the same direction.
Consequently, 1D LTE models of $\halpha$~can underestimate
the effective temperatures of metal-poor turn-off stars
such as HD 84937 by around $150\,\mathrm{K}$.

The ab initio 3D non-LTE 
model spectra are
generally able to reproduce the effective temperatures
of various benchmark stars to within the $1\sigma$~uncertainties
in the reference effective temperatures.
The solar analysis suggests that 
the error in the 3D non-LTE model spectra
are only of the order $50$-$65\,\mathrm{K}$~in
terms of the inferred effective temperature.

As demonstrated here, the use of 1D Balmer line profiles can lead
to significant systematic errors. 
We therefore provided 3D non-LTE model spectra
(\sect{access}) for the astronomy community to use 
to determine more reliable spectroscopic effective temperatures
of late-type stars.

%-------------------------------------------------------------------------------

\begin{acknowledgements}
We thank the anonymous
referee for their careful reading of and helpful suggestions 
on the manuscript, and
Andreas Korn for providing the FOCES spectra used in this work.
AMA and KL acknowledge funds from the Alexander von Humboldt Foundation in
the framework of the Sofja Kovalevskaja Award endowed by the Federal Ministry of
Education and Research, and KL also
acknowledges funds from the Swedish Research Council 
(grant 2015-004153) and Marie Sk{\l}odowska Curie Actions 
(cofund project INCA 600398).
TN and MA acknowledge funding from the Australian Research 
Council (grant DP150100250), and MA 
also acknowledges funding through ARC Laureate Fellowship (FL110100012).
Parts of this research were conducted by the 
Australian Research Council Centre of Excellence for 
All Sky Astrophysics in 3 Dimensions (ASTRO 3D),
through project number CE170100013.
PSB acknowledges financial support from the Swedish
Research Council and the project grant ``The New Milky Way'' 
from the Knut and
Alice Wallenberg Foundation.  
Funding for the Stellar Astrophysics Centre is provided by The
Danish National Research Foundation (grant DNRF106). 
This work made use of data from the UVES Paranal Observatory Project
(ESO DDT Program ID 266.D-5655).
This work was supported by computational resources provided by the Australian
Government through the National Computational Infrastructure (NCI)
under the National Computational Merit Allocation Scheme.
\end{acknowledgements}

%-------------------------------------------------------------------------------

\bibliographystyle{aa} 
\bibliography{/Users/ama51/Documents/work/papers/allpapers/bibl.bib}
%\bibliography{./bibl}

%-------------------------------------------------------------------------------

\label{lastpage}
\end{document}